\documentclass[useAMS,usenatbib]{mnras} 
\usepackage{amsmath}
\usepackage{graphicx} 
\usepackage{aas_macros}
\usepackage[usenames]{color} 

\usepackage{float} 
 
\title[Period spacings of 22 slowly-rotating $\gamma$ Dor stars]
{Period spacings of $\gamma$ Doradus pulsators in the \textit{Kepler} field: detection methods and application to 22 slow rotators}\label{title}
\author[]{Gang Li$^{1,2}$, Timothy R. Bedding$^{1,2}$, Simon J. Murphy$^{1,2}$, Timothy Van Reeth$^{1,2}$, 
\newauthor{Victoria Antoci$^2$, Rhita-Maria Ouazzani$^3$} 
\\
$^1$Sydney Institute for Astronomy (SIfA), School of Physics, 2006 University of Sydney, Australia\\
$^2$Stellar Astrophysics Centre, Department of Physics and Astronomy, Aarhus University, Ny Munkegade 120, DK-8000 Aarhus C, Denmark\\
$^3$LESIA, Observatoire de Paris, PSL Research University, CNRS, Sorbonne Universit\'{e}s, UPMC Univ. Paris 06, Univ. Paris Diderot, \\
~~Sorbonne Paris Cit\'{e}, 5 place Jules Janssen, 92195 Meudon, France
}

\date{Accepted XXX. Received YYY; in original form ZZZ}

\pubyear{2017}
 
\def\LaTeX{L\kern-.36em\raise.3ex\hbox{a}\kern-.15em 
 T\kern-.1667em\lower.7ex\hbox{E}\kern-.125emX} 
 
\begin{document} 
 
\setlength{\voffset}{-0.3in}
\maketitle 
 
\begin{abstract}\label{abstract}
In $\gamma$ Doradus stars, the g-mode period spacing shows an approximately linear relation with period. The slope is a new asteroseismic diagnostic, related to the rotation rate and the azimuthal order $m$. 
We report two automated methods, the `moving-window Fourier transform' and the `cross-correlation', to detect and measure the period spacings based on four-year light curves from the \textit{Kepler} satellite. The results show that the cross-correlation method performs better at detecting the period spacings and their slopes.
In this paper, we apply our method to 22 $\gamma$ Dor stars with g-mode multiplets split by rotation. The rotation periods are similar to the g-mode period spacings, causing the multiplets to overlap. To clarify the overlapping patterns, we use the \'{e}chelle diagram and introduce a `copy-shift' diagram to discern and measure the splittings. The first observational relation between slopes and splittings is shown. The slope deviates from zero when the splitting increases, as the theory predicts. We found that what appears to be rotational splittings in two stars is in fact caused by two nearly-identical overlapping patterns from binaries.
\end{abstract} 
 

\section{INTRODUCTION}
Gravity modes provide a successful way to study the internal structure of stars including the chemical gradient, differential rotation from the core to the surface, and angular momentum transport \citep[e.g.][]{2008MNRAS.386.1487M,2013MNRAS.429.2500B,2018arXiv180109228O}. The applications to white dwarfs and subdwarf B stars have been successful due to the suitable frequency range for both ground and space telescopes \citep[e.g.][]{SuJie2014, reedsdb2011}. However, detailed analysis of g modes in main-sequence stars (Slowly Pulsating B stars and $\gamma$ Doradus stars), only became available recently \citep[e.g.][]{2015ApJ...803L..25P, 2017A&A...598A..74P, 2014MNRAS.444..102K, 2017ApJ...851...39G, 2015ApJS..218...27V}, with great help from the 4-yr \textit{Kepler} mission \citep{2010Sci...327..977B}.

Gamma Doradus ($\gamma$\,Dor) stars are a class of pulsating variables \citep[e.g.][]{1994MNRAS.270..905B, 1999PASP..111..840K}. They have masses from about $1.3~\mathrm{M_\odot}$ to $2.0~\mathrm{M_\odot}$ and spectral types A7-F5 and are located near the red edge of the $\delta$ Sct instability strip in the Hertzsprung-Russell (H-R) diagram. The mass range covers the boundaries between low-mass stars possessing radiative cores with convective envelopes and high-mass stars which have convective cores and radiative envelopes.

Stellar rotation affects the evolution of all stars but is still poorly understood, such that it is usually neglected in 1D stellar computations. Due to the lack of magnetic braking, rapid rotators are typical among the intermediate-mass main-sequence stars \citep{2007A&A...463..671R}. F-type stars have typical projected rotation velocities from 50\,km\,s$^{-1}$ to more than 100\,km\,s$^{-1}$
and some A-type stars rotate very rapidly, with $v\sin i \approx 120 ~\mathrm{to}~ 300 \mathrm{\,km\,s^{-1}}$  \citep{1982PASP...94..271F,1996A&AS..118..545G}.
$\gamma$ Dor stars occupy these spectral types hence many should rotate very rapidly, and an accurate stellar model must account for this. The rotational frequency can be close to the pulsation frequency and to the Keplerian break-up rotation rate $\Omega_\mathrm{k}=\left( \mathrm{G}M/R_\mathrm{eq}^3\right)^{1/2}$, and rotation can therefore change the pattern of period spacings \citep[e.g.][]{2017A&A...598A.105P}. 

The pulsations in $\gamma$ Dor stars are mainly gravity (g) modes, which have buoyancy as the restoring force. The modes are driven by the convective flux blocking at the bottom of the convective envelope \citep{2000ApJ...542L..57G, 2004A&A...414L..17D, 2005A&A...435..927D}. They probe the internal stellar structure down to the edge of the convective core. The pulsations have typical periods between 0.3\,d and 3\,d, making them difficult to study from the ground. \cite{1979PASJ...31...87S} deduced that g modes have period spacings $(\Delta P\equiv P_{n+1,l}-P_{n,l})$ that are constant in a spherical chemically homogeneous non-rotating star if $l\ll n$, where $l$ is the degree and $n$ is the radial order \citep[e.g][]{2018MNRAS.475..359B}. However, breaking of either of these two assumptions, homogeneity and no rotation, results in deviations from the asymptotic formula. A more complete relation was deduced by \cite{2008MNRAS.386.1487M}. The authors found that a chemical gradient at the edge of the convective core led to a sharp variation (glitch) of the buoyancy frequency and is responsible for a series of dips in the period spacing pattern. Meanwhile, a homogeneous star, such as a pre-main sequence star, generally shows a regular period spacing \citep{2011A&A...531A.145B}.

To take the rotational effect into account, the first-order perturbative approach can be used in slowly-rotating stars \citep{1951ApJ...114..373L}. The frequency splitting is $\delta \nu=m\left(1-C_{n,l}\right)\Omega$, where $n$ is the radial order, $l$ is the degree, $m$ is the azimuthal order, $\Omega$ is the rotational frequency, and $C_{n,l}$ is the Ledoux constant. \cite{2010A&A...518A..30B} discussed the validity of the perturbative approach. They found that the first-order perturbative methods are suited to compute the gravity mode frequencies up to rotation speeds of $\sim50\,\mathrm{km\,s^{-1}}$ for typical $\gamma$ Dor stars within the $CoRoT$ frequency precision \citep{2006cosp...36.3749B}. The first-order approach is less valid when we take into account the very good frequency accuracy of the \textit{Kepler} data.  

Another method to treat the rotational effect is the traditional approximation of rotation \citep[TAR; e.g.][]{2003MNRAS.340.1020T}. Under the TAR, the period spacings are related to the variable eigenvalues from the Laplace Tidal Equation. The period spacing $\Delta P$ is generally a linear function of period $P$ in the inertial reference frame, and the slope $\Sigma\equiv \mathrm{d}\Delta P/\mathrm{d}P$ of a spacing pattern depends on the azimuthal order $m$. 
The retrograde modes (here with $m<0$) generally show upward patterns (i.e. $\Sigma>0$) while patterns of prograde ($m>0$) and zonal modes are downward \citep[][]{2013MNRAS.429.2500B}. 
The slope $\Sigma$ strongly correlates with the internal rotation. \cite{2017MNRAS.465.2294O} established the quantitative relation between $\Sigma$ and the internal rotational rate, and measured it successfully in several \textit{Kepler} stars, allowing them to determine the internal rotation rates.

The complex pattern of sloping period spacings with glitches makes the automated detection of period spacings in $\gamma$ Dor stars challenging. 
About 70 examples have already been presented \citep[e.g.][]{2015EPJWC.10101005B,2015ApJS..218...27V,2018arXiv180703707C,2018arXiv180603586V}, and a handful of objects have been analysed in detail \citep[see,][]{Chapellier 2012, 2014MNRAS.444..102K,2015MNRAS.447.3264S,2016MNRAS.459.1201M, 2015MNRAS.454.1792K, 2015A&A...584A..35S}.
To extract the slopes and period spacings of a much larger sample, we developed an automated method. Section~\ref{sec:LIGHT CURVE REDUCTION} presents the light curve reduction and the frequency extraction. Section~\ref{sec:wavelet} introduces the wavelet algorithm, which was only partly successful. Section~\ref{sec:algorithm} gives the details of the cross-correlation method and an MCMC algorithm. The algorithm in Section~\ref{sec:algorithm} works better, so we use the cross-correlation method to detect the pulsation patterns. In Section~\ref{sec: results}, the detection results are reported. We present two stars as examples to show how our code works, and then implemented it on 22 stars which have rotational splittings. For each star showing rotational splittings, we investigated their slopes and splittings to seek observational evidence that the rotation changes the period-spacing pattern.

\section{LIGHT CURVE REDUCTION}\label{sec:LIGHT CURVE REDUCTION}
We used 4-year \textit{Kepler} long-cadence (LC; 29.45-min sampling) light curves from the multi-scale MAP data pipeline \citep{2014PASP..126..100S}. 
In each quarter, the light curve was divided by a second-order polynomial fitting to remove the slow trend. The outliers were also deleted if the difference to the polynomial fitting was larger than $5\sigma$. We calculated the Fourier transforms (FTs) of the light curves to get the relation between the fractional amplitude and the frequency. The data have a remarkable frequency resolution of $\delta f=$0.00068\,d$^{-1}$, suitable for analysis of the g modes. 
Because g modes are asymptotically equally spaced in period, rather than frequency, we inverted the frequency domain to plot the FTs against period (in days), with the variable period resolution of $\delta P=P^2\delta f$. The period resolution is worse in the long-period region, and the pulsation period for prograde modes in the inertial frame can be shorter due to the fast rotation, hence we investigated the patterns of g modes between 0.2\,d and 2\,d, corresponding to the period resolution from 2.42\,s to 242\,s. 

We used an iterative prewhitening method to extract frequencies. 
In each iteration, the FT was calculated and the initial frequency corresponding to the largest amplitude was selected. We used a sine function with the initial frequency and amplitude to fit the light curve, in order to optimize the frequency in the time domain. Then the fitting residual became the input light curve in the next iteration. 
The signal to noise ratio (S/N) is defined as the amplitude divided by the median value of the amplitude spectrum within $0.2$\,d$^{-1}$. We extracted peaks until the S/N was smaller than 3, instead of 4 from the standard criterion \citep{1993A&A...271..482B}. When substracting high-amplitude peaks, the background level decreases significantly because the spectral window is removed along with the peak itself. The S/N was calculated based on the final noise level.

The frequency uncertainty is given by \begin{equation}
\sigma\left(f\right)=0.44\frac{\langle a \rangle}{a}\frac{1}{T},\label{equ:freq_uncertainty}
\end{equation}where $\langle a \rangle$ is the noise level in the amplitude spectrum, $a$ is the amplitude of peak, and $T$ is the total time span of the data \citep{1999DSSN...13...28M,2003Ap&SS.284....1K}. For the \textit{Kepler} data, $T\simeq1470$\,d. We extracted the peaks down to S/N=3, so the frequency uncertainty is smaller than $1\times10^{-4}\,\mathrm{c/d}$. The peaks with low S/N are still convincing if they follow clear pulsation patterns.

\section{WAVELET-LIKE ANALYSIS}\label{sec:wavelet}
\begin{figure}
\includegraphics[width=1\linewidth]{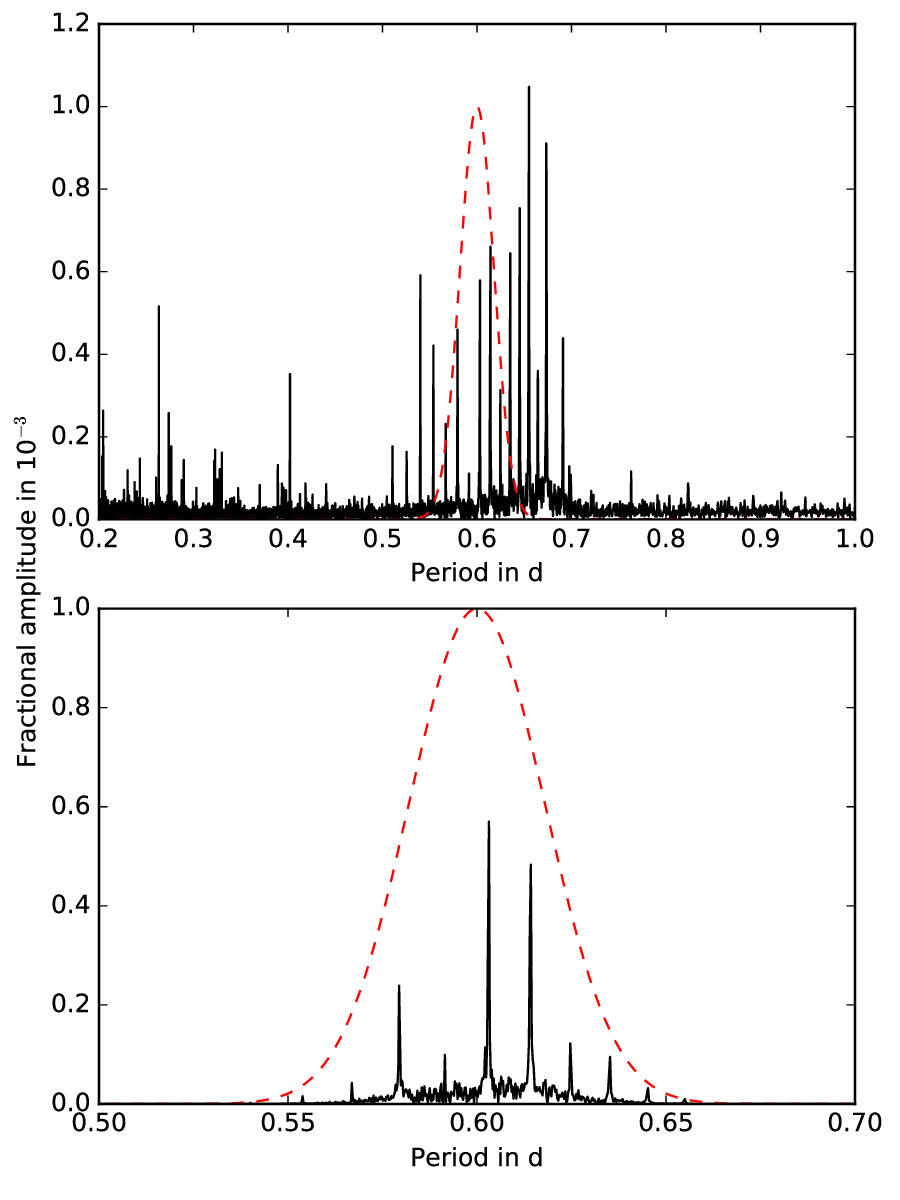}
\caption{Schematic explanation of the wavelet algorithm. KIC\,4253413 is used as a model. Top: the amplitude spectrum $O(P)$ (solid lines) and the window $G\left(P_c, \sigma, P \right)$ (dashed lines). Bottom: the product of spectrum and window. }\label{fig:wavelet_explanation}
\end{figure}

In this section, we briefly describe our attempt to use the moving-window Fourier transform to detect period spacing patterns. The algorithm is based on the quasi-linear variation of g-mode period spacings $\Delta P$ with period. Therefore, we can extract the quasi-periodic signal $\Delta P$ from peaks by calculating the FT in a moving window.

A Gaussian function $G\left(P_c, \sigma, P \right)=\exp\left[ -\frac{1}{2}\left( \frac{P-P_c}{\sigma} \right)^2 \right]$ was used as the window, which was specified by central period $P_c$ and the width $\sigma$. We calculated the discrete Fourier Transform (DFT) of the product of the window $G\left(P_c, \sigma, P \right)$ and the amplitude spectrum $O(P)$. This wavelet spectrum represents the periodic intensity $I(P_c, \Delta P)$ of the amplitude spectrum around the central period. We can inspect the periodicity by moving $P_c$. The schematic explanation of the process can be seen in Fig.~\ref{fig:wavelet_explanation}.
Windowing with a Gaussian decreases the length of the amplitude spectrum, and thereby permits period resolution, but it degrades the period spacing resolution. The width of the Gaussian must be chosen appropriately to balance these two resolutions, so we tried three choices of $\sigma$ (777\,s, 1555\,s and 3110\,s) for each star, which are 0.5\%, 1\%, and 2\% of the period span (0.2 to 2\,d). These choices cover the typical period spacing regions.

\begin{figure}
\includegraphics[width=1\linewidth]{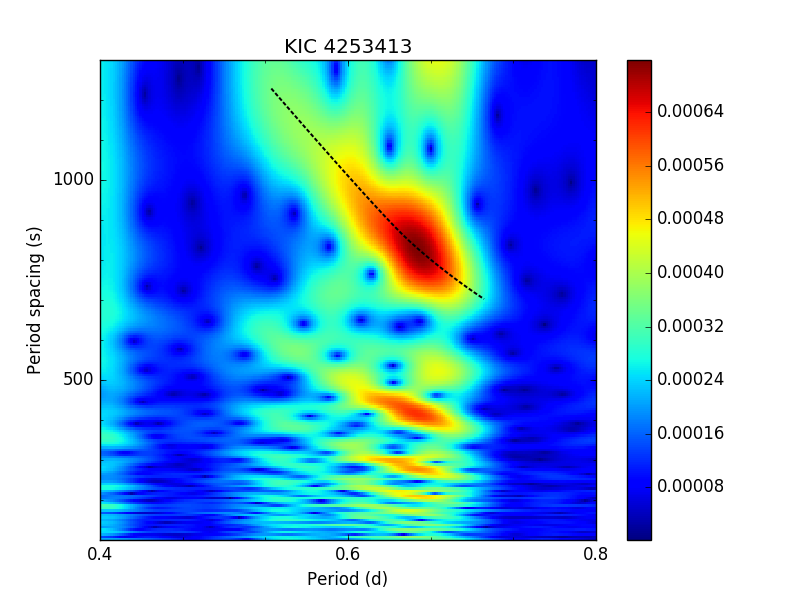}
\caption{Wavelet analysis of KIC\,4353413. The color denotes $\sqrt{I(P_c, \Delta P)}$, in which the square root is used to enhance the contrast. The significant signal appears around $P\sim$0.65\,d and $\Delta P \sim$800\,s as expected,  shown by the black dash line. Other islands of red are harmonics.}\label{fig:wavelet_results}
\end{figure}

The function $I(P_c, \Delta P)$ is two-dimensional and well represented by a heat map. Fig.~\ref{fig:wavelet_results} depicts the analysis of KIC\,4253413, for which \cite{2015EPJWC.10101005B} reported a linear period spacing pattern. Consequently, it was used as a test of our code. The width was set as $\sigma=$1555\,s. The expected signal appears around $P\sim$ 0.65\,d and $\Delta P \sim$ 800\,s seen as the dominant red area marked by the black dash line. Several harmonics at $\Delta P/2$, $\Delta P/3$ etc. are also seen. The red areas have a downward trend, indicating that the period spacing decreases with period. 

However, the disadvantage of the moving-window Fourier transform emerged after testing more targets: the resolution of period spacing is low, and the slope cannot be derived directly. In the next section, we will introduce the cross-correlation analysis, which shows a higher sensitivity and accuracy.

\section{CROSS-CORRELATION ANALYSIS}\label{sec:algorithm}

In this section, we describe an algorithm based on cross-correlation and MCMC to detect variable period spacing patterns. The main idea is to build a template pattern and calculate its product with the amplitude spectrum. After maximising the product, the parameters of the template reveal the period spacing and the slope. This method can give the period spacing and slope directly and accurately, minimising the impact of the missing peaks. This method is similar to the `comb response' algorithm used in the analysis of solar-like oscillations \citep[e.g.][]{1995AJ....109.1313K}. The aim of this work is to automatically identify $\gamma$ Dor stars that have regular patterns. Once they are found, a more detailed analysis can be done on individual stars.

\subsection{Combination frequencies}

\begin{figure}
\centering
\includegraphics[width=\linewidth]{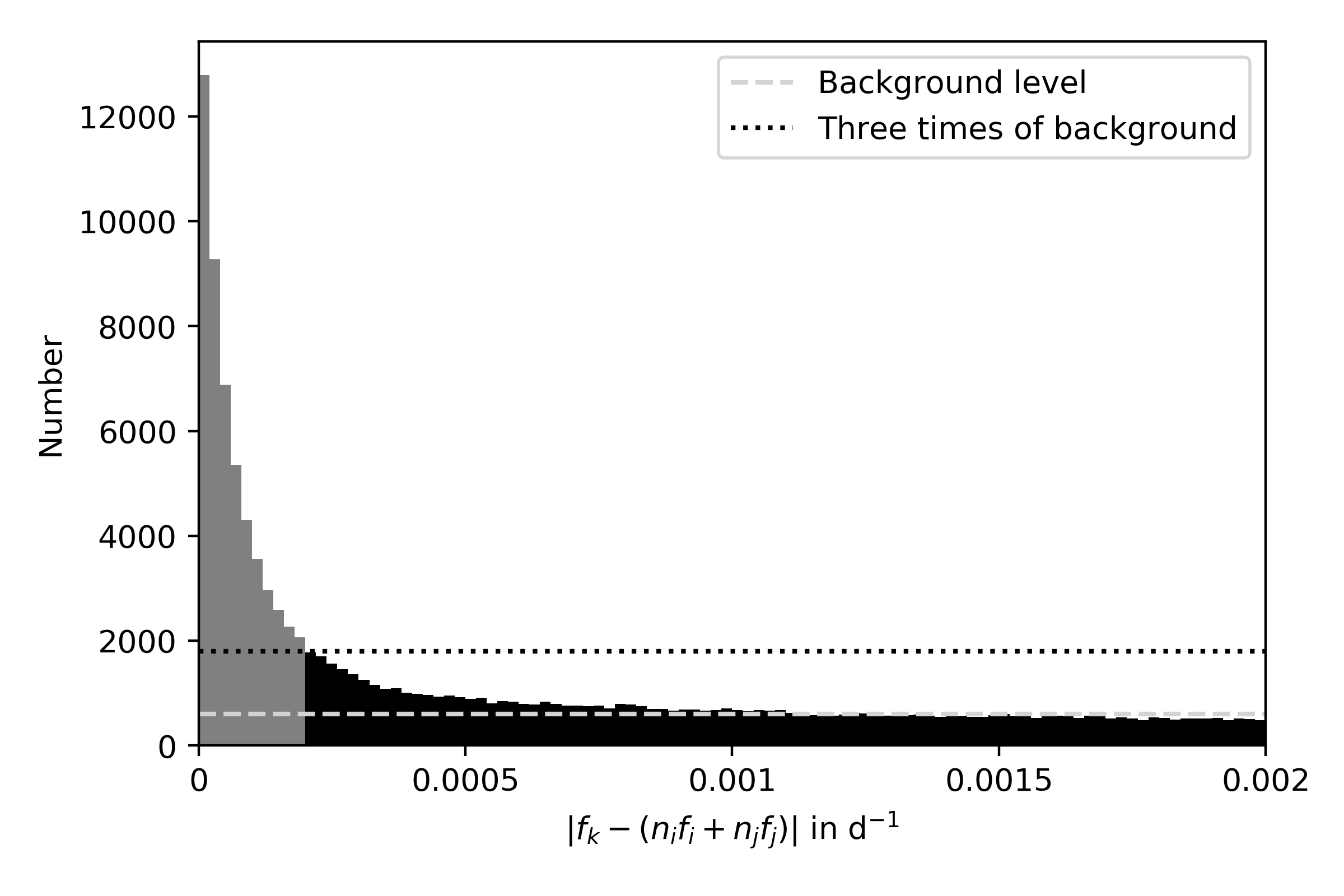}
\caption{The distribution of frequency differences $| f_{k}- \left( n_{i} f_{i}+n_{j} f_{j}\right) |$. The excess on the left is evidence of real combination frequencies over random matches. The percentage of the real combination frequencies over the random background in the grey area is larger than 67\%. }\label{fig: freq_difference_distribution}
\end{figure}

Combination frequencies are often seen in $\gamma$ Dor stars, which complicates the detection of period spacing patterns \citep[see an example in ][]{2015MNRAS.454.1792K}. 
So it is helpful to first remove these combinations. They can be identified using
\begin{equation}
| f_{k}- \left( n_{i} f_{i}+n_{j} f_{j}\right) |<\varepsilon,
\end{equation}
where $i$, $j$, $k$ are indices of peaks, $f_{i}$ and $f_j$ are parent frequencies, $f_k$ is the combination frequency candidate, $n_i$ and $n_j$ are the combination coefficients, and $\varepsilon$ is the criterion threshold. For each star, we considered the 20 highest peaks as the parent frequencies and only considered combinations where $|n_i|+|n_j|\leq2$. We show the distribution of the difference $| f_{k}- \left( n_{i} f_{i}+n_{j} f_{j}\right) |$ from 1371 \textit{Kepler} stars in Fig.~\ref{fig: freq_difference_distribution}. The distribution in Fig.~\ref{fig: freq_difference_distribution} has grey excess over the background. The background shown by the grey dashed line reflects the probability of the random match. The excess is most likely caused by the real combination frequencies. The black dotted line shows three times the background, intersecting the histogram at $0.0002\,\mathrm{d^{-1}}$. That means the probability of that a match is a genuine combination frequency in the grey area is at least 67\%. Based on this, we selected $\varepsilon=0.0002\,\mathrm{d^{-1}}$ as the threshold.

\begin{figure}
\centering
\includegraphics[width=0.5\textwidth]{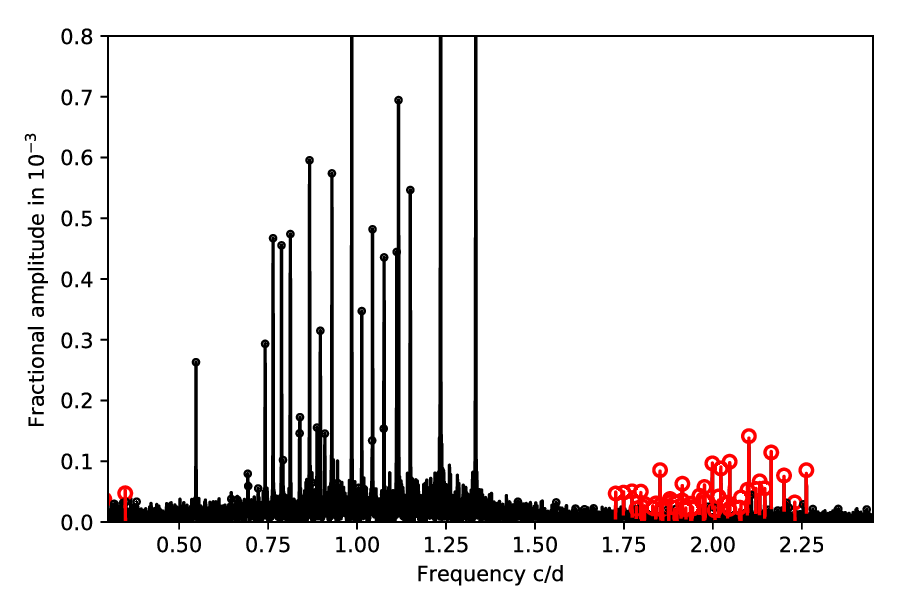}
\caption{The g-mode amplitude spectrum of KIC\,2450944. The solid dots denote the frequencies detected from prewhitening. The red lines and open circles stand for the frequencies which are combinations. Mode identification can be seen in Fig.~\ref{fig:KIC 2450944}. }\label{fig: combinations_example}
\end{figure}

Fig.~\ref{fig: combinations_example} presents the amplitude spectrum of KIC\,2450944, in which the likely combinations are plotted by red open circles and red lines. The peaks show two main `frequency groups', the left one is dominated by the parent frequencies while the right one is mainly composed of the combinations. However, because of the large number of detected frequencies, there is a chance that independent modes can also satisfy the combination criterion \citep{papics 2012}. As pointed out by \cite{Saio_com_real_2018}, this often happens in rapidly rotating stars where successive frequency groups can be prograde sectoral modes ($m=l$) of increasing degree, and make the detection of combinations tricky. The conclusions of this paper are not affected since the identification of likely combinations is only done to improve the process of correlating with the template. In the final analysis, we look at the overall patterns to decide which peaks are genuine modes.

After prewhitening the combination frequencies, the spectra were ready to be inspected for g-mode patterns.

\subsection{Mathematical formulaism of period spacings}\label{subsec:math}
In our work, the main goal is to measure the rate of change of the period spacing with respect to period, which we call the `slope'. 
In the observations, the signature of chemical gradients in the stellar interior is minor compared to that of the stellar rotation among published examples \citep[e.g.][]{2015ApJS..218...27V}. Even if there is a dip, the pattern can be recognised as two patterns separated by the dip. Hence we do not include dips caused by chemical gradients in our template. 

The asymptotic formula gives the relation between the period and radial order \citep{1979PASJ...31...87S, 1980ApJS...43..469T}
\begin{equation}
 P_n=\Delta P \left(n+\epsilon_g\right). \label{equ: traditional_asymptotic}
\end{equation}
In that case the slope is zero since the period spacing is constant. 
However, we assume that the period spacing $\Delta P$ depends linearly on the period $P$, then the relation between the period spacing and the period $P$ can be written as 
\begin{equation}
 \Delta P=\Sigma \cdot P+b,\label{equ:linear}
\end{equation}
in which $\Sigma$ is the slope and $b$ is the intercept. 

Now we present a new formula to calculate $P$ after eliminating $\Delta P$. The intercept $b$ in Equation~\ref{equ:linear} can be determined from one particular period $P_0$ and its corresponding period spacing $\Delta P_0$ so that the linear relation can be written as
\begin{equation}
 \Delta P=\Sigma \cdot P+\Delta P_0-\Sigma \cdot P_0\label{equ:linear_relation}
\end{equation}

Two adjacent period values $P_i$ and $P_{i+1}$ are related as
\begin{equation}
P_{i+1}=P_i+\Delta P_i=\left(1+\Sigma\right)P_i+\Delta P_0-\Sigma \cdot P_0\label{equ:iteration}, 
\end{equation}
in which $i$ is an integer. Therefore the analytic expressions of $\Delta P_i$ and $P_i$ are derived as
\begin{equation}
 \Delta P_i=\left(1+\Sigma\right)^i\Delta P_0 \label{equ:delta P_i}
\end{equation}
and
\begin{equation}
 P_i=\sum_{j=0}^{i-1} \Delta P_j+P_0=\Delta P_0\cdot \frac{(1+\Sigma)^i-1}{\Sigma}+P_0.\label{equ: p_i}
\end{equation}

Equation~\ref{equ:delta P_i} reveals that the series of period spacings varies as a geometric sequence with the common ratio of $1+\Sigma$, which is a new meaning of $\Sigma$.

Although Equation~\ref{equ: p_i} is invalid when $\Sigma=0$, which corresponds to the non-rotating case, the limit is still valid 
$$ P_i=\lim_{\Sigma \to 0} \Delta P_0\cdot \frac{(1+\Sigma)^i-1}{\Sigma}+P_0=i\cdot \Delta P_0+P_0.$$
Hence, the limit when $\Sigma \to 0$ can describe the non-rotating equally-spaced case.

Equation~\ref{equ: p_i} can be written as 
\begin{equation}
 P_i=\Delta P_0\cdot \left( \frac{(1+\Sigma)^i-1}{\Sigma}+\epsilon_\mathrm{g} \right), \label{equ: general asymptotic}
\end{equation}
which is an extension of traditional asymptotic relation $P_n=\Delta P \left( n'+\epsilon_\mathrm{g} \right)$, where $n'=\frac{(1+\Sigma)^i-1}{\Sigma}$. Using Equation~\ref{equ: p_i} with given periods and their integers $i$, the first period spacing $\Delta P_0$ and the slope $\Sigma$ can be derived by fitting. The advantage is that we can fit a period series by giving the indices $i$, even when a lot of peaks are missing and the period spacing cannot be calculated directly. However, when too many peaks are missing, assignment of the series index becomes ambiguous and a unique fit is not possible.

\subsection{The Template}\label{subsec:Template}
\begin{figure}
\centering
\includegraphics[width=\linewidth]{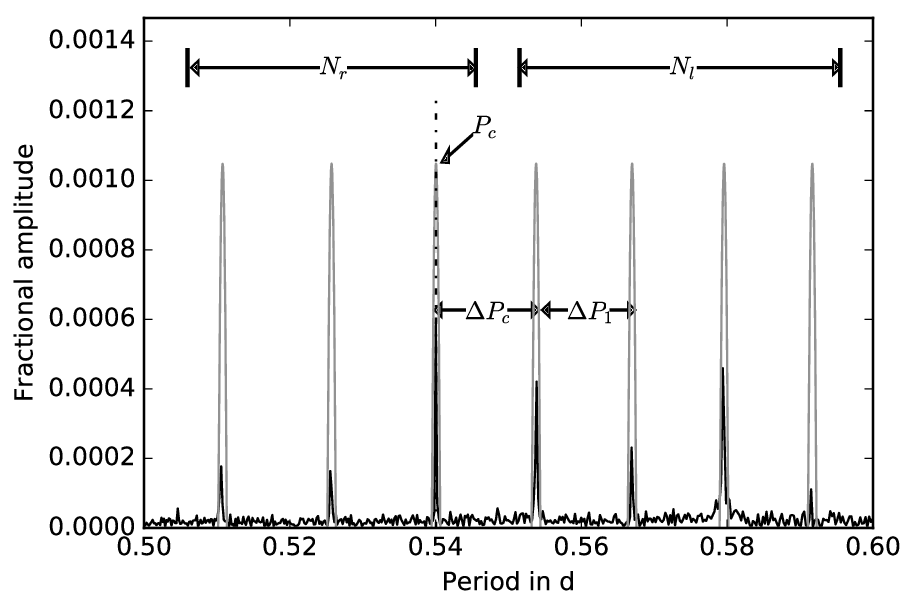}
\caption{The example of template for KIC\,4253413 (the grey lines), which best matches the power spectrum (black lines). The parameters are: $P_c=0.54$\,d, $\Delta P_c=1183$\,s, $\Sigma=\Delta P_1 / \Delta P_c-1=-0.041$, $N_l=3$, and $N_r=4$.}\label{fig:template}
\end{figure}

In order to avoid the problem of ambiguous index assignment caused by the missing peaks, we used a template to find the parameters of the pattern. Considering that the peaks in $\gamma$ Dor stars generally cluster into several groups \citep{2015MNRAS.450.3015K}, the template $T(\nu)$ was designed to detect the pattern centred around the highest amplitude peak in each group. Fig.~\ref{fig:template} shows the template. It is comprised of a series of narrow parabola functions with uniform amplitude. We chose a parabola over a more complicated function, such as a Gaussian or a sinc, to improve computation speed. Five parameters are needed to specify the template: the central period $P_c$, the period spacing $\Delta P_c$ at $P_c$, the slope $\Sigma$, the number of peaks $N_l$ on the left of $P_c$ (including $P_c$), and the number of peaks $N_r$ on the right of $P_c$ (exluding $P_c$). The full-width at zero-amplitude of each parabola is fixed as 100 s, which becomes the lower limit of the period spacing we can detect. The lower limit is deliberately much larger than the period resolution mentioned in Section~\ref{sec:LIGHT CURVE REDUCTION} because they have different meanings: the full width allows the template to tolerate some dips and scatters in the observed period spacing patterns, considering the linear relation between the period spacing and period is not strict. The amplitude of the parabola is arbitrary since it cancels in the fraction of the likelihood function (Equation~\ref{equ:likelihood_initial}).

\subsection{Cross-correlation and MCMC}\label{subsec:tmp}
The cross-correlation of the template with the observed amplitude spectrum represents the goodness of fit. If the template matches the observed power spectrum very well, the value should be large, otherwise, it will be close to zero because the peaks in the template line up with the noise in the spectrum, rather than the observed peaks. In this method, any missing peaks do not influence the results greatly.

We define the likelihood function, given the observed spectrum and five parameters, as
\begin{equation}
 L(\mathrm{data}|P_c, \Delta P_c, \Sigma, N_r, N_l)=\left[\frac{\int T(\nu) O(\nu)\, \mathrm{d}\nu}{\int T(\nu)\, \mathrm{d}\nu}\right]^{\beta}, \label{equ:likelihood_initial}
\end{equation}
where $T(\nu)$ stands for the template, $O(\nu)$ is the power spectrum in unit of frequency $\nu$, and $\beta$ is the tampering parameter. The parameter $\beta$ changes the gradient of the likelihood function, which can enhance the convergence or guide the chain to the global minimum. We tested several choices of $\beta$ and found the convergence to be best when $\beta=7$.

A special prior was designed as
\begin{equation}
 P(P_c, \Delta P_c, \Sigma, N_r, N_l)=\frac{N_r N_l}{\Delta P}.\label{equ:prior}
\end{equation}
There are two reasons for such a design: to find the fundamental period spacing rather than a harmonic, and to find longer patterns. Considering that a series of periods $\{P_i,i=0,1,2...\}$ forms a pattern that satisfies the Equation~\ref{equ: p_i}, a new series $P_i'=P_{2i}$ that takes every second point also roughly obeys the Equation~\ref{equ: p_i} with twice spacing $\Delta P'=2\Delta P$ and twice slope $\Sigma'=2\Sigma$. The prior in Equation~\ref{equ:prior} is designed to distinguish the fundamental spacing $\Delta P$ and the harmonic $\Delta P'$.
They have the same likelihood value but can be clarified by the prior.

Finally, the posterior probability, given certain parameters $(P_c, \Delta P_c, \Sigma, N_r, N_l)$, is
\begin{equation}
 P(P_c, \Delta P_c, \Sigma, N_r, N_l|\mathrm{data})=\frac{N_r N_l}{\Delta P_c}\cdot \left[\frac{\int T(\nu) O(\nu)\, \mathrm{d}\nu}{\int T(\nu)\, \mathrm{d}\nu}\right]^\beta.\label{equ:posterior}
\end{equation}

In order to enhance the convergence of MCMC chains, we set the amplitudes of all the independent peaks in the power spectrum equal to the maximum amplitude and set the amplitudes of combination frequencies and noise to be zero.
Since the peaks are generally clustered into groups, we analysed them with a 1-D $k$-means clustering algorithm \citep{Kmeans_reference, 2012book_advances_in_k_means}, which is a common and powerful clustering algorithm used in unsupervised machine learning \citep[see two applications in][]{2018MNRAS.479.5596G, 2018MNRAS.478.4416R}. We used six frequency groups and in each group we included the pattern around the highest peak if the number of independent frequencies within the group was larger than four, since there are three free parameters in Equ~\ref{equ: p_i} ($P_0$, $\Delta P_0$, and $\Sigma$).
The sampling range of the central peak was set to be $P_p\pm 3\delta P$, where $P_p$ is the highest peak in each frequency group and $\delta P$ is the period resolution. 
The ranges for the other four free parameters were $\Delta P~\mathrm{from}~100\,\mathrm{s}~\mathrm{to}~4000\,\mathrm{s}$, 
$\Sigma~\mathrm{from}~0.2~\mathrm{to}$ $-0.2$, $N_l$ and $N_r$ from 2 to 10.
The {\sc emcee} package in python was used to implement the MCMC algorithm \citep{2013PASP..125..306F}. We used 20 parallel chains and the chain length was set to be 2400. The parameters were derived after maximising the posterior. Since the posterior distributions may be multi-modal (see the example in Fig.~\ref{fig:KIC3331147distribution}), we used $k$-means clustering again in 5-dimensional space ($P_c,~ \Delta P_c, ~\Sigma,~ N_l,~ N_r$) to cluster the chain points into two groups. The medians of the group with the most points were used as the MCMC results.

The results from MCMC gave a good initial guess on the series indices $i$. However, our likelihood function is not defined by the conditional probability, so the distributions do not represent the uncertainties. To derive the uncertainties of $P_c$, $\Delta P$ and $\Sigma$, the parameters were optimized further based on minimizing $\chi^2$ between the observed period series and the prediction from the Equation~\ref{equ: p_i}, defined as
\begin{equation}
\chi^2=\sum_i \left(P_i^\mathrm{cal}-P_i^\mathrm{obs} \right)^2,
\end{equation}
where $P_i^\mathrm{cal}$ is the period calculated from the template and $P_i^\mathrm{obs}$ the observed one. The patterns were finally checked and confirmed by eye. 

\section{ANALYSIS OF \textit{Kepler} $\gamma$ DOR STARS}\label{sec: results}

\begin{figure}
	\centering
	\includegraphics[width=0.5\textwidth]{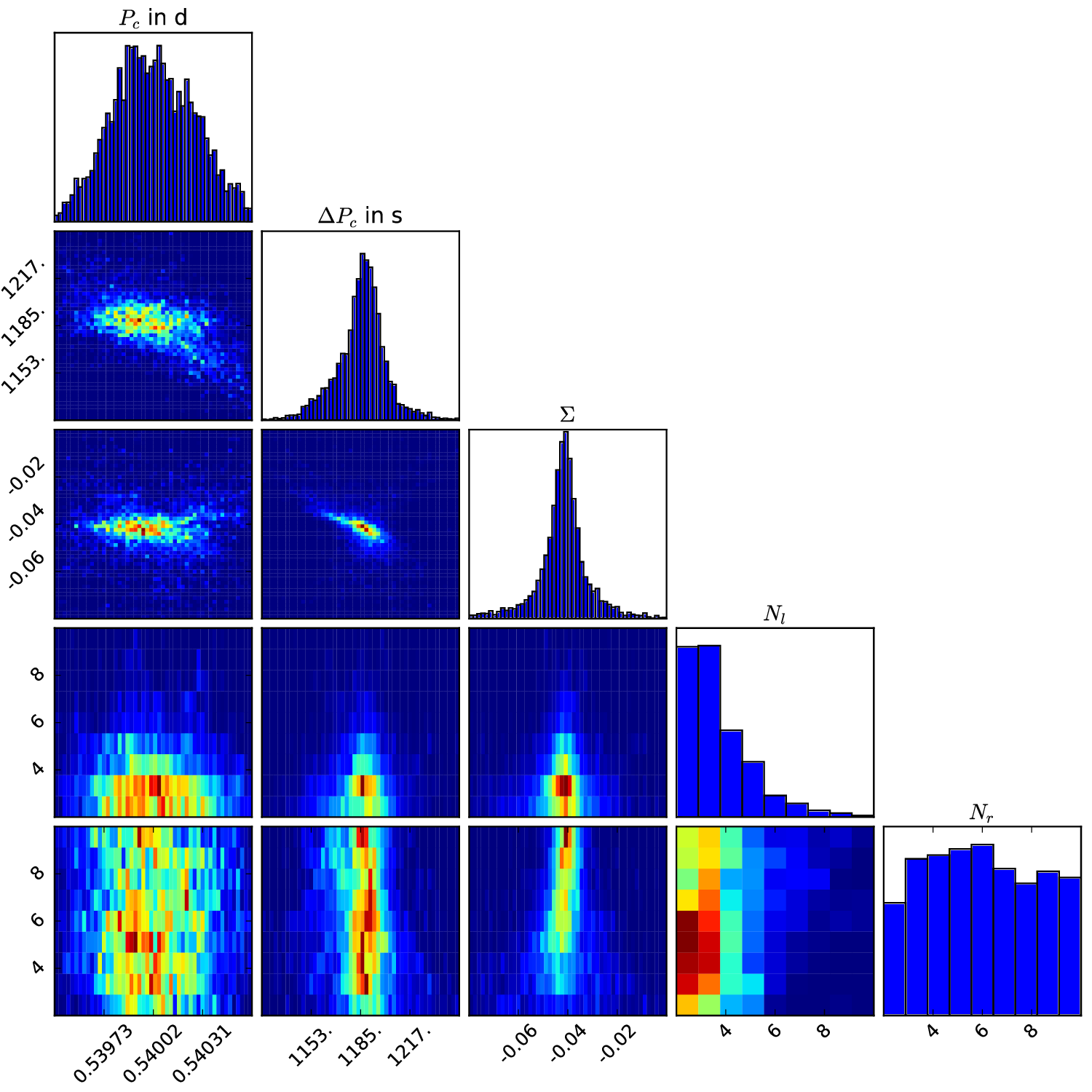}
	\caption{The posterior distributions of KIC\,4253413 based on Equation~\ref{equ:posterior}. The significant signal appears around $\Delta P\approx1100~\mathrm{s}$ and $\Sigma=-0.04$. Axis labels are given above each column. }\label{fig:distribution}
\end{figure}

\begin{figure}
	\centering
	\includegraphics[width=0.5\textwidth]{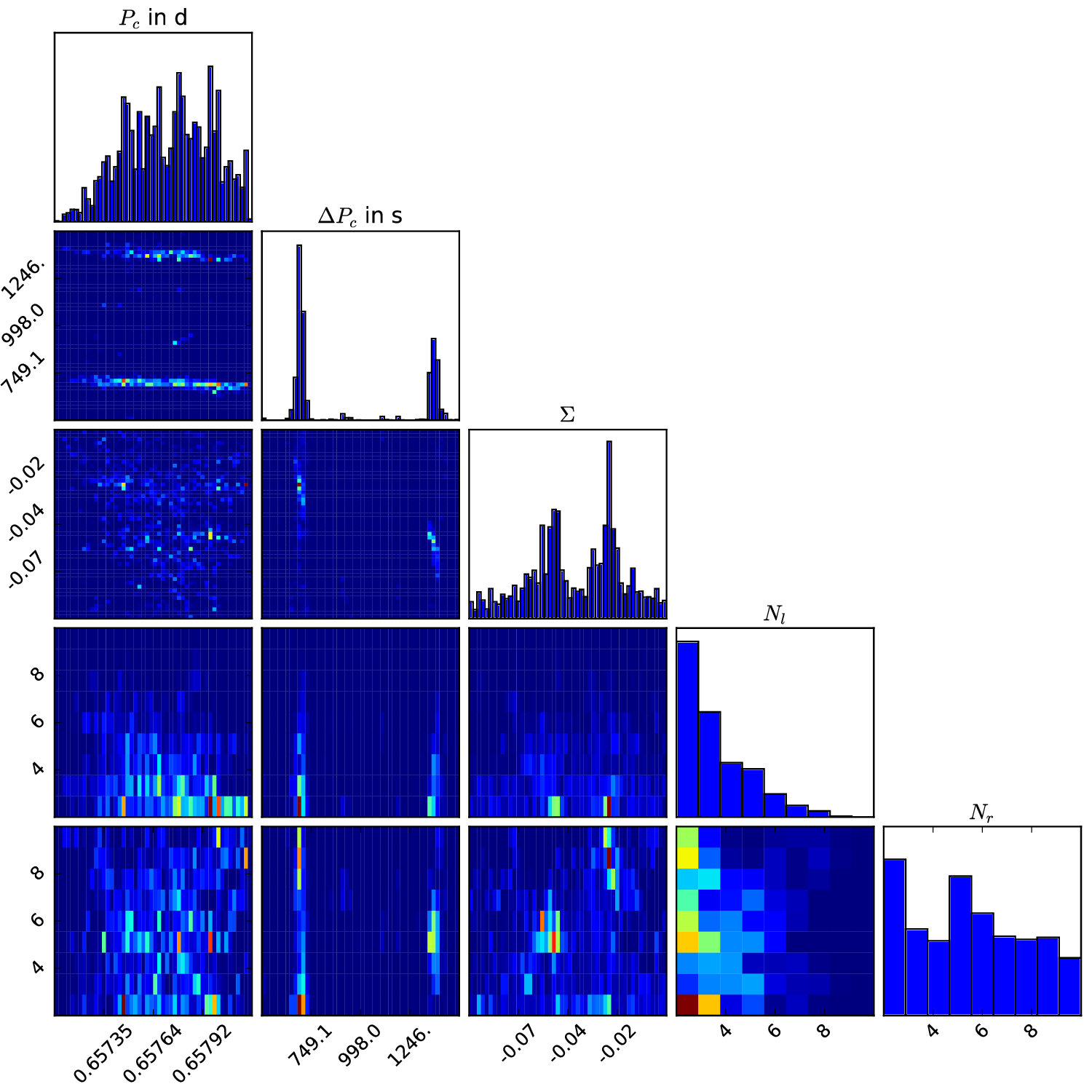}
	\caption{Same as Fig.~\ref{fig:distribution} but for KIC\,3331147. Note that there are two peaks in the frames of period spacing and the slope. The one with higher posterior ($\Delta P=740$\,s, $\Sigma=-0.025$) corresponds to the correct solution. }\label{fig:KIC3331147distribution}
\end{figure}

We visually inspected the Fourier transforms of four-year light curves of 1371 \textit{Kepler} targets with an effective temperature from 6600\,K to 10000\,K, including 339 $\delta$ Scuti binaries detected by the pulsation timing method \citep{2018MNRAS.474.4322M}. Among the binary systems, either the $\delta$ Scuti companion or its companion could show g mode pulsations, and in some cases both stars do \citep[e.g.][]{2015MNRAS.454.1792K, 2017ApJ...851...39G}. We found $\sim500$ stars showing clear period spacing patterns. We identified 22 stars in which possible rotational splittings are seen and applied the algorithm to measure the slopes. The splittings are approximately proportional to the rotational rates, therefore the relation between the slope and the rotational rates can be revealed in different $m$.

\begin{figure*}
	\centering
	\includegraphics[width=1\textwidth]{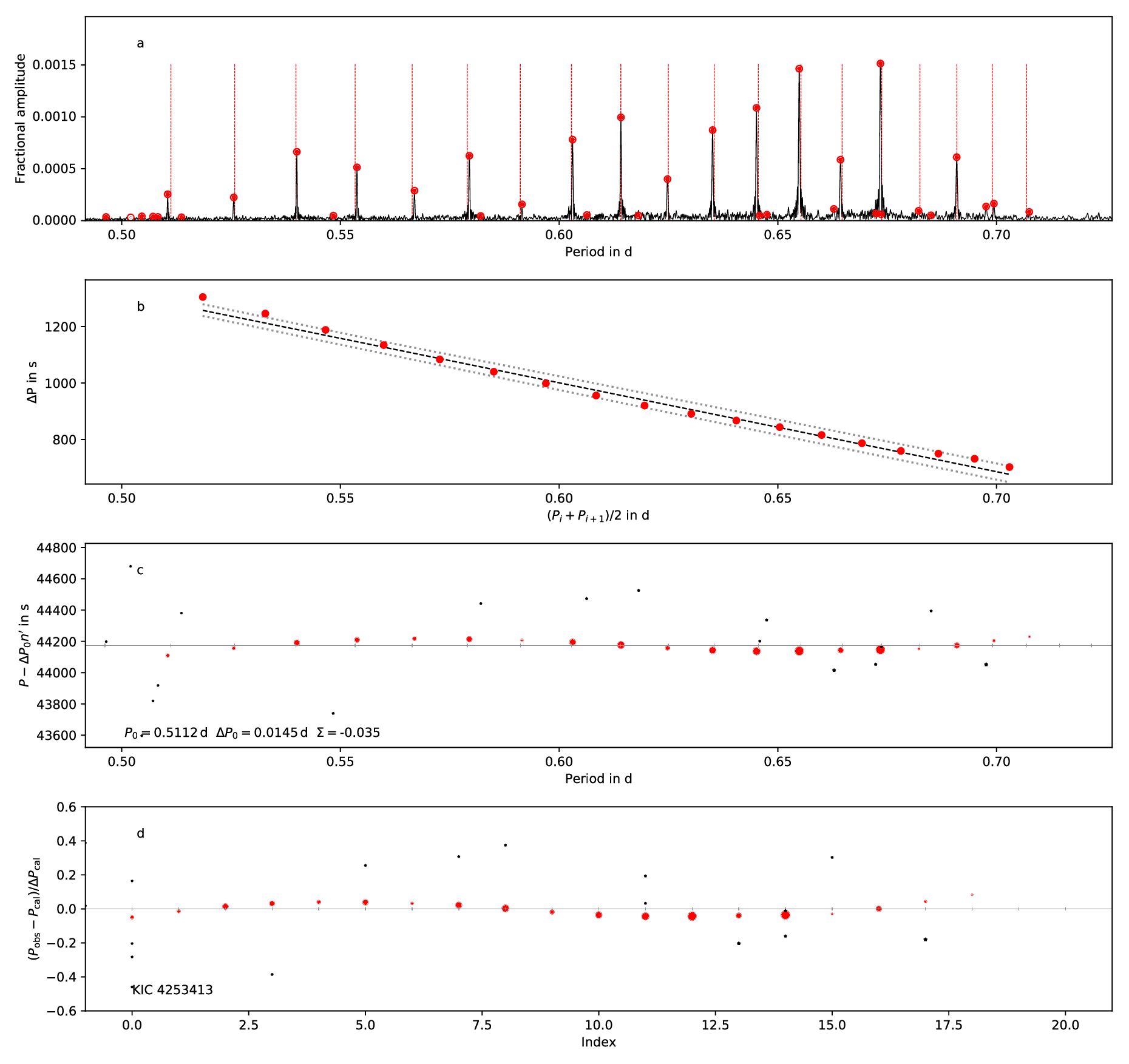}
	\caption{The amplitude spectrum and period spacing pattern of KIC\,4253413. Top panel: the amplitude spectrum. The grey dashed lines show the best-fitting results for a linearly-varying period spacing. The solid red circles are the peaks from prewhitening and the open red circles show combinations. Second panel: the period spacing pattern. The red circles are the observed data, for each circle two adjacent peaks $P_i$ and $P_{i+1}$ are detected so that $\Delta P_i=P_{i+1}-P_{i}$ can be calculated. The $x$ coordinate of each solid red circle is $\frac{P_i+P_{i+1}}{2}$. The slope of this pattern is $\Sigma=-0.0358\pm0.0006$. Third panel: sideways \'{e}chelle diagram, where circles are peaks that fit the period-spacing patterns and black stars show peaks that do not. Symbol size is proportional to peak amplitude. Fourth panel: same as the third panel, but with $x$ axis rescaled so that the points are equally spaced. }\label{fig:KIC4253413pattern}
\end{figure*}

\begin{figure*}
	\centering
	\includegraphics[width=1\textwidth]{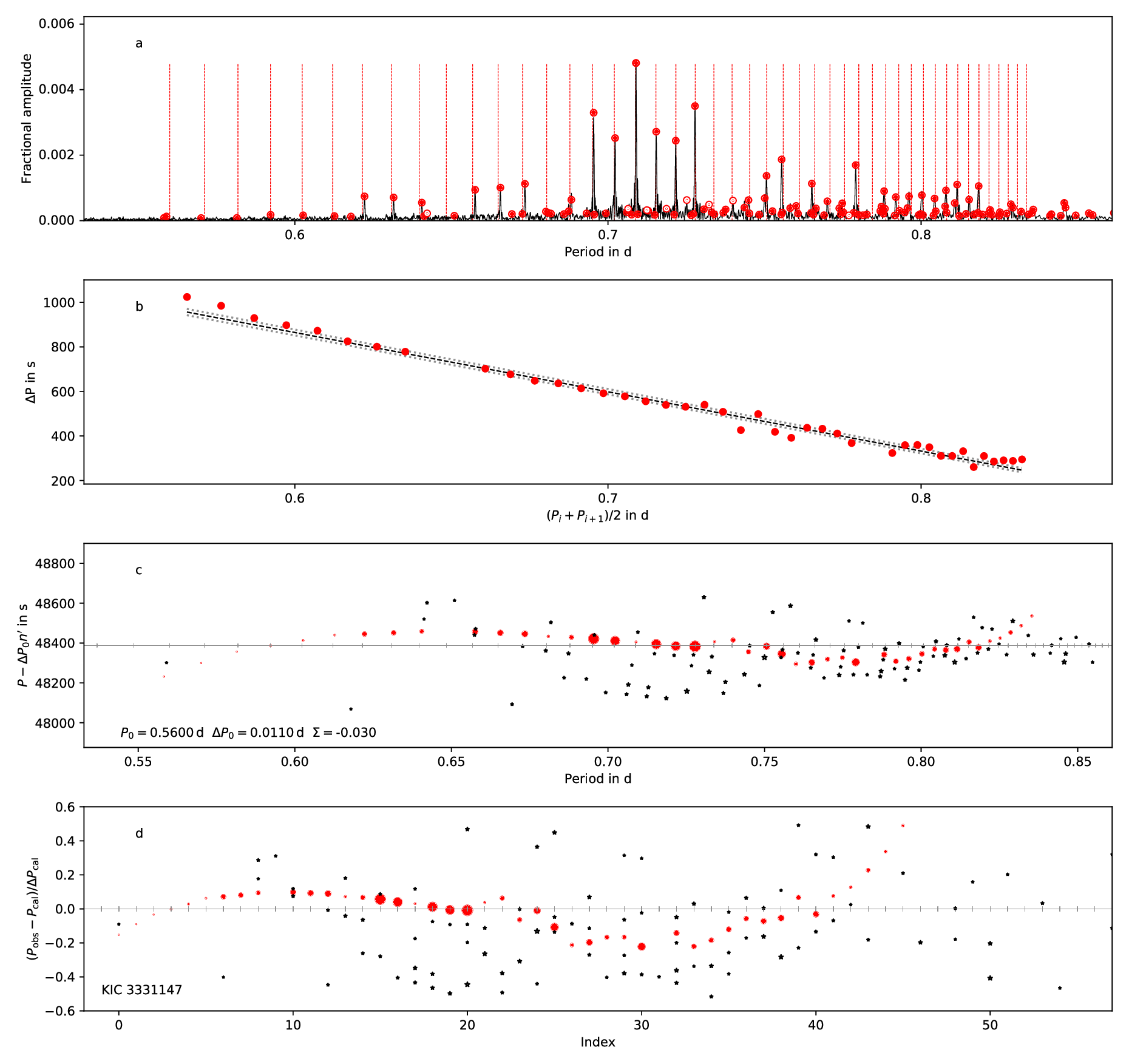}
	\caption{Same as Fig.~\ref{fig:KIC4253413pattern} but for KIC\,3331147. The slope of this pattern is $\Sigma=-0.0307\pm0.00002$. }\label{fig:KIC3331147pattern}
\end{figure*}

\subsection{Posterior distributions of two examples}\label{subsec:Two examples of the posterior distributions}

KIC\,4253413 is again used for illustrating our code. Figure \ref{fig:distribution} displays the one- and two-dimensional projections of the posterior distributions. It shows that there is a significant detection around $\Delta P\approx1185\,\mathrm{s}$ and $\Sigma=-0.04$. The distributions of $N_r$ and $N_l$ are asymmetric because the strongest peak does not lie in the centre of the excited distribution. $N_r$ and $N_l$ cannot cover the whole pattern in some cases, which is caused by the deviation from the linear model (Eq.~\ref{equ: p_i}), similar to the curvature in the \'{e}chelle diagram of solar-like oscillators \citep[e.g. Figure 1.2 in][]{2014aste.book...60B}.

Fig.~\ref{fig:KIC3331147distribution} shows our results for KIC\,3331147. The period spacing harmonics are significant so that it is an opportunity to see how the MCMC algorithm performs when searching for the fundamental period spacing. The posterior distributions in Fig.~\ref{fig:KIC3331147distribution} give two peaks, corresponding to $\Delta P$ and $2\Delta P$. Thanks to the prior $1 / \Delta P$ included in Equation~\ref{equ:prior}, the ones with higher probabilities are the correct period spacing and slope. From this example, we see that our code has the ability to avoid the influence of missing peaks and to find the correct period spacing in an incomplete pattern.

The corresponding period spacing patterns of these two stars can be seen in Fig.~\ref{fig:KIC4253413pattern} and \ref{fig:KIC3331147pattern}. The first panel shows the amplitude spectrum. The vertical dashed lines are the peaks' locations from the best-fitting model assuming the period spacing varies linearly (Equation~\ref{equ: p_i}). The second panel depicts the observed and best-fitting period spacings. We only plot the period spacings that can be calculated by two adjacent peaks, as shown in red solid circles in the second panel. Since these spacings are calculated from two peaks, we used the mean period $\frac{P_i+P_{i+1}}{2}$ as their x-coordinates in the second panels of Figs.~\ref{fig:KIC4253413pattern} and \ref{fig:KIC3331147pattern}. The black dashed line is the best-fitting result, while the grey dotted lines show the $\pm1\sigma$ region. The values for the slopes are $\Sigma=-0.0358\pm0.0006$ for KIC\,4253413 and $-0.0307\pm0.00002$ for KIC\,3331147. 

Considering the formula $P_i=\Delta P \left(n'+\epsilon_g \right)$ from the Equation~\ref{equ: general asymptotic}, we can plot the period \'{e}chelle diagrams analogous to the traditional \'{e}chelle diagram, which is often used for solar-like oscillators. In the third panel of Fig.~\ref{fig:KIC4253413pattern}, the y-axis is the difference between the observed period $P_{\mathrm{obs}}$ and the term $\Delta P n'$ while the x-axis is the period. This is therefor a `sideways' \'{e}chelle diagram. For the peaks that fit the pattern, we can calcuate the value $P_{\mathrm{obs}}-\Delta P n'$ accurately and plot them as red circles. For the peaks which do not belong to any pattern, we find the value $n'$ that minimises $|P_{\mathrm{obs}}-(\Delta P n'+\Delta P \epsilon_g)|$, shown in black stars. The \'{e}chelle diagram of KIC\,4253413 (third panel of Fig.~\ref{fig:KIC4253413pattern}) shows a clear fluctuation around $\Delta P \epsilon_g$, implying deviations from the linear functional form. 

The fluctuation indicates the change of slope, which is not an exact constant for a real star. Some of the fluctuations are most likely caused by the Coriolis force and the change from the co-rotating to the inertial reference frame. In addition, pulsation mode trapping, caused by chemical gradients in the stellar interior, also results in changes in the slope of the pattern. Usually, fluctuations in a period spacing pattern caused by mode trapping are the most pronounced \citep[e.g.][]{2013MNRAS.429.2500B}.

The y-axis term $P_{\mathrm{obs}} - \Delta P n'$ in the third panel is always smaller than half of the local spacing and decreases with period. Also, the points along x-axis become denser. To remove the influence from the changing period spacings, we plot the normalised diagram in the fourth panel in Fig.~\ref{fig:KIC4253413pattern}. The y-axis is normalised by the period spacing $\Delta P$ hence the meaning of the y-axis is the percentage deviation. The x-axis is the index of peaks, counted from the first peak in the pattern. This normalised diagram makes it easy to compare between different stars since the effect from changing period spacing is eliminated.

\subsection{Slow rotators}\label{sec:star rotation}
Several examples of slowly-rotating $\gamma$ Dor stars have been reported such as KIC\,11145123 \citep{2015MNRAS.450.3015K}, KIC\,9244992 \citep{2015EPJWC.10101005B,2015MNRAS.447.3264S}, KIC\,3127996 \citep{2015EPJWC.10101005B}, KIC\,7661054 \citep{2016MNRAS.459.1201M}, and KIC\,10080943 \citep{2015MNRAS.454.1792K}. In these cases, clear doublets or triplets are observed and the rotation rates of their cores can be calculated with model independence. The observed rotation rate is smaller than expected from both observations and theories for A-type stars, implying some new mechanism of angular momentum transport operates in these stars \citep[][submitted]{2018arXiv180109228O}. It is a challenge for current stellar structure and evolution theory.

As pointed out by \cite{2017MNRAS.465.2294O}, the slope defined as $\Sigma=\mathrm{d}\Delta P/\mathrm{d}P$ is a diagnostic of the internal rotation. To investigate this diagnostic without model dependence, we applied our algorithm to 24 $\gamma$ Dor stars in which splittings are seen. The internal rotation can be calculated from the splitting, hence the slope-rotation relation can be tested. Our results show that there are 22 stars whose splittings are caused by rotation, while there are two stars, KIC\,6862920 and KIC\,8458690, whose apparent splittings are two overlapping patterns from a binary, instead of the rotational effect. Results for all 24 stars are shown in the Appendix.

\subsubsection{Splitting identification}

\begin{figure}
\centering
 \includegraphics[width=1\linewidth]{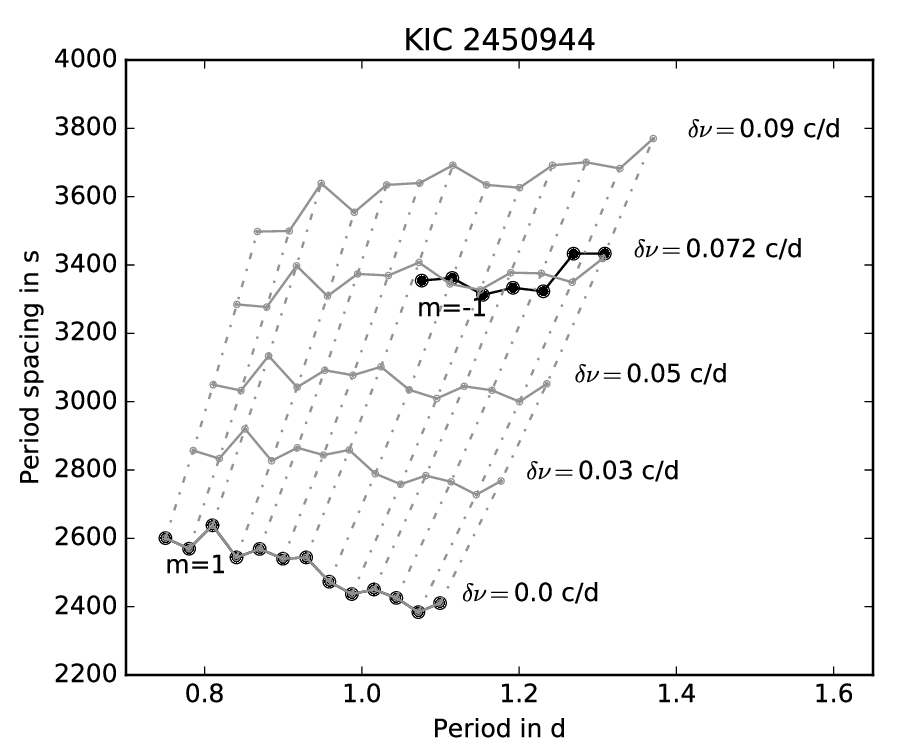}
 \caption{This copy-shift diagram shows how we identify the unclear splittings, using KIC\,2450944 as example. The black circle show the two observed patterns, while the grey points show the shifted copies of $m=1$ modes with different splittings. The dashed lines connect the modes with same radial order.}\label{fig:KIC2450944_process}
\end{figure}

We use KIC\,2450944 as the example to illustrate the process. Fig.~\ref{fig:KIC 2450944} depicts two uniformly-spaced overlapping patterns. It seems likely that these two patterns are the result of rotational splittings. However, due to the relatively large splitting and the absence of zonal modes, it is hard to identify the doublets. In order to solve this problem, we identified their degrees as $l=1$ and azimuthal orders as $m=\pm1$ respectively, based on the mean period spacings of the observed patterns \citep[e.g.][]{2016A&A...593A.120V} and the known effects of geometric mode cancellation \citep[e.g.][]{2001A&A...375..113C}. We selected the $m=1$ pattern as the template, tried several splitting values and calculated the shifted copy. Fig.~\ref{fig:KIC2450944_process} is a `copy-shift diagram' and it displays the process, in which the grey line with $\delta \nu=0.072$\,$\mathrm{d}^{-1}$ fits the $m=-1$ (retrograde) pattern. The retrograde pattern has larger period spacings than those of the prograde pattern. The grey dashed lines in Fig.~\ref{fig:KIC2450944_process} join the modes with same radial order $n$. We see that the prograde modes extend to lower $n$ than the retrograde modes.

Since the rotation rates are sufficiently low, we can use the first-order perturbative treatment to obtain the rotational period of the core \citep{2010aste.book.....A},
\begin{equation}
P_c=\frac{1-C_{n,l}}{\overline{\delta \nu}},
\end{equation}
where $\overline{\delta \nu}$ is the observed frequency splitting of the g modes. The Ledoux constant $C_{n,l}\approx \left[l(l+1)\right]^{-1}$ is fixed to 0.5 without considering its variation with radial order $n$ \citep[][]{1981ApJ...244..299S}. For KIC\,2450944, for example, $\overline{\delta \nu}=0.072$\,$\mathrm{d}^{-1}$. The rotation period near the boundary of the convective core is therefore determined as 6.9\,d. 

We identified the doublets from the splitting $\delta \nu=0.072$\,$\mathrm{d^{-1}}$, plotting them in the top panel of Fig.~\ref{fig:KIC 2450944}. The left panel of Fig.~\ref{fig:KIC 2450944echelle_splitting} presents the \'{e}chelle diagram of KIC\,2450944, where two series are seen. Considering that the Ledoux `constant' changes slightly and it may give some information about differential rotation, we also investigated the splitting variation with frequency. The right panel of Fig.~\ref{fig:KIC 2450944echelle_splitting} shows that the splitting increases with frequency and their variation reaches $\sim 6\times 10^{-4}$\,$\mathrm{d}^{-1}$, one order of magnitude bigger than the typical frequency uncertainty \citep[for other examples, see also][]{2016MNRAS.459.1201M, 2015MNRAS.454.1792K}. 

Based on 22 stars with rotational splittings, we find that the triplets are slightly asymmetric as expected. The value $f_{m=0}-f_{m=-1}$ is smaller than $f_{m=1}-f_{m=0}$ (see Fig.~\ref{fig:KIC 4919344echelle_splitting} and \ref{fig:KIC 5038228echelle_splitting} for examples). This asymmetry is due to the second-order perturbation \cite[as stated in ][]{2015MNRAS.454.1792K}. On the other hand, the splitting generally rises or does not change with increasing frequency for $l=1$ modes (like Fig.~\ref{fig:KIC 2450944echelle_splitting} and \ref{fig:KIC 10080943Aechelle_splitting}). This trends reflect that the Ledoux constant varies slightly as a function of radial order.

\subsubsection{Quadrupole modes}

In some stars $l=2$ modes are detected. KIC\,3127996 shows splittings both in $l=1$ and $l=2$ modes \citep{2015EPJWC.10101005B}, but assigning the azimuthal orders to $l=2$ is not straightforward. Meanwhile, the frequencies of $l=2$ modes deviate from the perturbative approach at a smaller rotation rate than for $l=1$ modes. To solve this problem, we used the traditional approximation of rotation (TAR) to build theoretical patterns for KIC\,3127996. In the TAR, the period in the corotating frame $P_n^\mathrm{co}$ is calculated by the formula
\begin{equation}
 P_n^\mathrm{co}=\frac{\Pi_0}{\sqrt{\lambda}}\left(n+\alpha_g\right),
\end{equation}
where $\Pi_0=2\pi^2\left(\int\frac{N}{r}\mathrm{d}r\right)^{-1}$ is the asymptotic spacing, $N$ is the Brunt-V\"{a}is\"{a}l\"{a} frequency, $r$ is radius, $n$ is the radial order, and the phase term $\alpha_g$ is assumed to be 0.5. Note that $\lambda$ is the eigenvalue of Laplace's tidal equation, which is related to the rotational rate $\nu_{\mathrm{rot}}$, the pulsation frequency in the corotating frame $\nu^{\mathrm{co}}$ and the quantum numbers $l$ and $m$ \citep{1987MNRAS.224..513L, 2003MNRAS.340.1020T,2013MNRAS.429.2500B}.

\begin{figure}
\centering
 \includegraphics[width=1\linewidth]{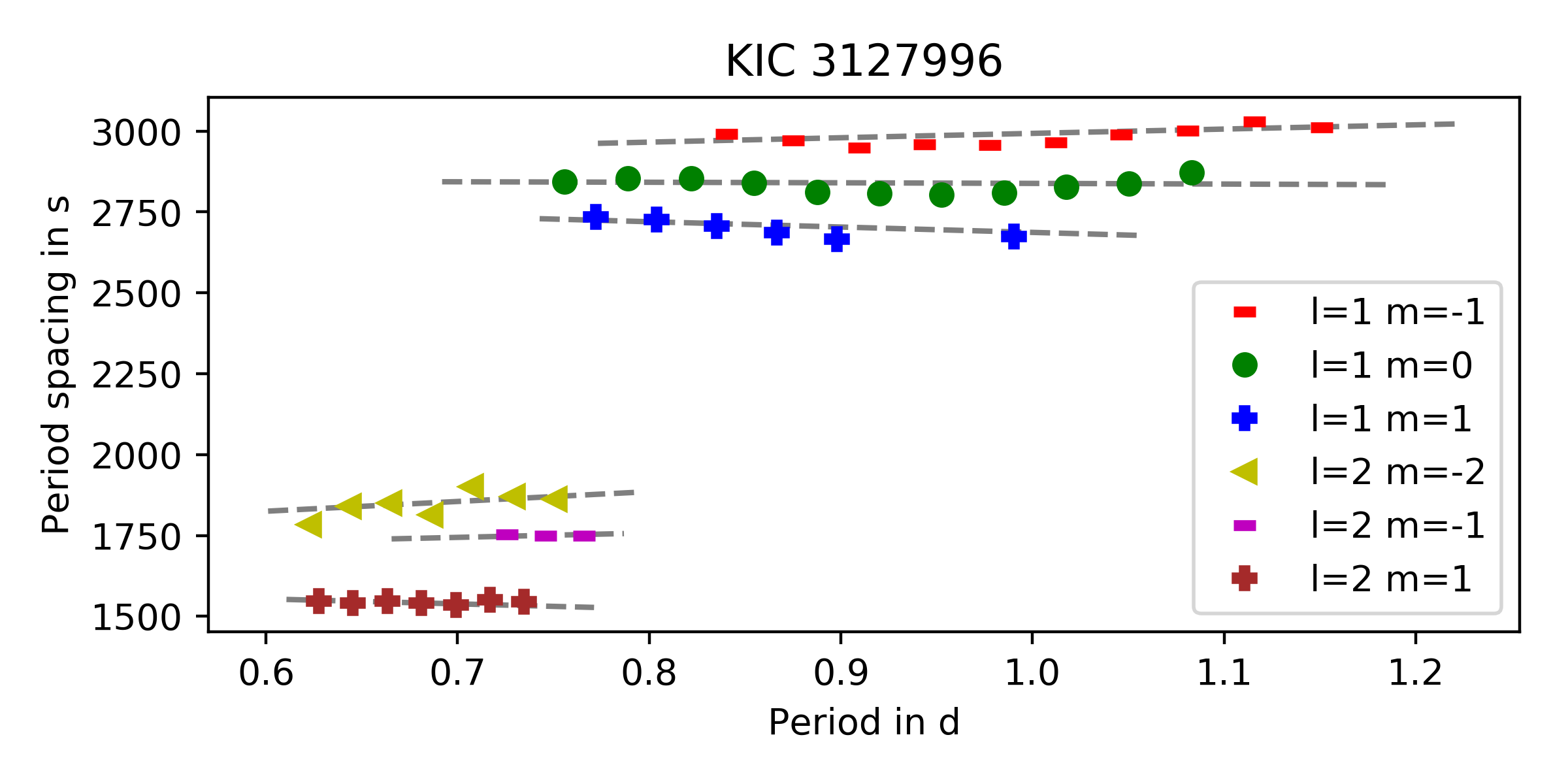}
 \caption{The observational and theoretical period spacing patterns of KIC\,3127996.}\label{fig:obs_tar_pattern_3127996}
\end{figure}

The quadrupole patterns in KIC\,3127996 were calcuated based on the asymptotic spacing $\Pi_0=4028$\,s and the rotational rate $\Omega=0.053$\,$\mathrm{d}^{-1}$, which were calculated from the dipole triplets. The eigenvalues $\lambda$ were read from {\sc gyre}'s table \citep{2013MNRAS.435.3406T}. Fig.~\ref{fig:obs_tar_pattern_3127996} presents the results from the TAR, in which several possible $m$ choices were tried and $m=\pm1$ fitted the observational pattern best. Fig.~\ref{fig:KIC 3127996} shows the amplitude spectrum and the period spacing patterns of KIC\,3127996. The \'{e}chelle diagrams and the splitting variations can be seen in Fig.~\ref{fig:KIC 3127996_l=1_echelle_splitting} and~\ref{fig:KIC 3127996_l=2_echelle_splitting} for $l=1$ and $l=2$ modes, respectively. The splittings are about 0.27\,$\mathrm{d}^{-1}$ for $l=1$ and 0.45\,$\mathrm{d}^{-1}$ for $l=2$ modes, whose ratio obeys the theoretical value $0.6$ from the Ledoux constant $C_{n,l}=\left[l(l+1)\right]^{-1}$. 

KIC\,5038228 (Fig.~\ref{fig:KIC 5038228}), KIC\,5459805 (Fig.~\ref{fig:KIC 5459805}), and KIC\,6937123 (Fig.~\ref{fig:KIC 6937123}) also show $l=2$ patterns. We used the TAR to identify them (see Fig.~\ref{fig:KIC5038228mode_identification}, Fig.~\ref{fig:KIC5459805mode_identification}, and Fig.~\ref{fig:KIC6937123mode_identification}). 

\subsubsection{Binaries}

KIC\,6862920 is a PB2 binary system detected by phase modulation by \cite{2018MNRAS.474.4322M}. Here, `PB2' means that two stars are pulsating in pressure modes and we can detect the phase motions of both pulsators. The orbital parameters are listed as follows: orbital period=$61.54^{+0.07}_{-0.07}$\,d, eccentricity=$0.58^{+0.14}_{-0.11}$, $a_1\sin i/c=94^{+10}_{-8}$\,s, $a_2\sin i/c=88^{+14}_{-14}$\,s, and mass ratio $M_2/M_1=1.1^{+0.20}_{-0.15}$, where $a_1$ and $a_2$ are the semi-major axes for two components, $i$ is the inclination and $c$ is the speed of light.

The gravity mode patterns are seen in Fig.~\ref{fig:KIC 6862920}. The amplitude spectrum shows `doublets' clearly, causing two parabola curves in the \'{e}chelle diagram (the left panel of Fig.~\ref{fig:KIC 6862920echelle_splitting}). These two parallel period spacings have slopes of -0.037, which implies that the core rotation is fast enough to deviate from the asymptotic relation. However, as shown in the right panel of Fig.~\ref{fig:KIC 6862920echelle_splitting}, the mean splitting is only $\sim0.0058$\,d$^{-1}$, which would imply a rotational period of 86\,d. 

To explain the conflict between the slope and the `splitting', we assume that those two components with nearly equal masses ($M_2/M_1=1.1^{+0.20}_{-0.15}$) have the almost same evolutionary stages. The period spacing patterns come from two identical components so they look like rotational splittings but actually come from two different stars. 

We find that KIC\,8458690 is another binary that shows similar patterns. This spectrum has a very small splitting ($\delta \nu = 0.0015$\,d$^{-1}$) but extremely steep slopes $\Sigma=-0.08$ (Fig.~\ref{fig:KIC 8458690} and Fig.~\ref{fig:KIC 8458690echelle_splitting}). These are incompatible and we deduce that there are two $\gamma$ Dor stars contributing to the spectrum. The mass ratio is $M_2/M_1=0.83\pm0.29$ \citep{2018MNRAS.474.4322M}, also implying that these two patterns are from two components, rather than rotational splittings.

\begin{table*}
\caption{The slopes and splittings of 22 slowly-rotating stars collected from this paper and previous literature. KIC\,10080943 is a binary, whose components are marked by A and B. Only $l=1$ splittings are listed. $\delta \nu$ are the mean splittings. $l$ are the degrees. $\Sigma$ are the slopes for each $m$. The uncertainty on the last digit is given between brackets. The final column gives references to additional studies on the targets.}\label{tab:slow rotation data}
\begin{tabular}{lllllll}
\hline\hline
KIC number &$\delta \nu$ in $\mathrm{d}^{-1}$ & $l$ & $\Sigma_{m=1}$ & $\Sigma_{m=0}$ & $\Sigma_{m=-1}$ & Reference\\
\hline
2450944     &   0.0731(3)    & 1  & $-0.0084(4)$               &             & \phantom{$-$}$0.003(1)$       & \\
3127996     &   0.0271(1)    & 1  & $-0.0043(4)$               & $-0.0007(5)$  &  \phantom{$-$}$0.0024(4)$     & {\cite{2015EPJWC.10101005B}} \\
3222854     &   0.0059(1)    & 2  & $-0.0003(3)$               &             & $-0.0004(3)$     & \\           
4480321     &  0.0041(1)     & 1  &  \phantom{$-$}$0.0006(3)$  & $-0.0002(1)$  &  $-0.0001(4)$    & {\cite{2018A&A...610A..17L}}\\
4919344     &  0.1055(4)     & 1  &  $-0.0221(7)$              & $-0.0167(3)$  &  $-0.0108(4)$    & \\
5038228     &   0.0801(4)    & 1  & $-0.0116(8) $              & $-0.0006(7)$  &  \phantom{$-$}$0.0041(2)$     & \\
5459805     &  0.06009(6)    & 1  &  $-0.000(2$)               &             &  \phantom{$-$}$0.011(3)$      & \\
5557072     & 0.0020(3)      & 1  & \phantom{$-$}$0.0013(3)$   & \phantom{$-$}$0.0020(3)$   &  \phantom{$-$}$0.0020(5)$     & \\
5810197     &   0.0427(1)    & 1  &  $-0.0066(4)$              &             &  $-0.006(2)$     & \\
6302589     &   0.0561(1)    & 1  &  $-0.006(1)$             & $-0.012(4)$   &  \phantom{$-$}$0.000(2)$     & \\
6467639     &   0.1179(6)    & 1  &   $-0.0228(4)$      & $-0.001(1)$   &  \phantom{$-$}$0.0082(2)$     & \\
6937123     &   0.0638(4)    & 1  &  $-0.0039(2)$       &  $-0.0007(9)$ &  \phantom{$-$}$0.0003(4)$     & \\
7661054     &  0.01853(4)    & 1  &  \phantom{$-$}$0.0000(4)$       &             &  \phantom{$-$}$0.0014(4)$     & {\cite{2016MNRAS.459.1201M}}\\ 
7697861     &   0.0262(1)    & 1  &  $-0.0078(7)$       &             &  $-0.0018(8)$    & \\
8197761     & 0.00168(6)     & 1  & \phantom{$-$}$0.002(1)$          & \phantom{$-$}$0.000(2)$    & $-0.005(2)$      & {\cite{2017MNRAS.467.4663S}}\\
9028134     &    0.093(1)    & 1  &  \phantom{$-$}$0.001(2)$         & $-0.000(3)$   &    $-0.0024(6)$  & \\
9244992     &  0.00789(7)    & 1  & $-0.0022(2)$        & $-0.0022(2)$   &  \phantom{$-$}$0.0002(2)$     & {\cite{2015MNRAS.447.3264S}}\\
9751996     &  0.0351(1)     & 1  & $-0.0042(2)$        & \phantom{$-$}$0.001(1)$   &  \phantom{$-$}$0.0036(5)$     & {\cite{2015ApJS..218...27V}}\\
10080943(A) & 0.04555(6)     & 1  & $-0.0028(4)$        &             & \phantom{$-$}$0.0019(5)$      & {\cite{2015MNRAS.454.1792K}}\\ 
10080943(B) &  0.0706(2)     & 1  &   $-0.0171(8)$      & $-0.008(1)$   & \phantom{$-$}$0.0024(7)$      & {\cite{2015MNRAS.454.1792K}}\\
10468883    &  0.075(1)      & 1  & \phantom{$-$}$0.003(2)$          & \phantom{$-$}$0.011(1)$    & \phantom{$-$}$0.024(1)$       & \\
11145123    &  0.0048(1)     & 1  &  \phantom{$-$}$0.0026(4)$        & \phantom{$-$}$0.001(1)$    & \phantom{$-$}$0.0031(6)$      & {\cite{2014MNRAS.444..102K}}\\
\hline
\end{tabular}
\end{table*}

\subsubsection{Slope vs splitting} 

 \cite{2017MNRAS.465.2294O} gave a theoretical relationship between the slope and rotation rate. They found that the slope will decrease to negative values for prograde and zonal modes, and increase for retrograde modes. In order to investigate the observational evidence, we selected 22 slowly-rotating targets listed in Table~\ref{tab:slow rotation data}. 13 of them are firstly reported in our sample, while others are from previous literature. 
 
\begin{figure}
  \includegraphics[width=1\linewidth]{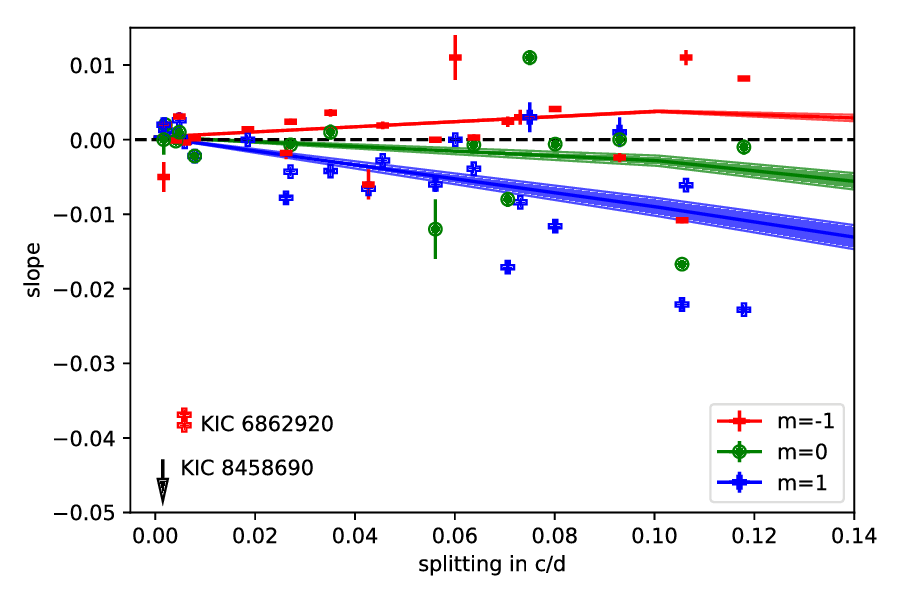}
  \caption{The slope versus splitting diagram from 22 $\gamma$ Dor stars whose splittings are caused by the rotation effects. The shaded curves are the theoretical prediction from \protect \cite{2017MNRAS.465.2294O}. KIC\,6862920 and KIC\,8458690 shows two outliers, because their splittings are the two overlapped patterns from both components. We use a downward arrow to note KIC\,8458690 since its slope is too small ($\Sigma=-0.08$). }\label{fig: slope_vs_splitting}
 \end{figure}

After identifying their period spacing patterns and rotational splittings, we present the observational slope versus splitting relation in Fig.~\ref{fig: slope_vs_splitting}. 
When the splitting is close to zero, i.e. when the rotation rate is very low, the period spacing is well reproduced by the \cite{1979PASJ...31...87S} asymptotic relation. This predicts a period spacing that is constant a function of period, hence the slope near to zero that we see in Fig.~\ref{fig: slope_vs_splitting}. 
When the splitting rises, the diagram shows more spread. The rotation breaks the asymptotic relation and changes the spacing. The prograde and zonal patterns (shown as circles and squares) generally have negative slopes. For the retrograde patterns, the slopes are larger and more likely to be positive. 

\begin{figure}
\centering
\includegraphics[width=\linewidth]{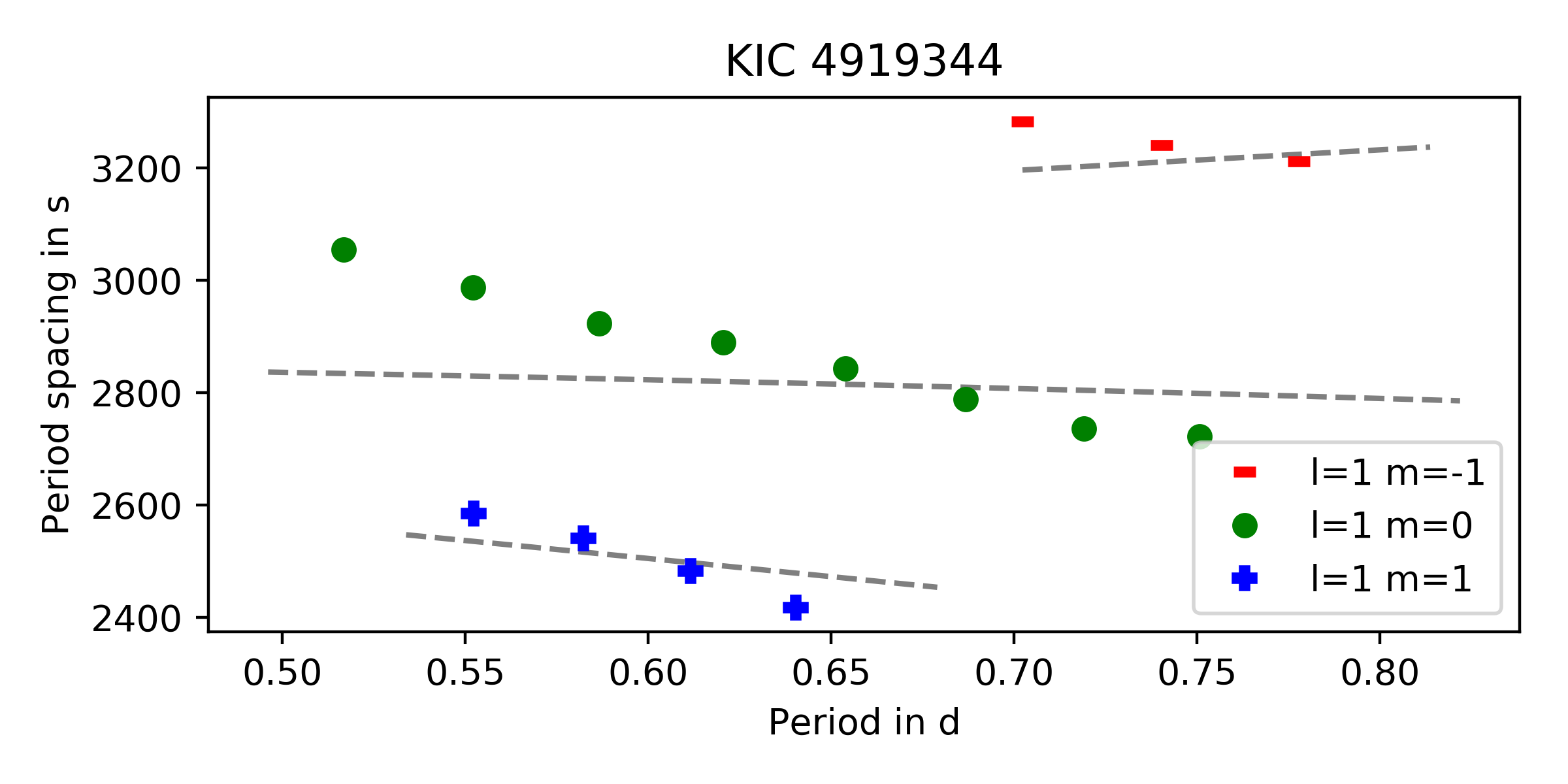}
\caption{The observational and theoretical period spacing patterns of KIC\,4919344. The fits are poor because these three patterns are parts of dips. }\label{fig:KIC4919344_TAR}
\end{figure}

In Fig.~\ref{fig: slope_vs_splitting}, the observational data follow the general trend expected from theory but there is a large spread, which is due to the intrinsic glitches and dips in the period spacing patterns.
To include the deviations from the linear model, we used the residuals rather than the frequency uncertainties to estimate the uncertainty of slope. This process includes the uncertainty caused by the small glitches. 
However, there is another contribution to the slope uncertainty that comes from partially-observed dips in the pattern. Fig.~\ref{fig:KIC4919344_TAR} shows the theoretical fit of KIC\,4919344, where the measured slopes are significantly smaller than theory predicts. This systematic discrepancy is caused by broad dips in the period-spacing pattern that affect many radial orders, more, in fact, than are observable in this star. Another explanation for the discrepancy is that the global parameters of our sample ($\log g$, $T_\mathrm{eff}$ and metallicity) are not in the regime that was used by \cite{2017MNRAS.465.2294O} to estimate the error margins (shaded area in Fig.~\ref{fig: slope_vs_splitting}).

\section{CONCLUSIONS}\label{sec:conclusion}

Gravity-mode pulsations of $\gamma$ Doradus stars act as the probe of the stellar interior. The patterns showing equally-spaced periods are distorted due to the molecular weight gradient and fast rotation. The dips, varying period spacings, as well as the missing peaks allow us to understand the physics of rotation and angular momentum transport in main sequence stars and can place much-needed constraints on diffusive mixing and chemical gradients, whilst also providing stellar ages.

In this paper, we tried two methods to parametrize the patterns. Firstly, the `moving-window Fourier transform' was used to detect the local periodicity in the period spectrum. It was partly successful since it can only give rough information about the period spacing pattern. Secondly, the `cross-correlation' optimizes the match between an artificial template and the observed spectrum. It can give the posterior distribution of the period, period spacing and slope of a detected spacing pattern. We mainly used the second method to detect g-mode patterns.

To build a relation between the slope and the rotational rate, we applied our algorithm to 22 stars with observational rotational splitting. However, the splittings generally have the same magnitude as the period spacings, hence their patterns overlap and are hard to distinguish. We used the \'{e}chelle diagram and the `copy-shift' diagram to clarify their patterns and splittings and reported the first slope-vs-splitting diagram containing 22 slowly-rotating targets. The observational evidence follows the theory, that the slope deviates from zero when the splitting and hence the rotation rate increases. 
We also investigated the splitting variations for these stars. They show that for dipole modes the splitting typically rises or remains unchanged when the frequency increases. There are two stars whose apparent splittings and slopes are incompatible. These are explained by binaries with two similar period spacing patterns from almost identical components.

\section*{Acknowledgement}
We gratefully acknowledge support from the Australian Research Council, and from the Danish National Research Foundation (Grant DNRF106) through its funding for the Stellar Astrophysics Centre (SAC). We also thank the referee for very helpful comments.







\appendix
\section{RESULTS FOR 24 STARS}

In the appendix, we display the amplitude spectra, the period spacing patterns, the \'{e}chelle diagrams and splitting variations for 24 multiplet stars. The splittings of KIC\,6862920 and KIC\,8458690 are just the binary effects, while other 22 stars show the rotational effects. 

\clearpage
\begin{figure*}
\centering
\includegraphics[width=0.9\textwidth]{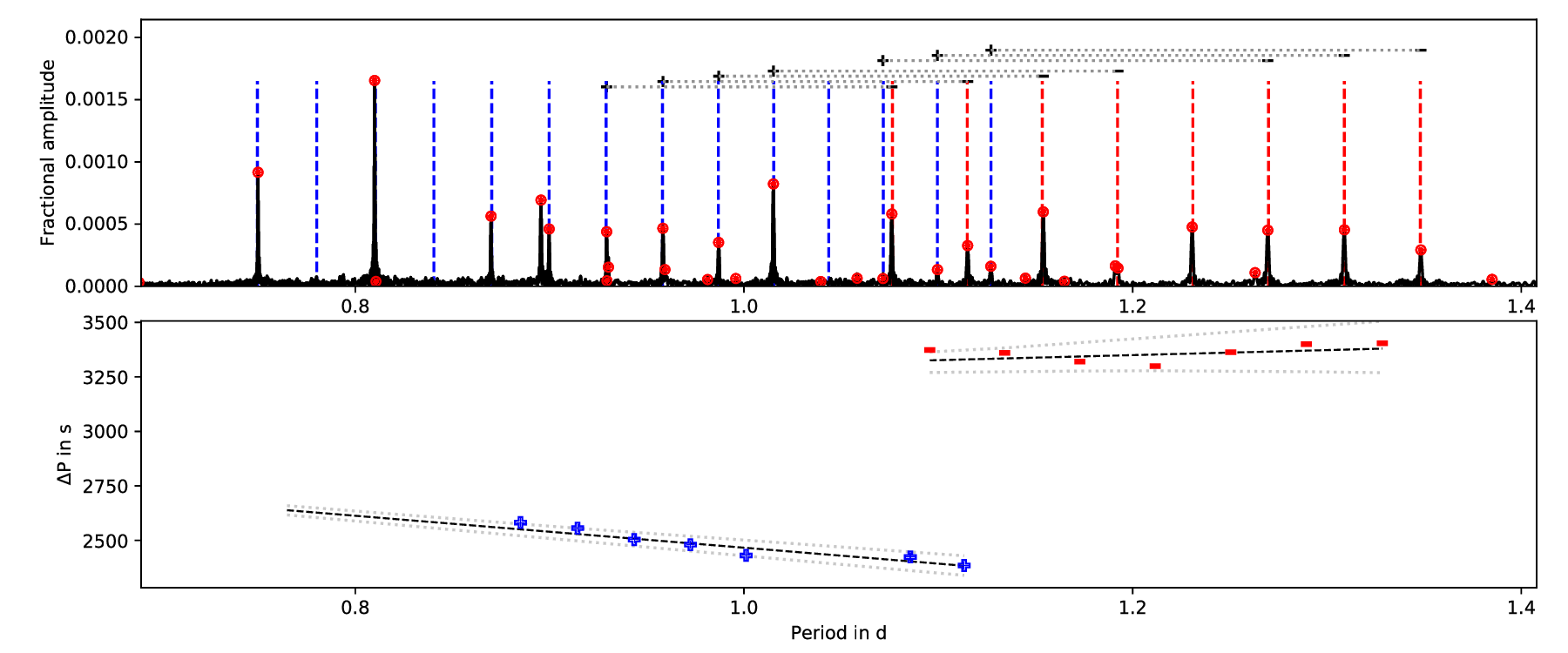}
\caption{Rotationally split doublets in KIC\,2450944. Upper panel: the period spectrum and doublets. The solid red circles show the independent frequencies, while open red circles are likely combination frequencies (none present in this figure). The vertical dashed lines mark the linear fit. The horizontal dashed lines show the splittings. The symbol `+' means $m=+1$. `-' denotes $m=-1$. `o' will be used in other figures to show $m=0$ modes. Lower panel: the period spacing patterns. The markers are the period spacings, using $\left(P_i+P_{i+1}\right)/2$ as their x-coordinates. The black dashed lines show the linear fit while the grey dotted lines around them present the uncertainties. The blue pluses present $l=1,~m=1$ modes. The red minuses are $l=1, m=-1$ modes. }\label{fig:KIC 2450944}
\end{figure*}

\begin{figure*}
\centering
\includegraphics[width=0.85\textwidth]{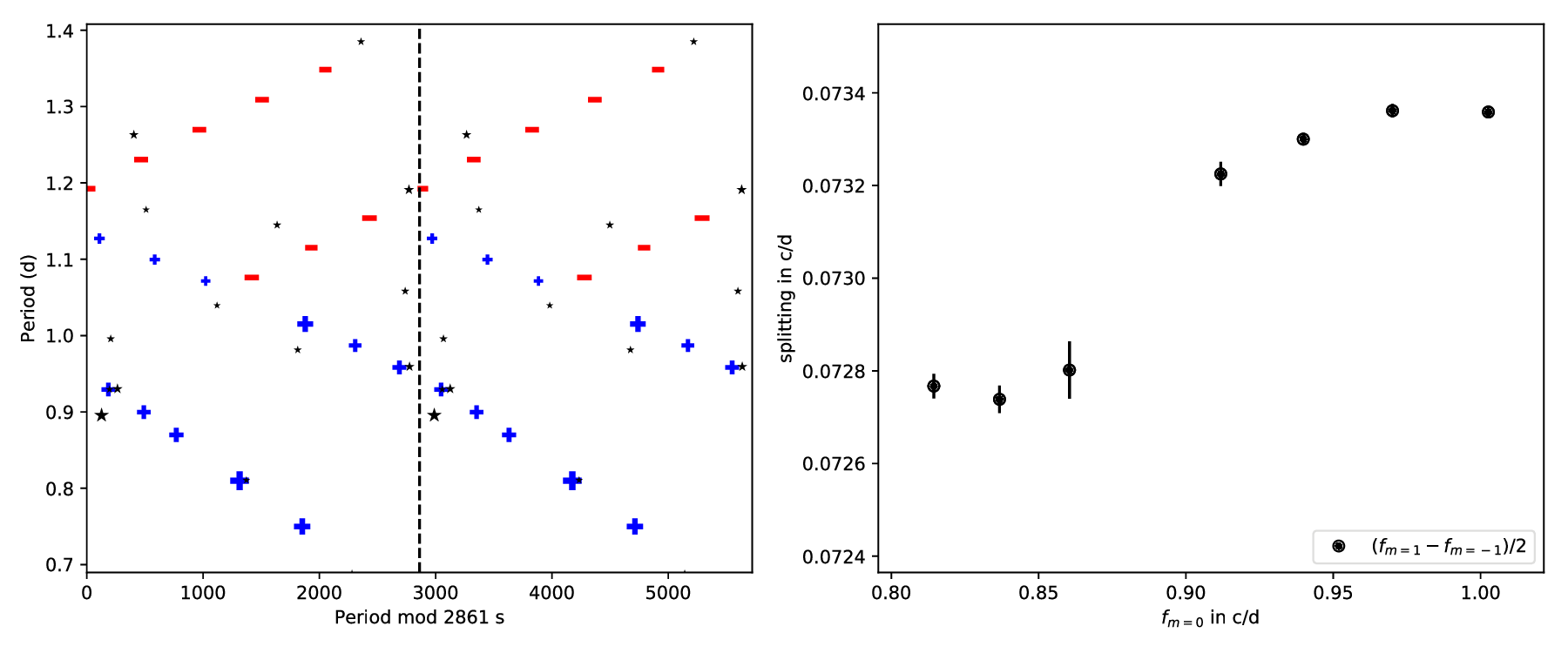}
\caption{The \'{e}chelle diagram and the splitting variations of KIC\,2450944. Left panel: the \'{e}chelle diagram. Different symbols and colors represent different azimuzal orders, as in Fig.~\ref{fig:KIC 2450944}. \textcolor{red}{The black crosses show the peaks that do not belong to any pattern.} Right panel: splitting variation with $m=0$ frequency. The errorbars show the frequency uncertainties calculated by Eq.~\ref{equ:freq_uncertainty}. }\label{fig:KIC 2450944echelle_splitting}
\end{figure*}

\clearpage
\begin{figure*}
\centering
\includegraphics[width=0.9\textwidth]{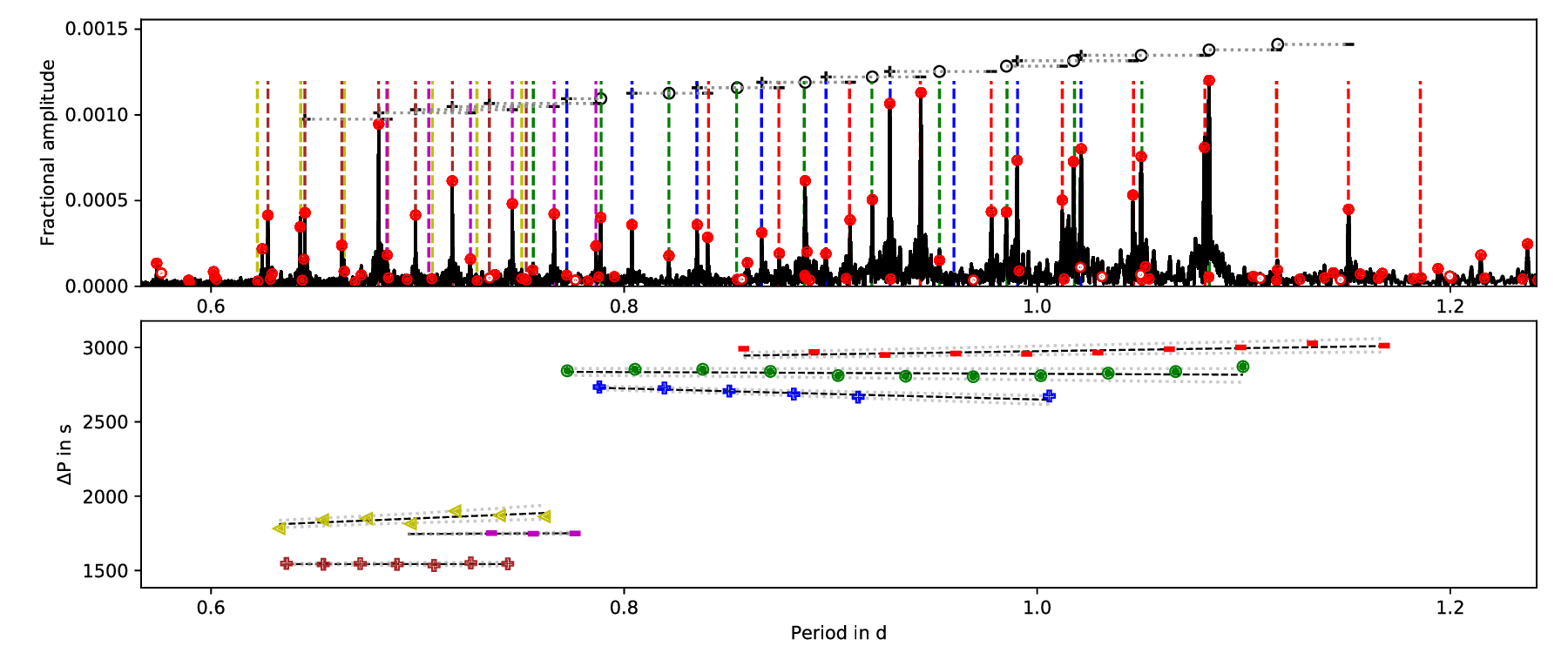}
\caption{The period spacing patterns of KIC\,3127996. The period spacing patterns of $l=1$ and $l=2$ modes are seen in the same time, shown in the second panel. The blue pluses show $l=1,~m=1$ modes, the green circles show $l=1, m=0$ modes, the red minuses are $l=1, m=-1$ modes, the brown pluses are $l=2, m=1$ modes, the purple minus are $l=2, m=1$ modes, and the yellow triangles represent $l=2, m=-2$ modes.}\label{fig:KIC 3127996}
\end{figure*}

\begin{figure*}
\centering
\includegraphics[width=0.85\textwidth]{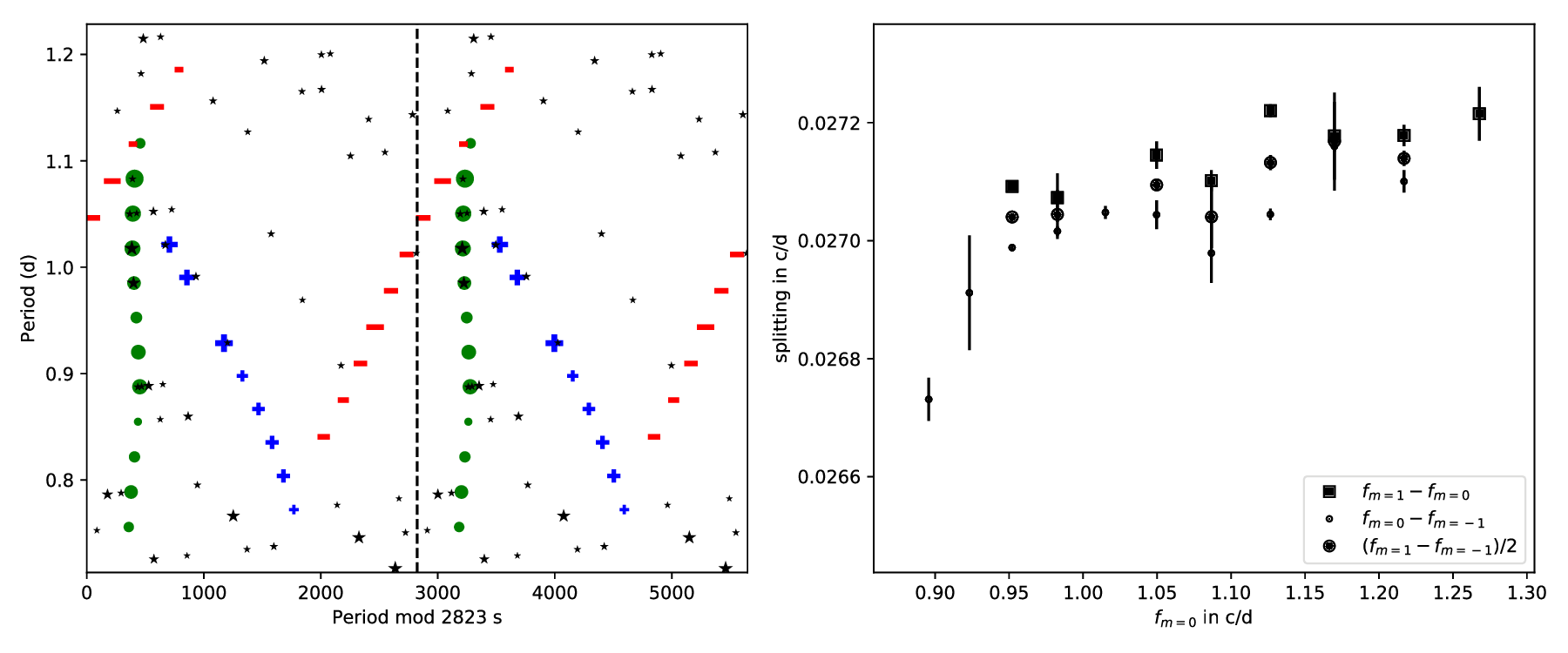}
\caption{The \'{e}chelle diagram and the splitting variations of $l=1$ modes of KIC\,3127996.}\label{fig:KIC 3127996_l=1_echelle_splitting}
\end{figure*}

\begin{figure*}
\centering
\includegraphics[width=0.85\textwidth]{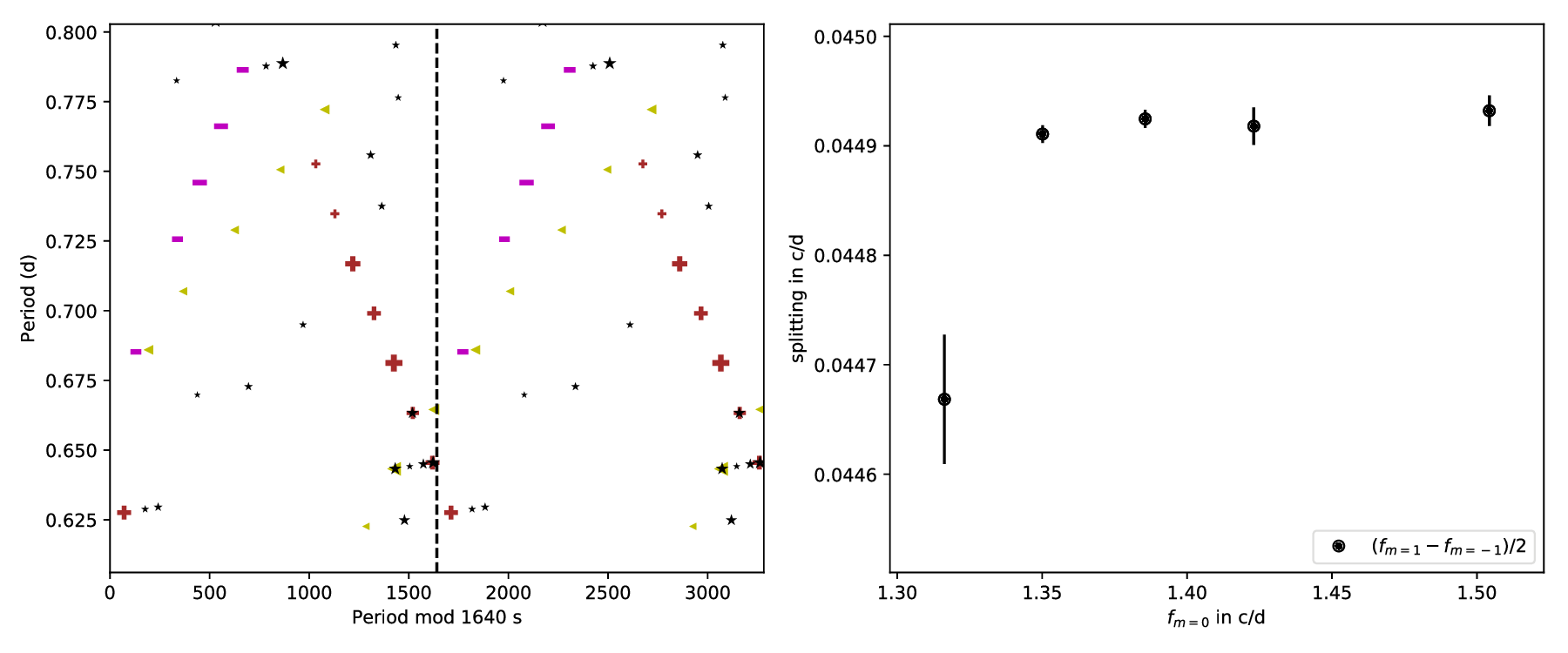}
\caption{The \'{e}chelle diagram and the splitting variations of $l=2$ modes of KIC\,3127996.}\label{fig:KIC 3127996_l=2_echelle_splitting}
\end{figure*}

\clearpage
\begin{figure*}
\centering
\includegraphics[width=0.9\textwidth]{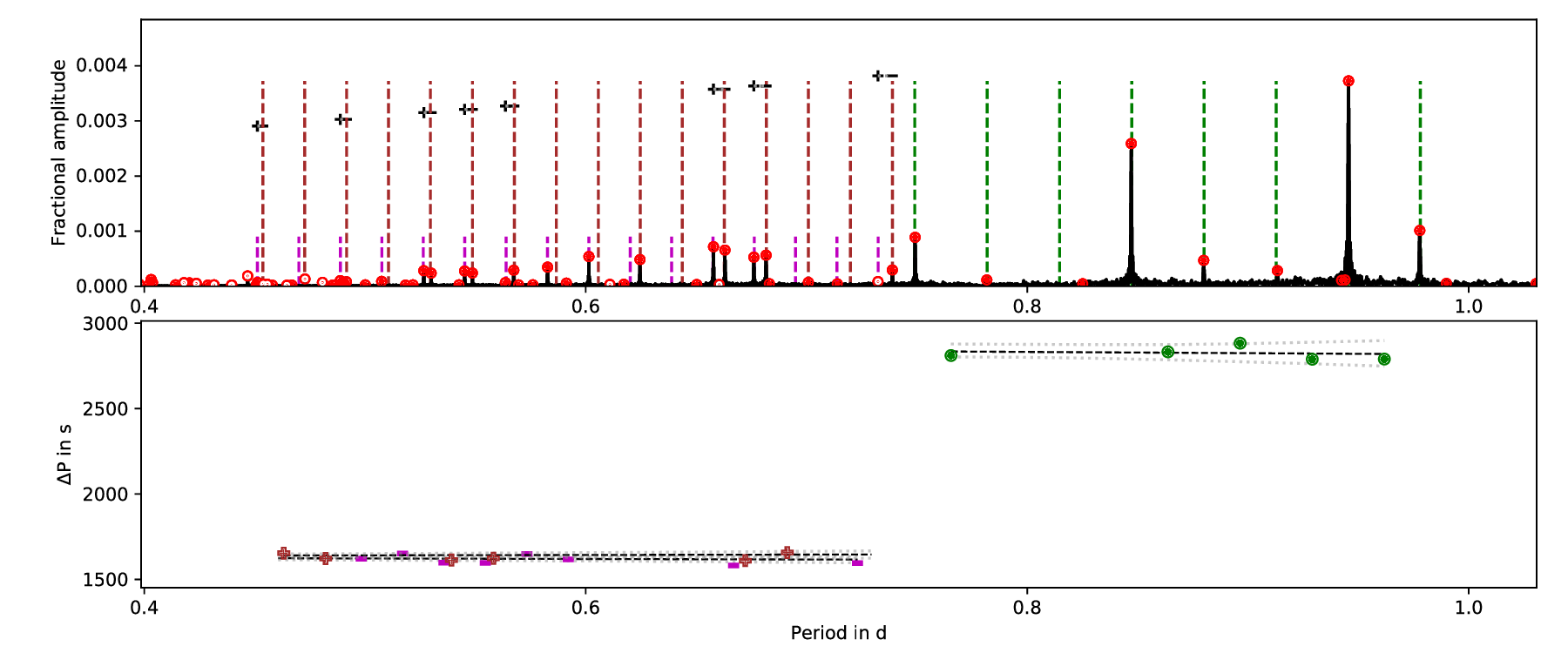}
\caption{The period spacing patterns of KIC\,3222854. Note that the splittings are $l=2$ modes. The azimuthal orders for two $l=2$ patterns cannot been determined. We assume they are $m=\pm1$ modes. }\label{fig:KIC 3222854}
\end{figure*}

\begin{figure*}
\centering
\includegraphics[width=0.85\textwidth]{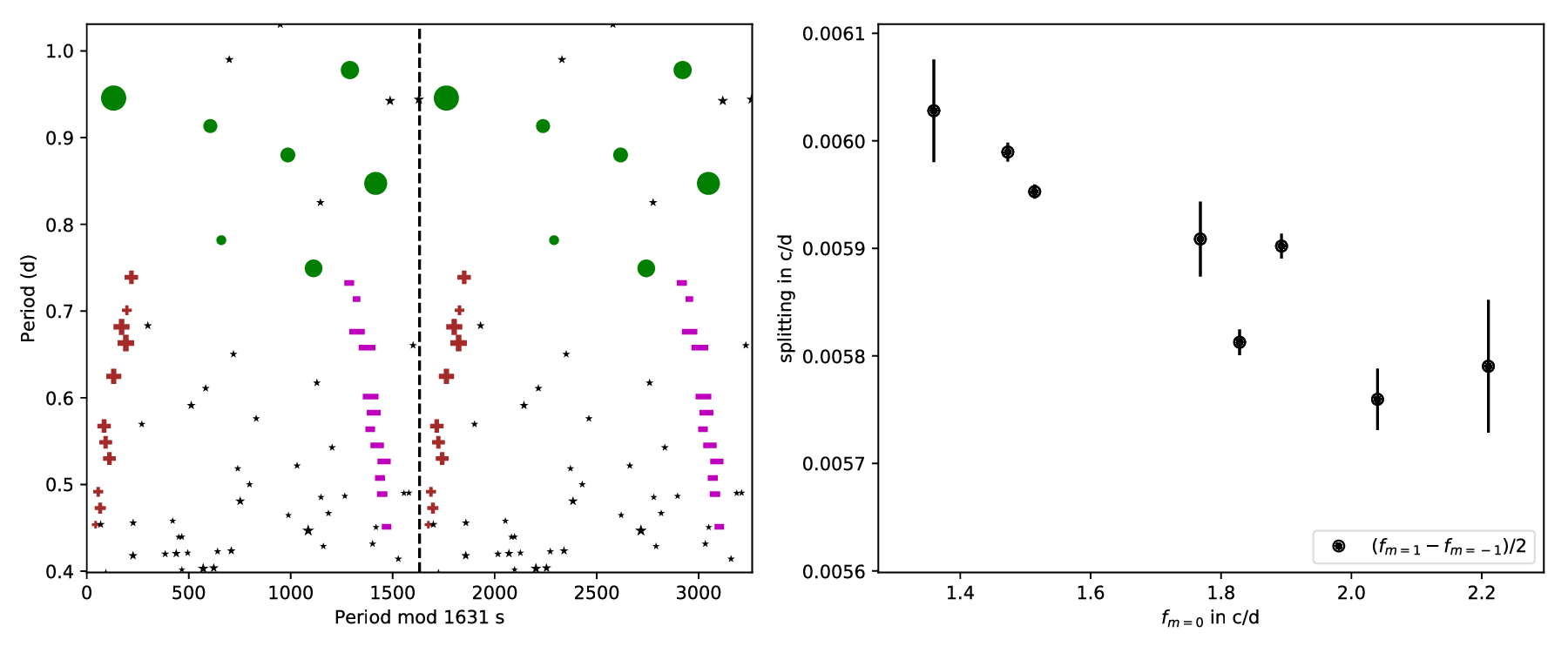}
\caption{The \'{e}chelle diagram and the splitting variations of KIC\,3222854. Note that the splittings are $l=2$ modes. }\label{fig:KIC 3222854echelle_splitting}
\end{figure*}

\clearpage
\begin{figure*}
\centering
\includegraphics[width=0.9\textwidth]{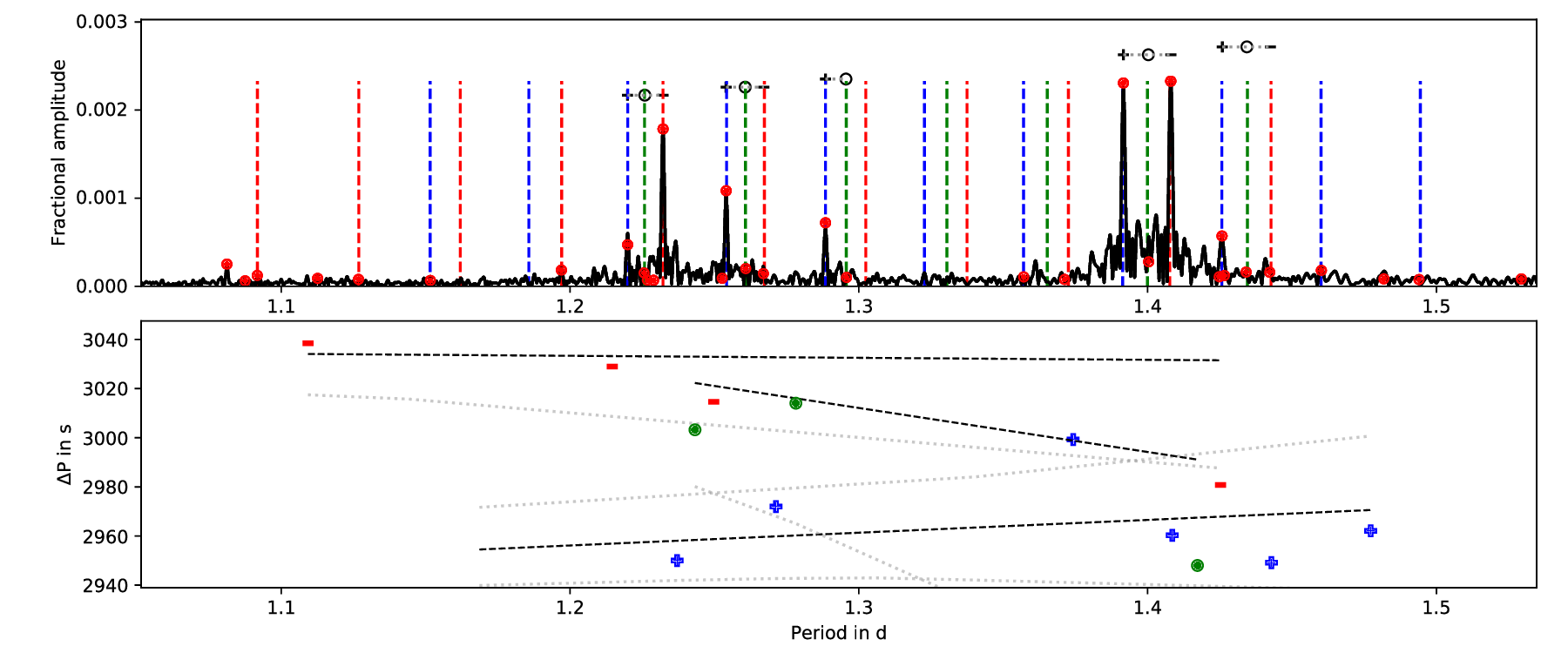}
\caption{The period spacing patterns of KIC\,4480321.}\label{fig:KIC 4480321}
\end{figure*}

\begin{figure*}
\centering
\includegraphics[width=0.85\textwidth]{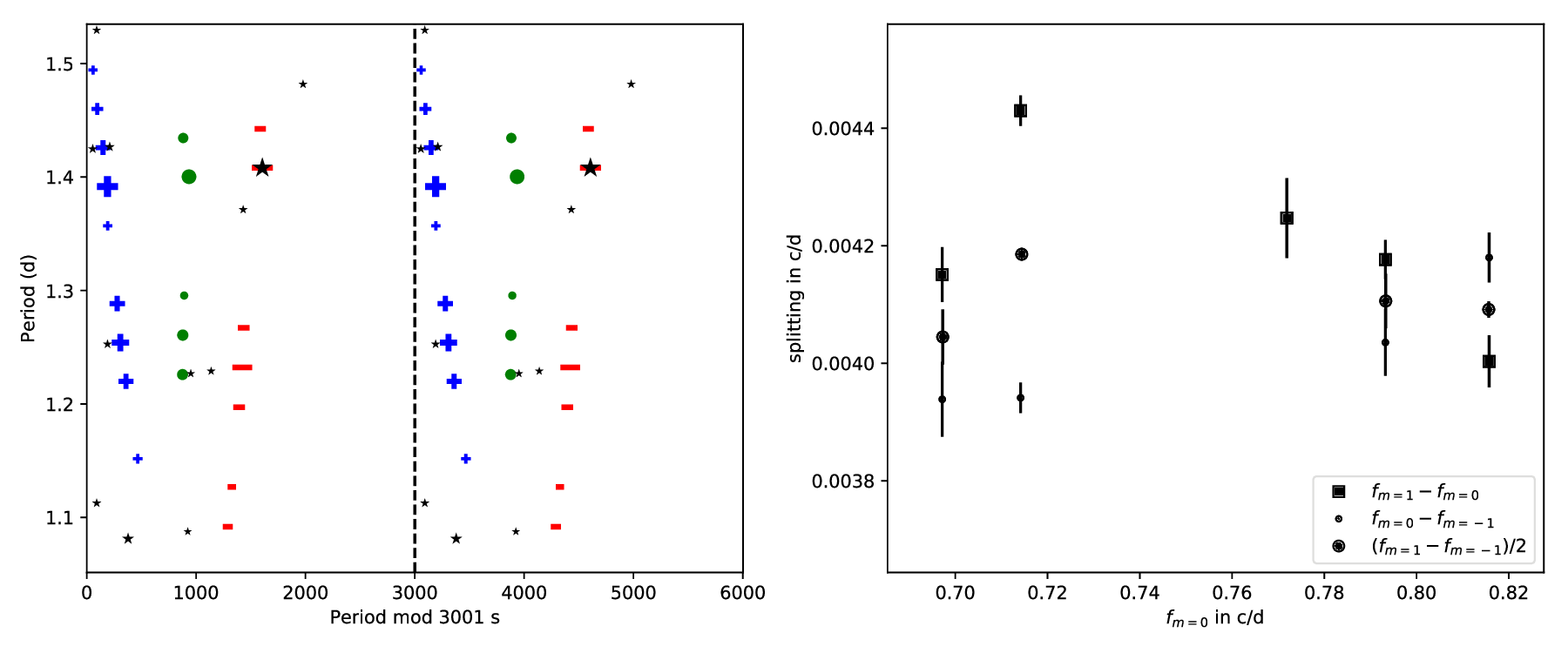}
\caption{The echelle diagram and the splitting variations of KIC\,4480321.}\label{fig:KIC 4480321vertical echelle}
\end{figure*}

\clearpage
\begin{figure*}
\centering
\includegraphics[width=0.9\textwidth]{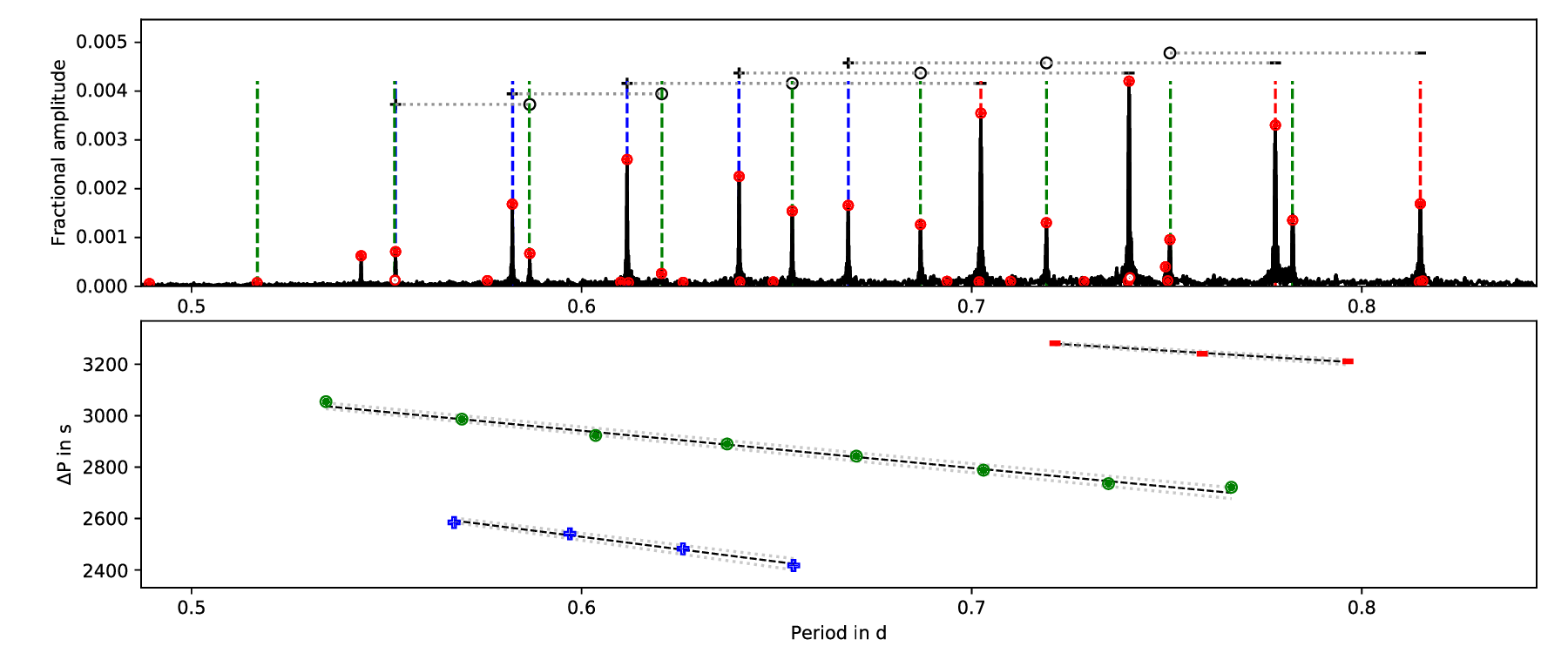}
\caption{The period spacing patterns of KIC\,4919344.}\label{fig:KIC 4919344}
\end{figure*}

\begin{figure*}
\centering
\includegraphics[width=0.85\textwidth]{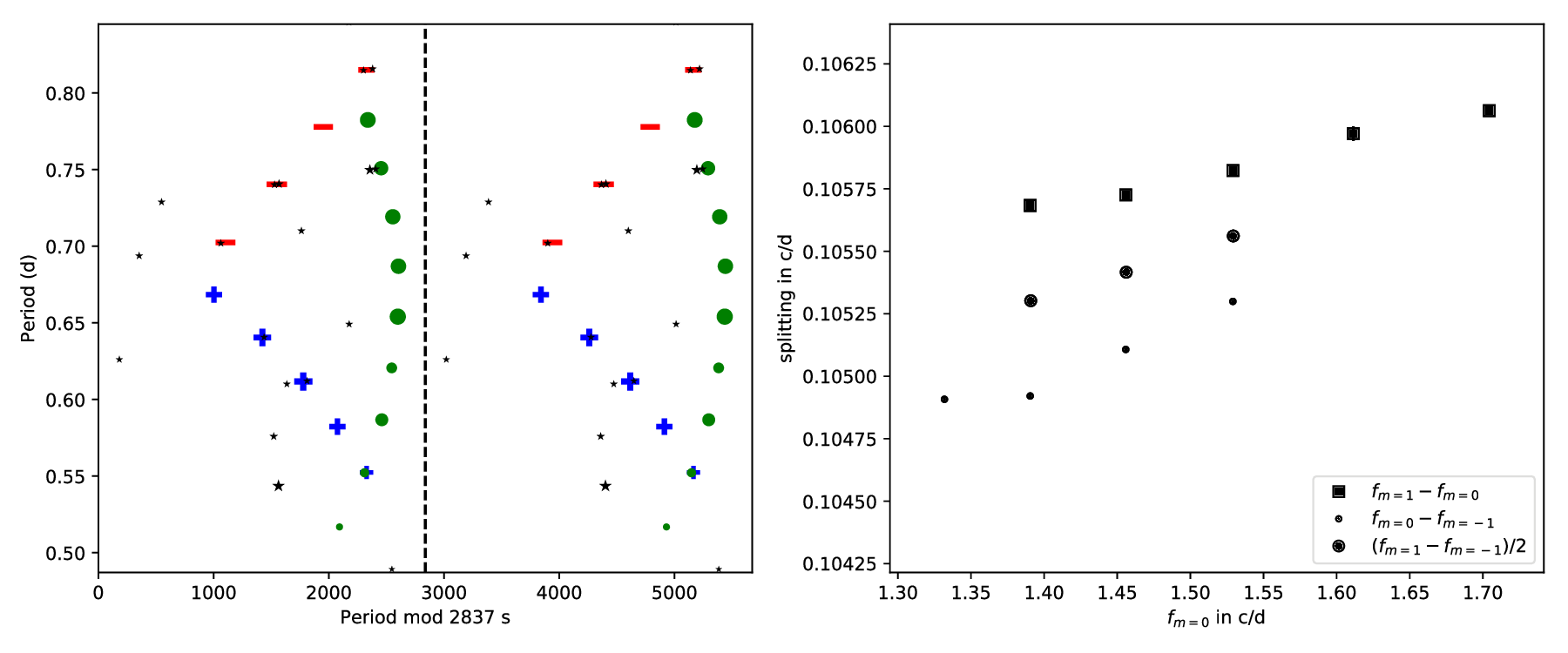}
\caption{The \'{e}chelle diagram and the splitting variations of KIC\,4919344.}\label{fig:KIC 4919344echelle_splitting}
\end{figure*}

\clearpage
\begin{figure*}
\centering
\includegraphics[width=0.9\textwidth]{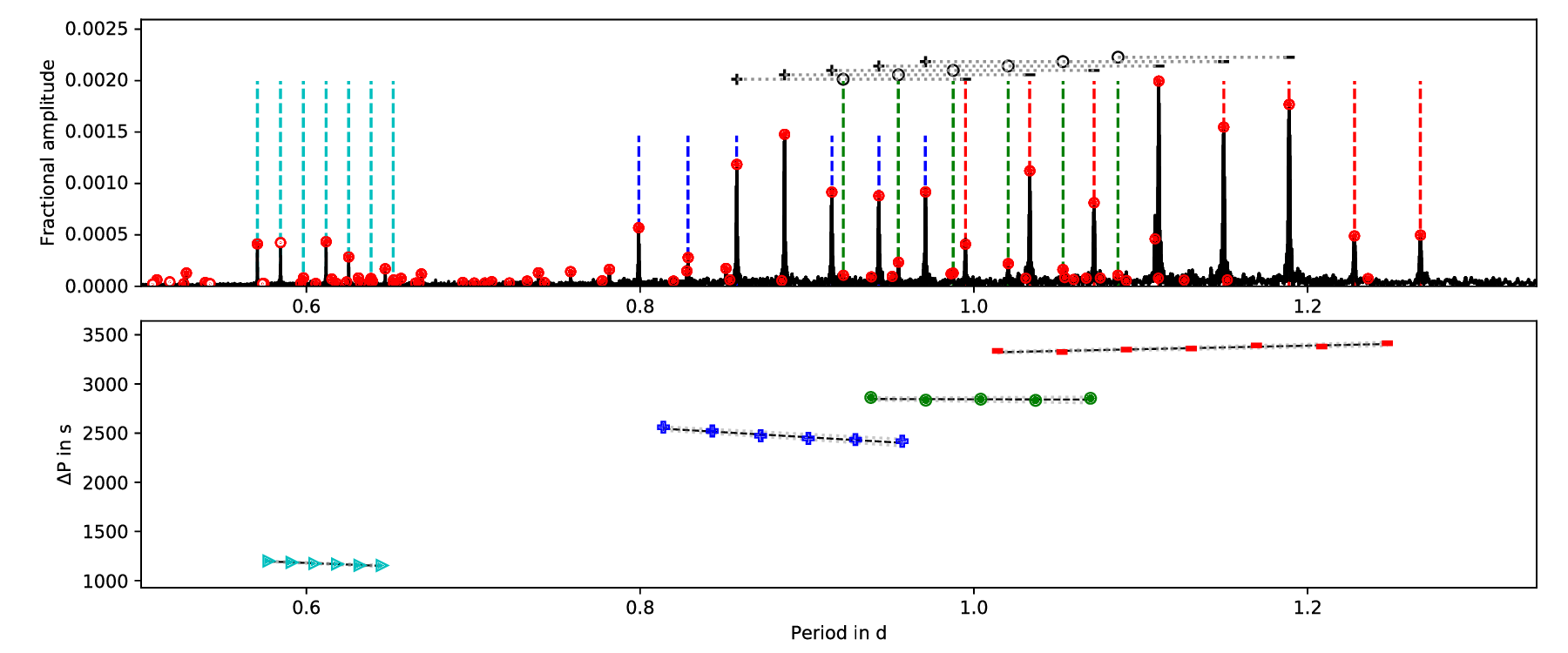}
\caption{The period spacing patterns of KIC\,5038228. $l=2, m=2$ patterns are seen, shown as cyan right triangles. }\label{fig:KIC 5038228}
\end{figure*}

\begin{figure*}
\centering
\includegraphics[width=0.85\textwidth]{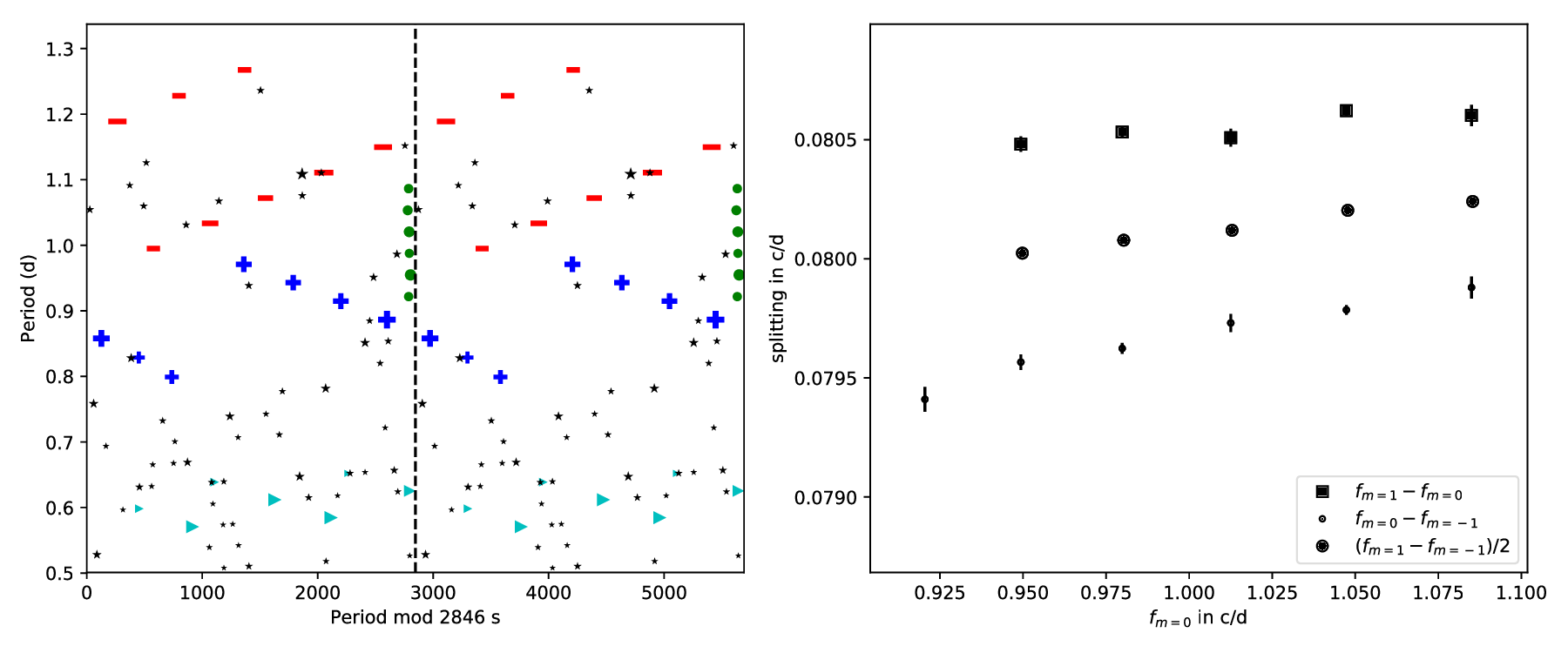}
\caption{The \'{e}chelle diagram and the splitting variations of KIC\,5038228.}\label{fig:KIC 5038228echelle_splitting}
\end{figure*}

\begin{figure*}
 \centering
 \includegraphics[width=0.8\textwidth]{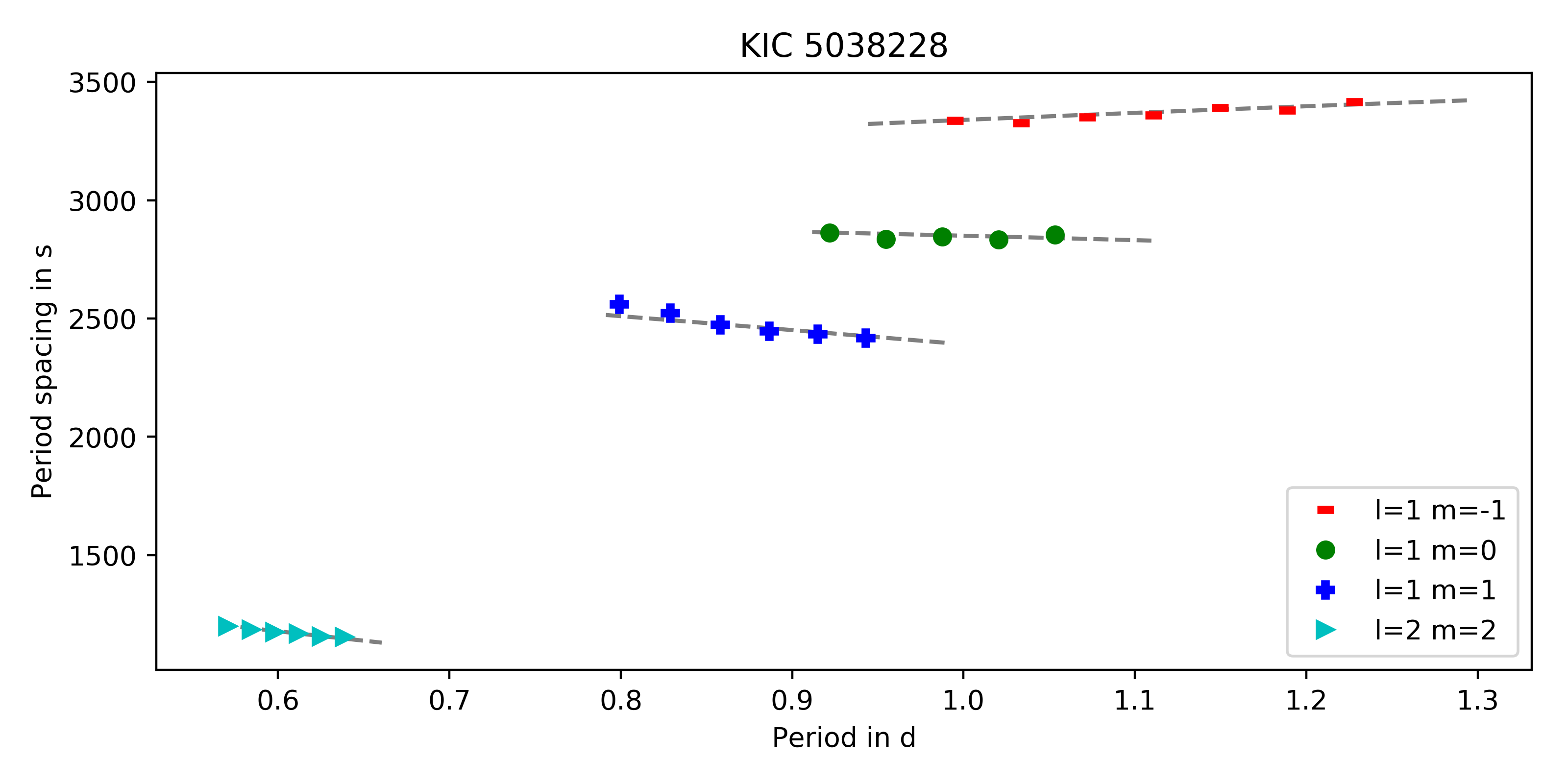}
 \caption{The modes identification of KIC\,5038228 from the TAR.}\label{fig:KIC5038228mode_identification}
\end{figure*}

\clearpage
\begin{figure*}
\centering
\includegraphics[width=0.9\textwidth]{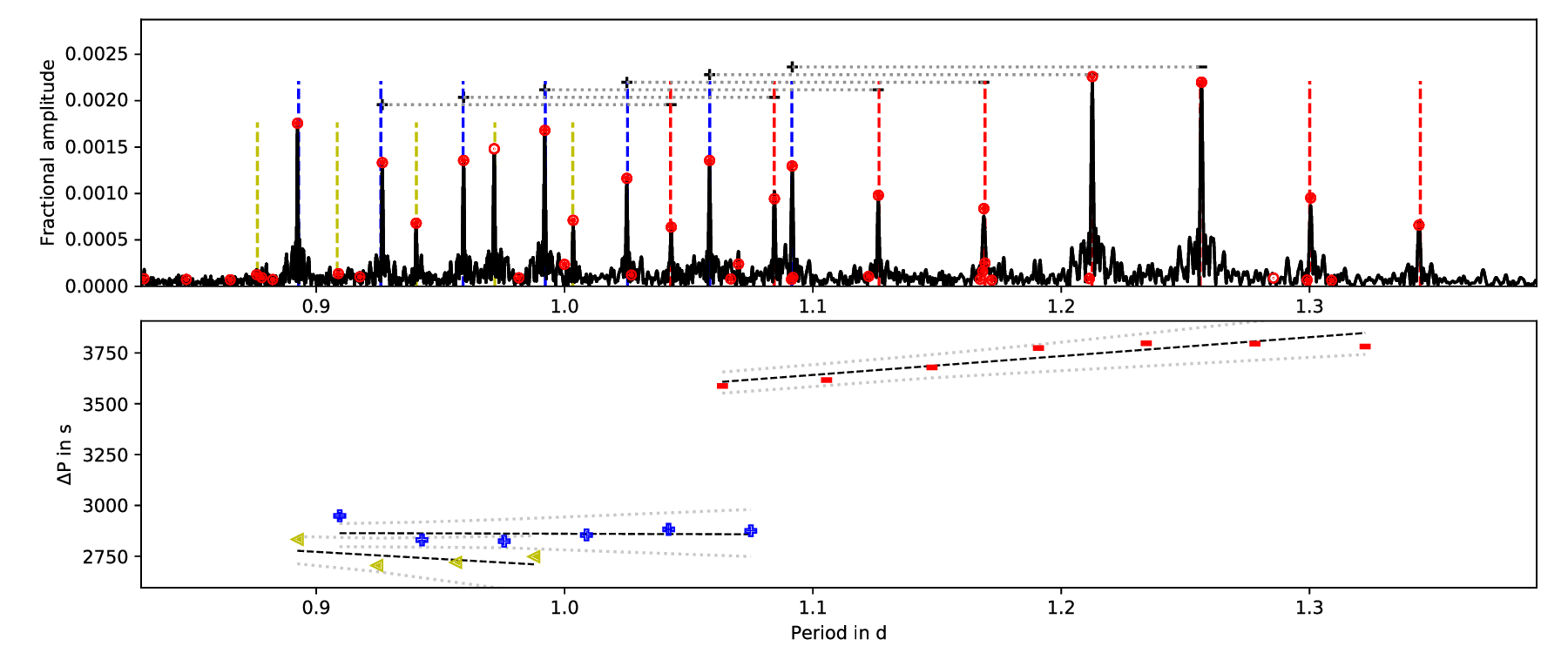}
\caption{The period spacing patterns of KIC\,5459805. Note that the splitings are not the effect of rotation. }\label{fig:KIC 5459805}
\end{figure*}

\begin{figure*}
\centering
\includegraphics[width=0.85\textwidth]{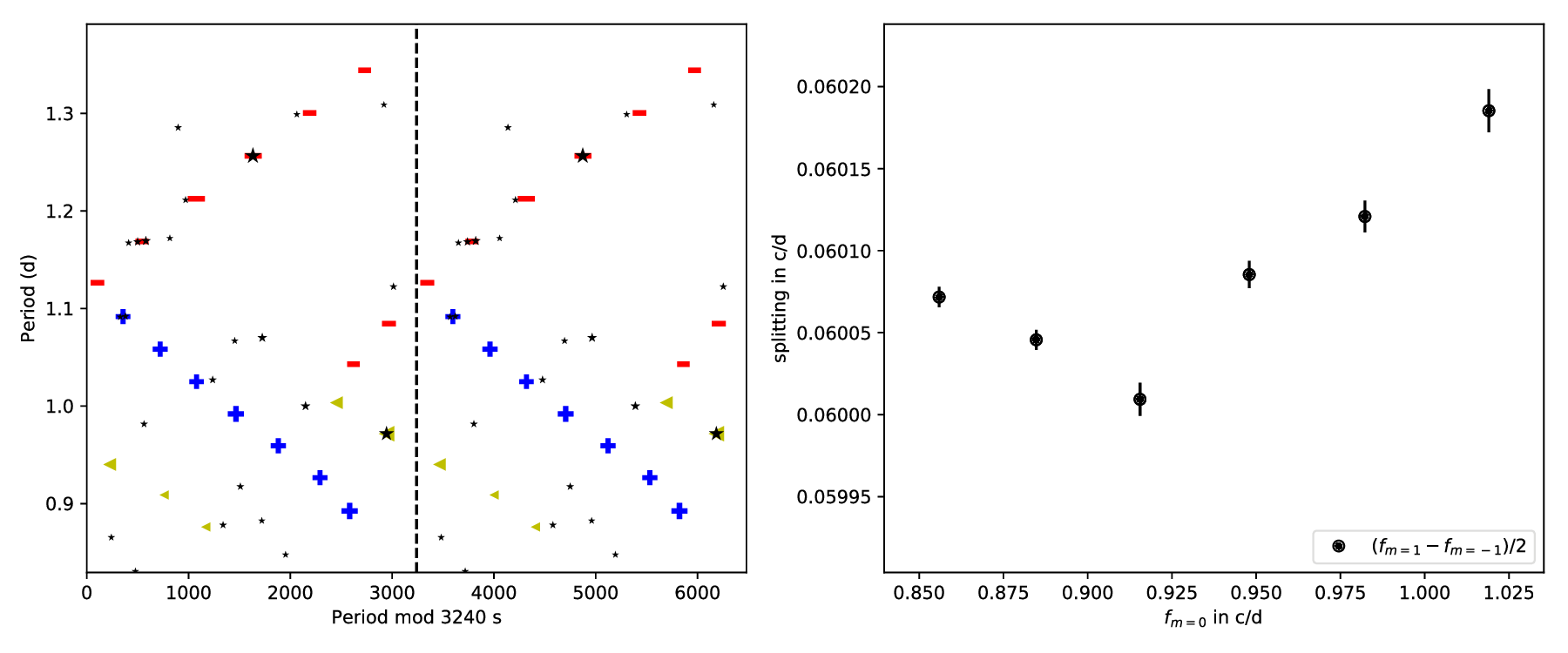}
\caption{The \'{e}chelle diagram and the splitting variations of KIC\,5459805.}\label{fig:KIC 5459805echelle_splitting}
\end{figure*}

\begin{figure*}
 \centering
 \includegraphics[width=0.8\textwidth]{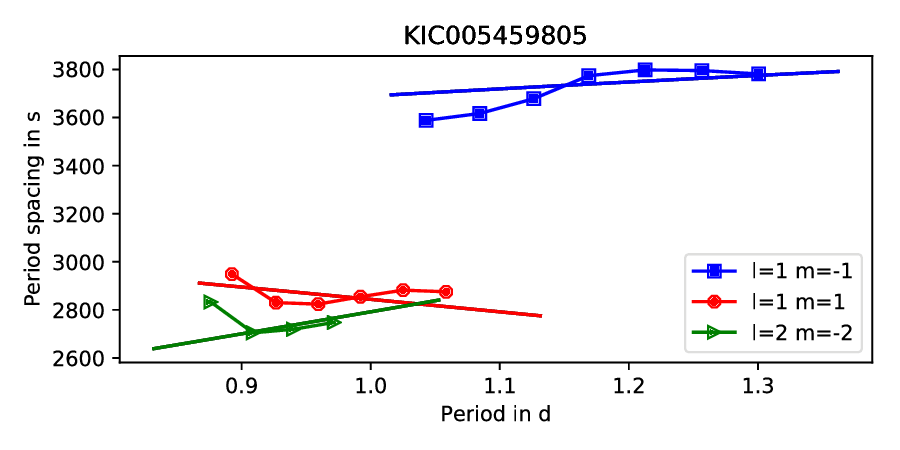}
 \caption{The modes identification of KIC\,5459805.}\label{fig:KIC5459805mode_identification}
\end{figure*}

\clearpage
\begin{figure*}
\centering
\includegraphics[width=0.9\textwidth]{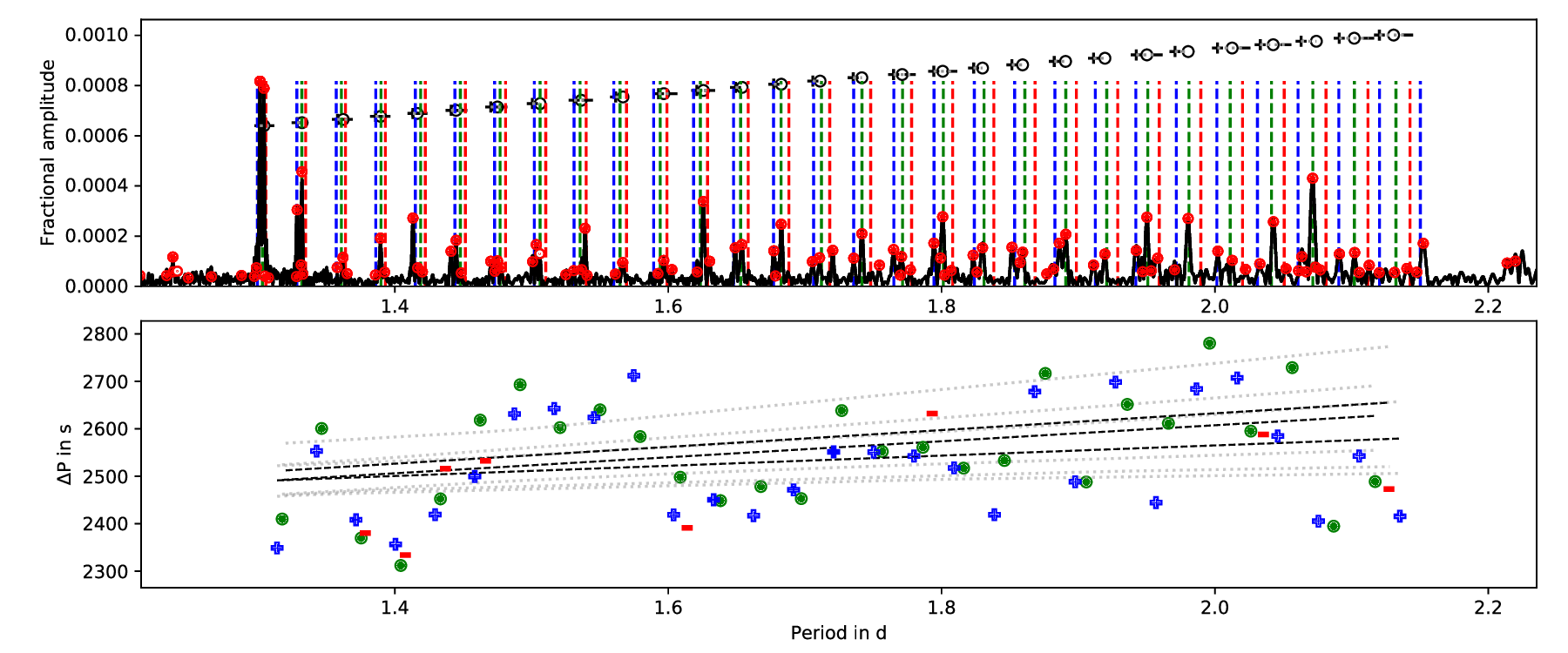}
\caption{The period spacing patterns of KIC\,5557072.}\label{fig:KIC 5557072}
\end{figure*}

\begin{figure*}
\centering
\includegraphics[width=0.85\textwidth]{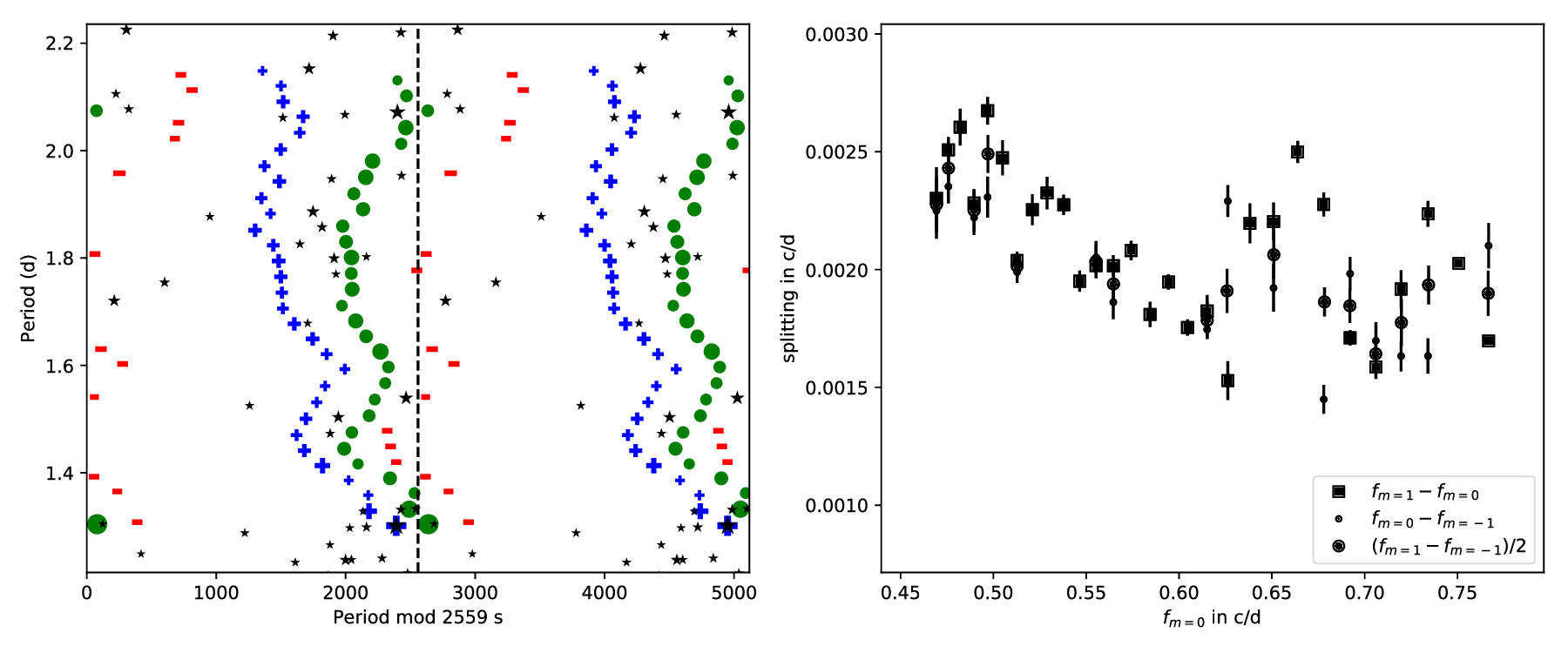}
\caption{The \'{e}chelle diagram and the splitting variations of KIC\,5557072.}\label{fig:KIC 5557072echelle_splitting}
\end{figure*}

\clearpage
\begin{figure*}
\centering
\includegraphics[width=0.9\textwidth]{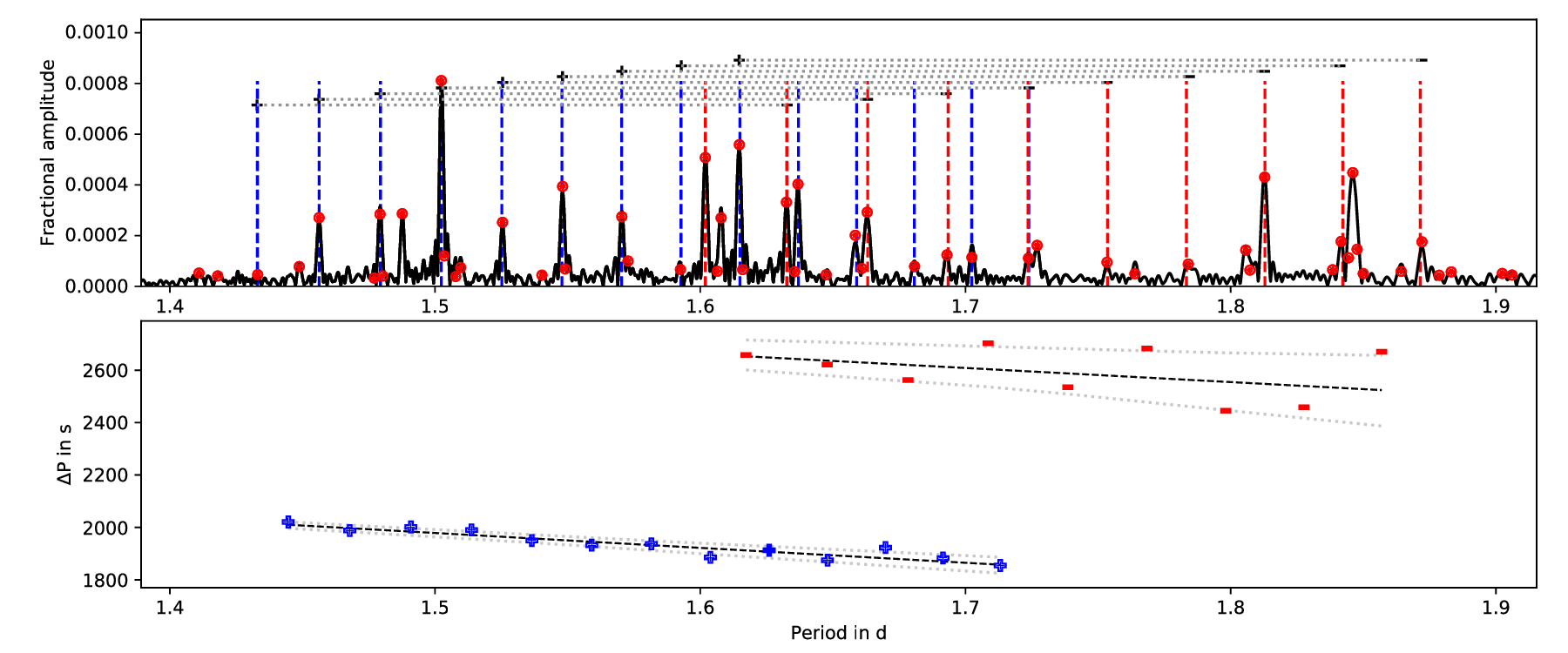}
\caption{The period spacing patterns of KIC\,5810197.}\label{fig:KIC 5810197}
\end{figure*}

\begin{figure*}
\centering
\includegraphics[width=0.85\textwidth]{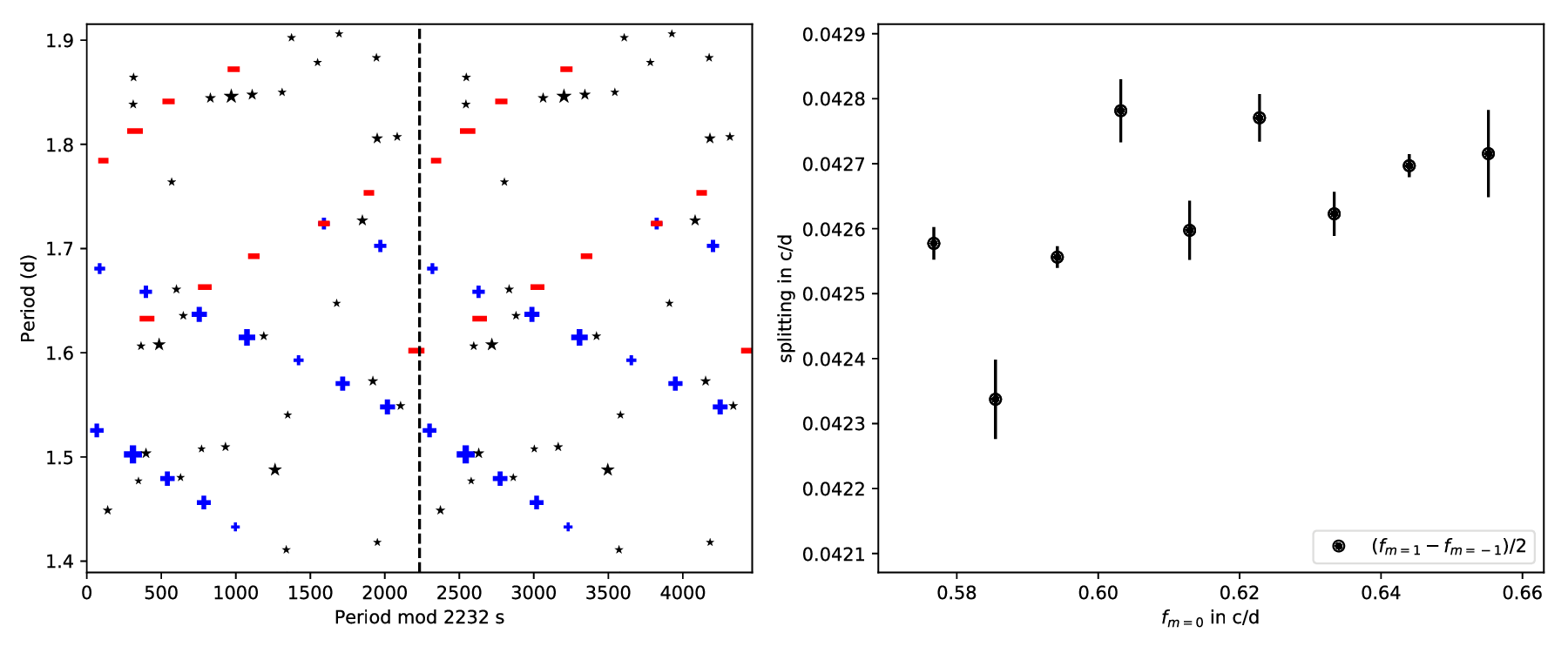}
\caption{The \'{e}chelle diagram and the splitting variations of KIC\,5810197.}\label{fig:KIC 5810197echelle_splitting}
\end{figure*}

\clearpage
\begin{figure*}
\centering
\includegraphics[width=0.9\textwidth]{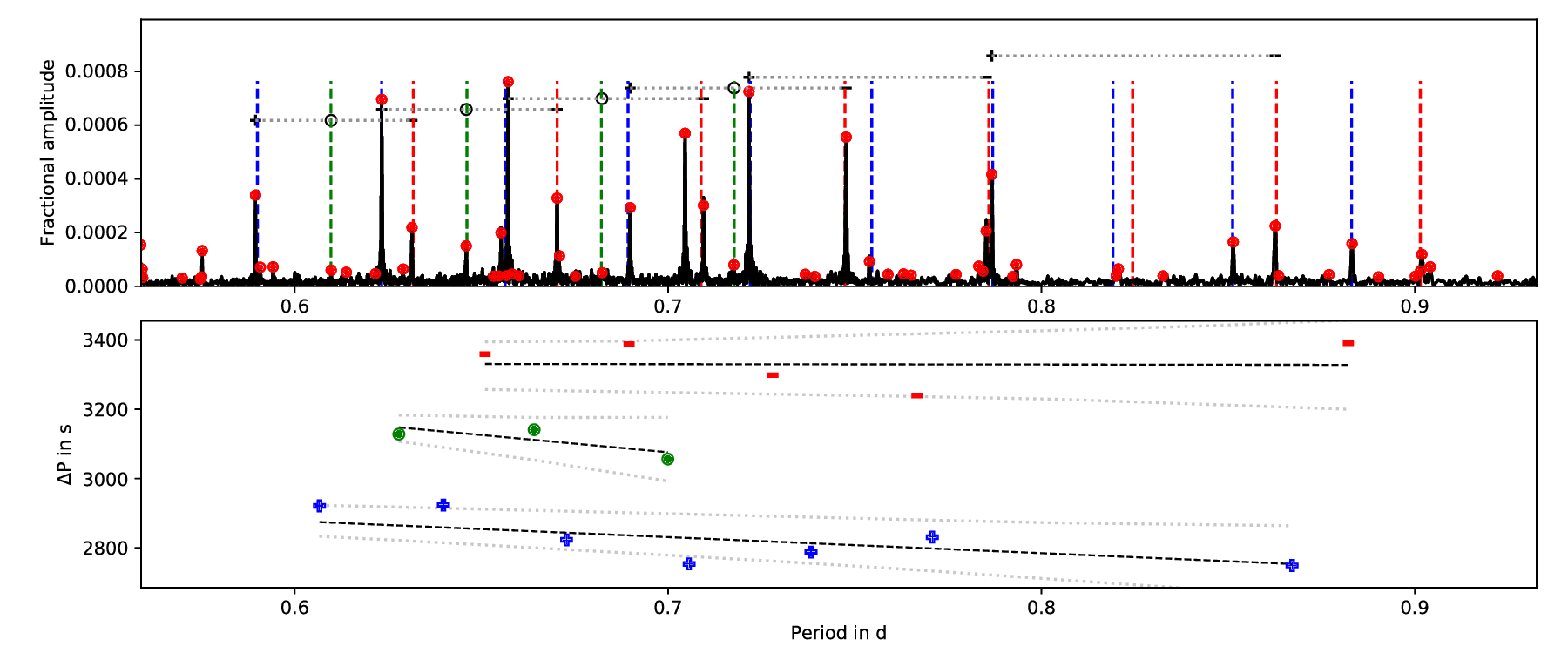}
\caption{The period spacing patterns of KIC\,6302589.}\label{fig:KIC 6302589}
\end{figure*}

\begin{figure*}
\centering
\includegraphics[width=0.85\textwidth]{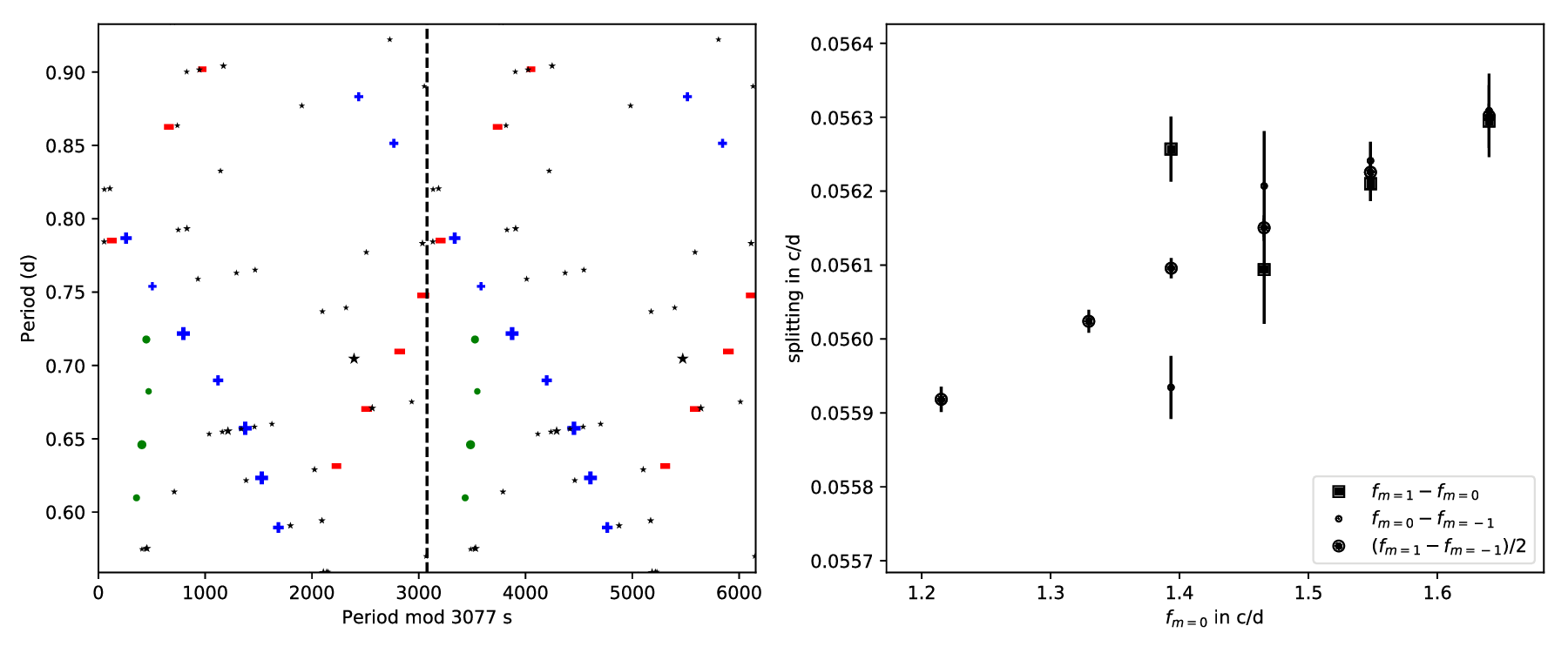}
\caption{The \'{e}chelle diagram and the splitting variations of KIC\,6302589.}\label{fig:KIC 6302589echelle_splitting}
\end{figure*}

\clearpage
\begin{figure*}
\centering
\includegraphics[width=0.9\textwidth]{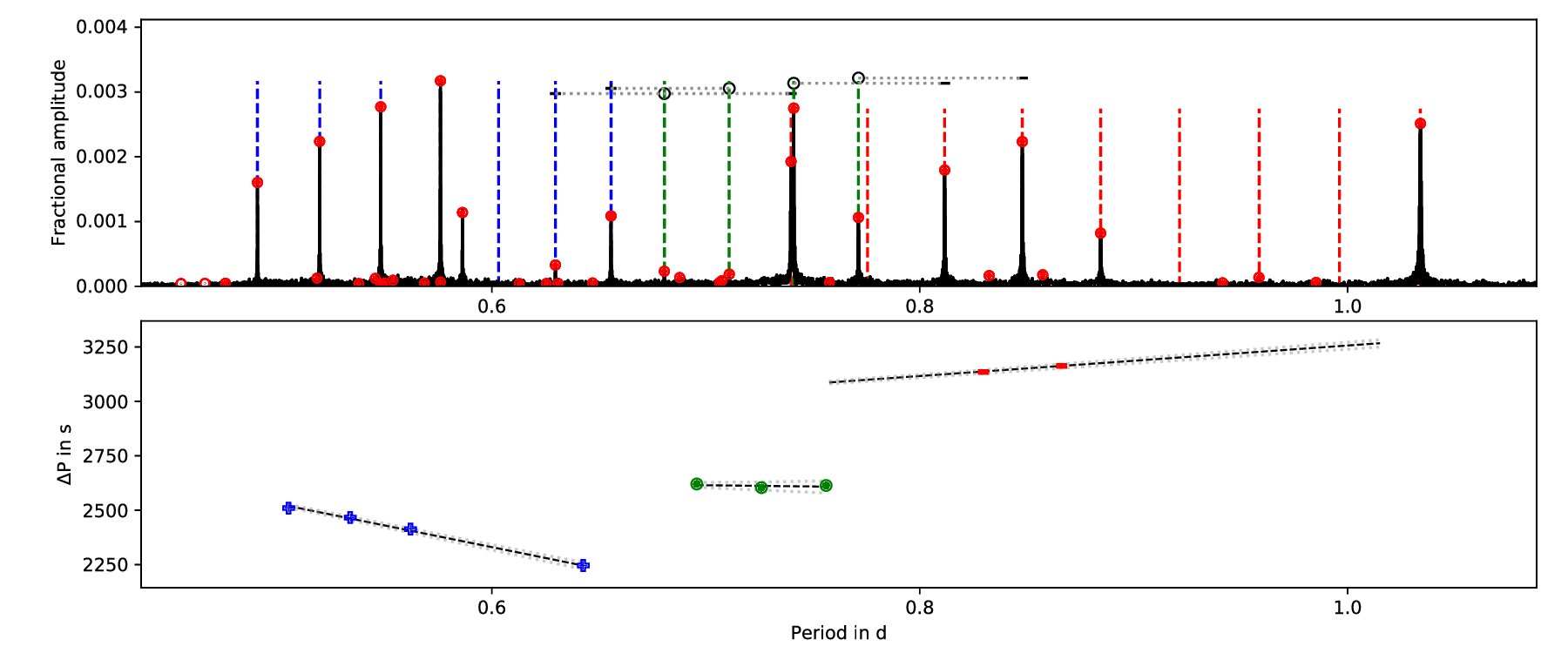}
\caption{The period spacing patterns of KIC\,6467639.}\label{fig:KIC 6467639}
\end{figure*}

\begin{figure*}
\centering
\includegraphics[width=0.85\textwidth]{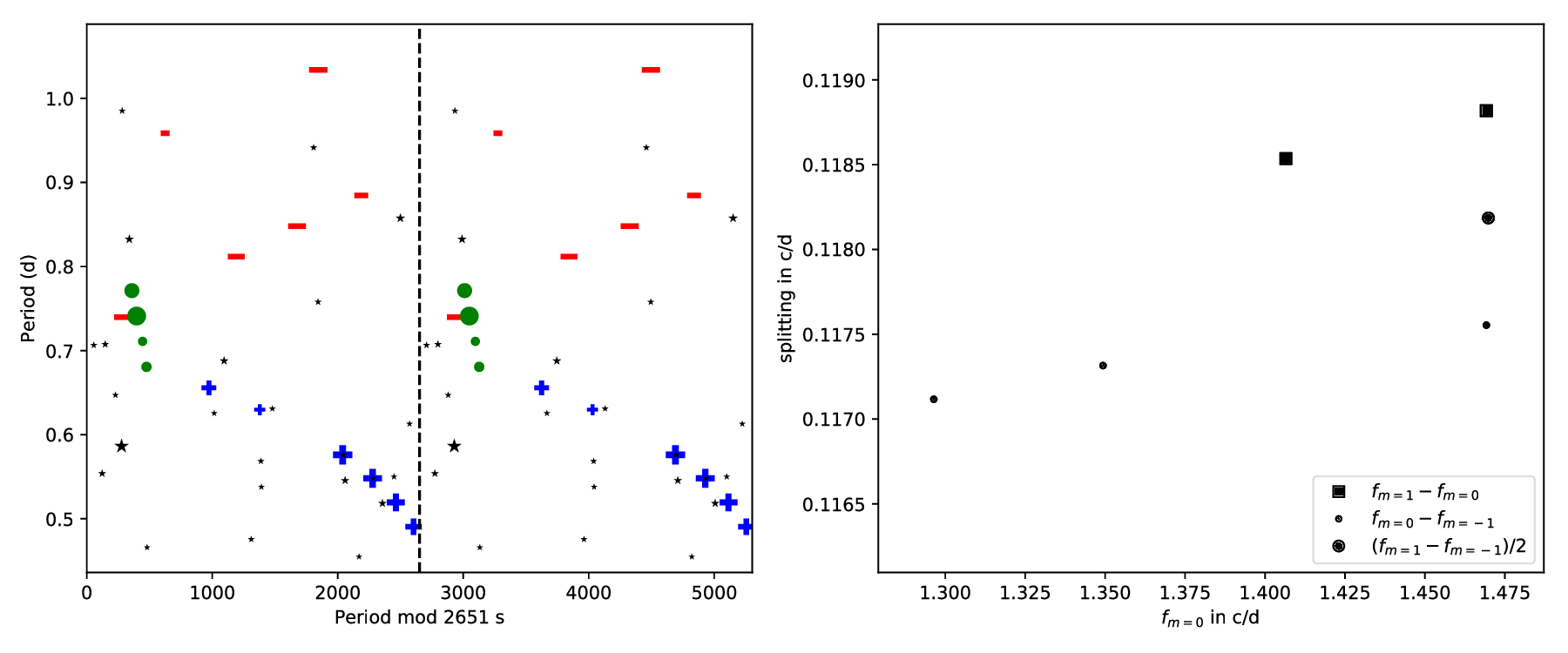}
\caption{The \'{e}chelle diagram and the splitting variations of KIC\,6467639.}\label{fig:KIC 6467639echelle_splitting}
\end{figure*}

\clearpage
\begin{figure*}
\centering
\includegraphics[width=0.9\textwidth]{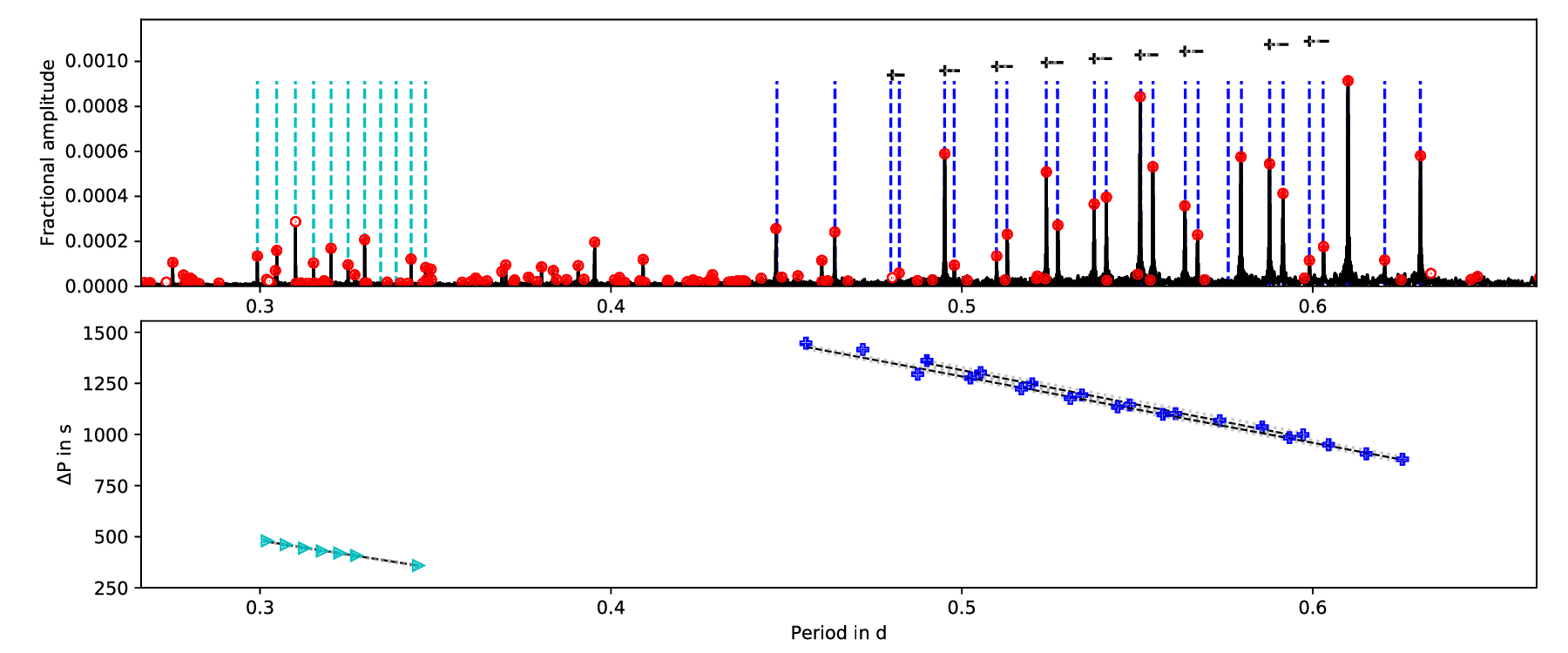}
\caption{The period spacing patterns of KIC\,6862920. Note that the splitings are not the effect of rotation. }\label{fig:KIC 6862920}
\end{figure*}

\begin{figure*}
\centering
\includegraphics[width=0.85\textwidth]{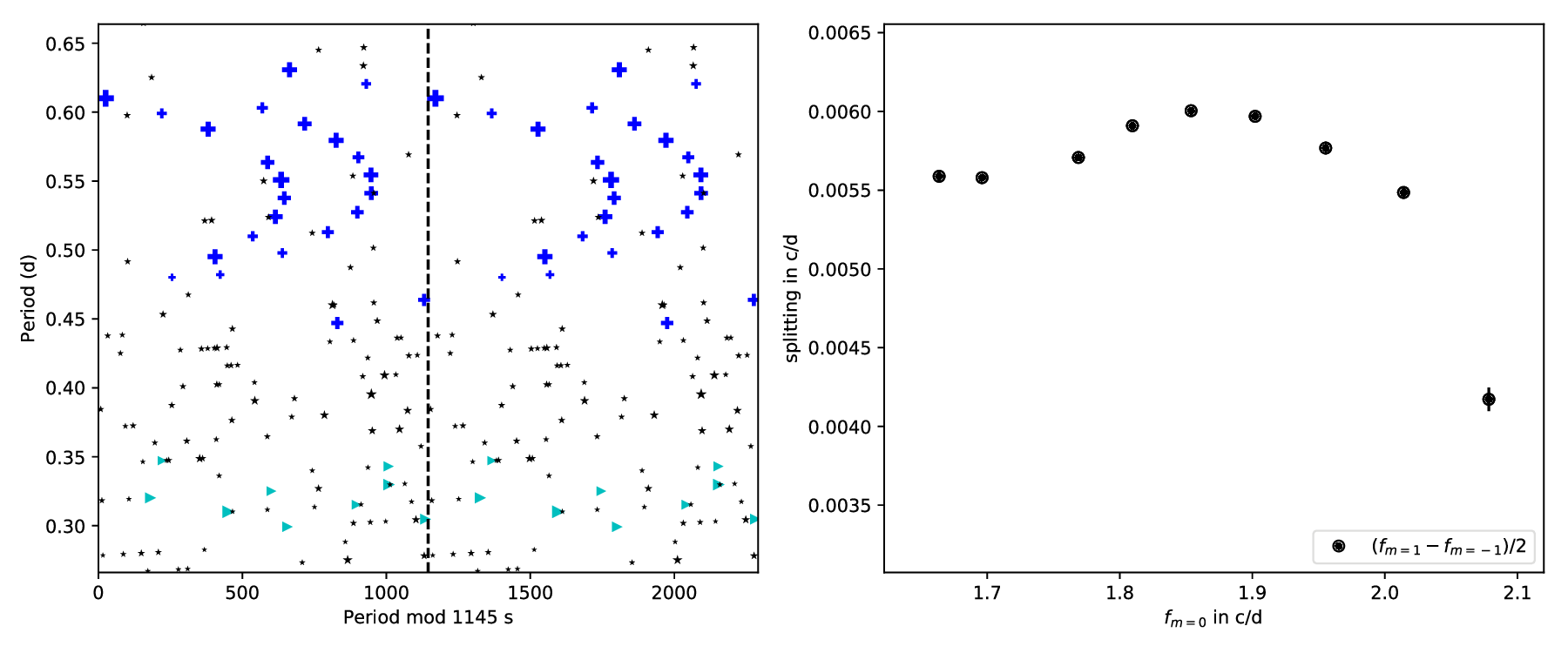}
\caption{The \'{e}chelle diagram and the splitting variations of KIC\,6862920. Note that the splittings are not caused by the rotational effect.}\label{fig:KIC 6862920echelle_splitting}
\end{figure*}

\clearpage
\begin{figure*}
\centering
\includegraphics[width=0.9\textwidth]{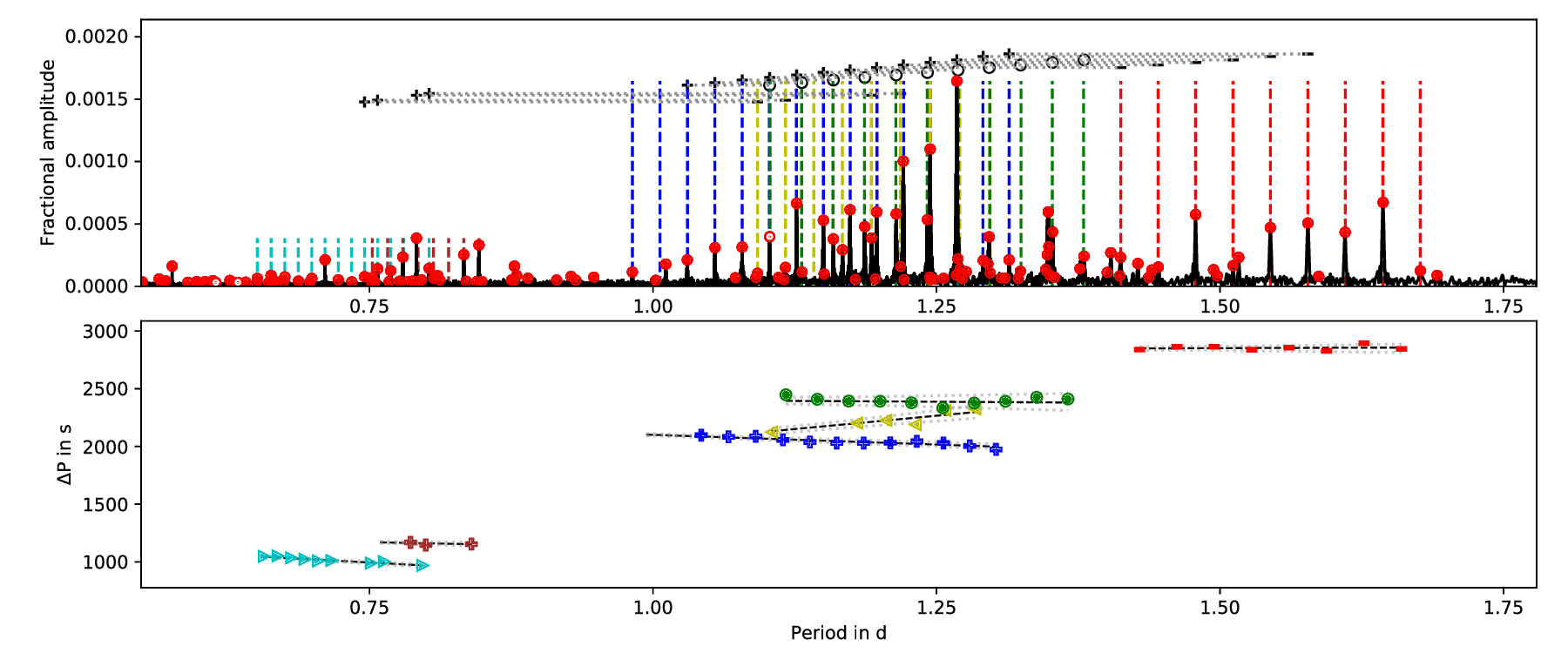}
\caption{The period spacing patterns of KIC\,6937123. Note that the splittings are not caused by the rotational effect.}\label{fig:KIC 6937123}
\end{figure*}

\begin{figure*}
\centering
\includegraphics[width=0.85\textwidth]{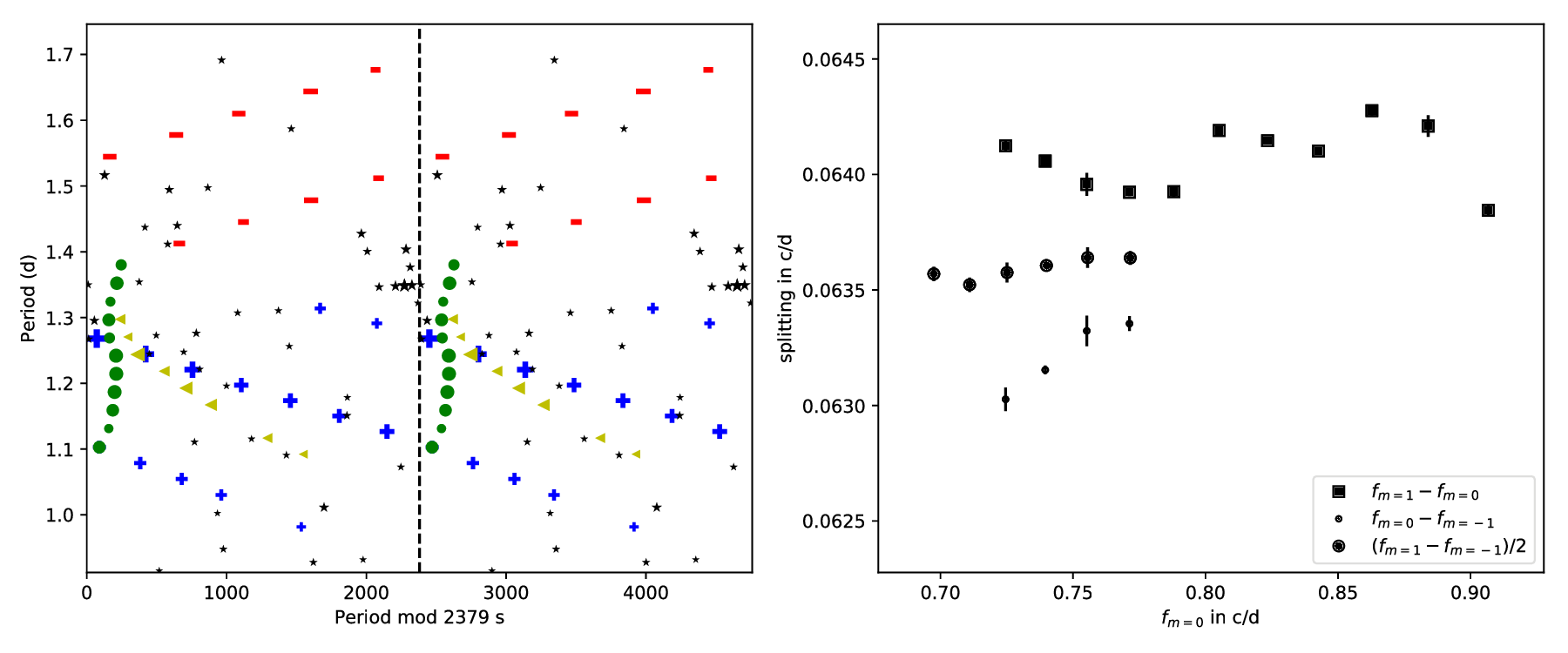}
\caption{The \'{e}chelle diagram and the splitting variations of KIC\,6937123 of $l=1$ modes.}\label{fig:KIC 6937123echelle_splitting_l_1}
\end{figure*}

\begin{figure*}
\centering
\includegraphics[width=0.85\textwidth]{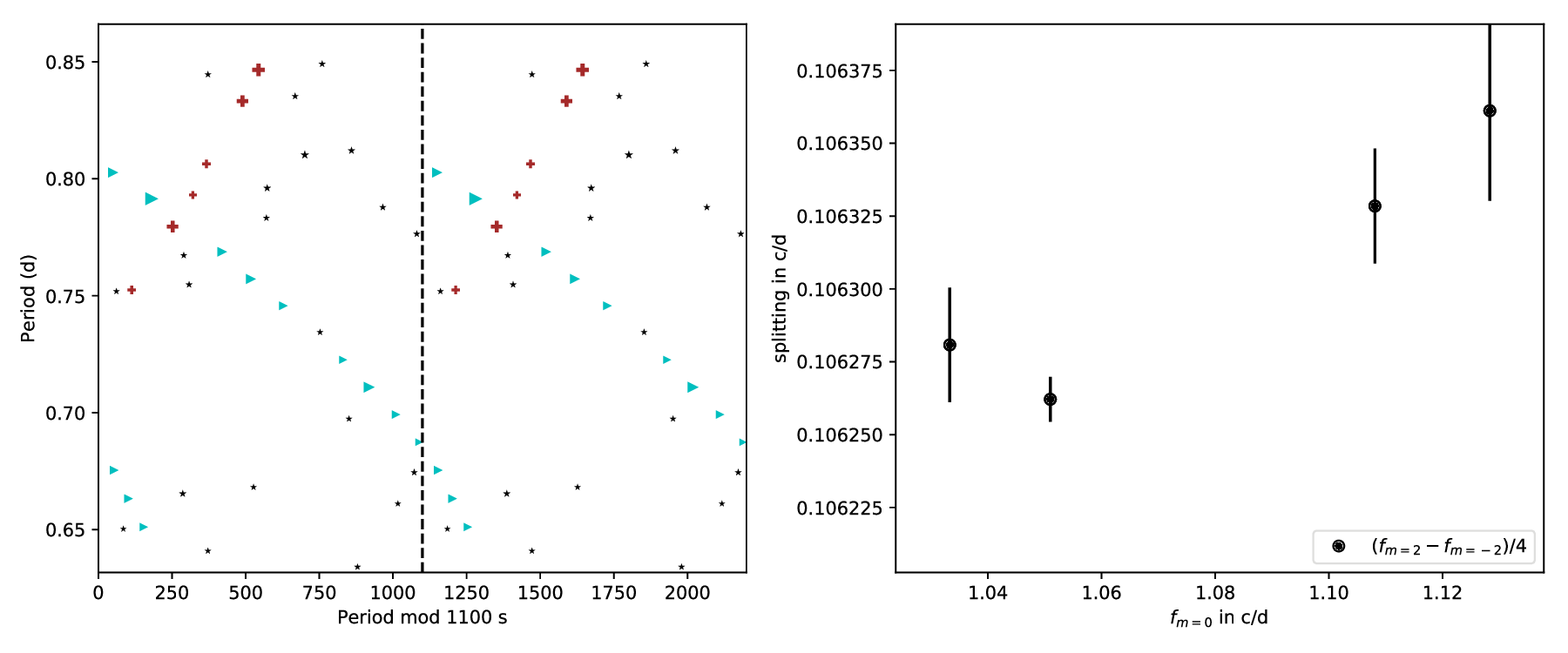}
\caption{The \'{e}chelle diagram and the splitting variations of KIC\,6937123 of $l=2$ modes.}\label{fig:KIC 6937123echelle_splitting_l_2}
\end{figure*}

\begin{figure*}
 \centering
 \includegraphics[width=0.8\textwidth]{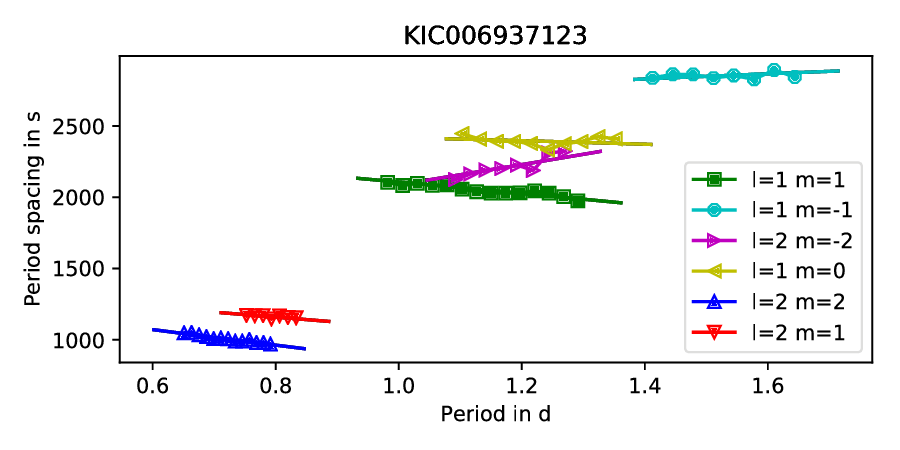}
 \caption{The modes identification of KIC\,6937123.}\label{fig:KIC6937123mode_identification}
\end{figure*}

\clearpage
\begin{figure*}
\centering
\includegraphics[width=0.9\textwidth]{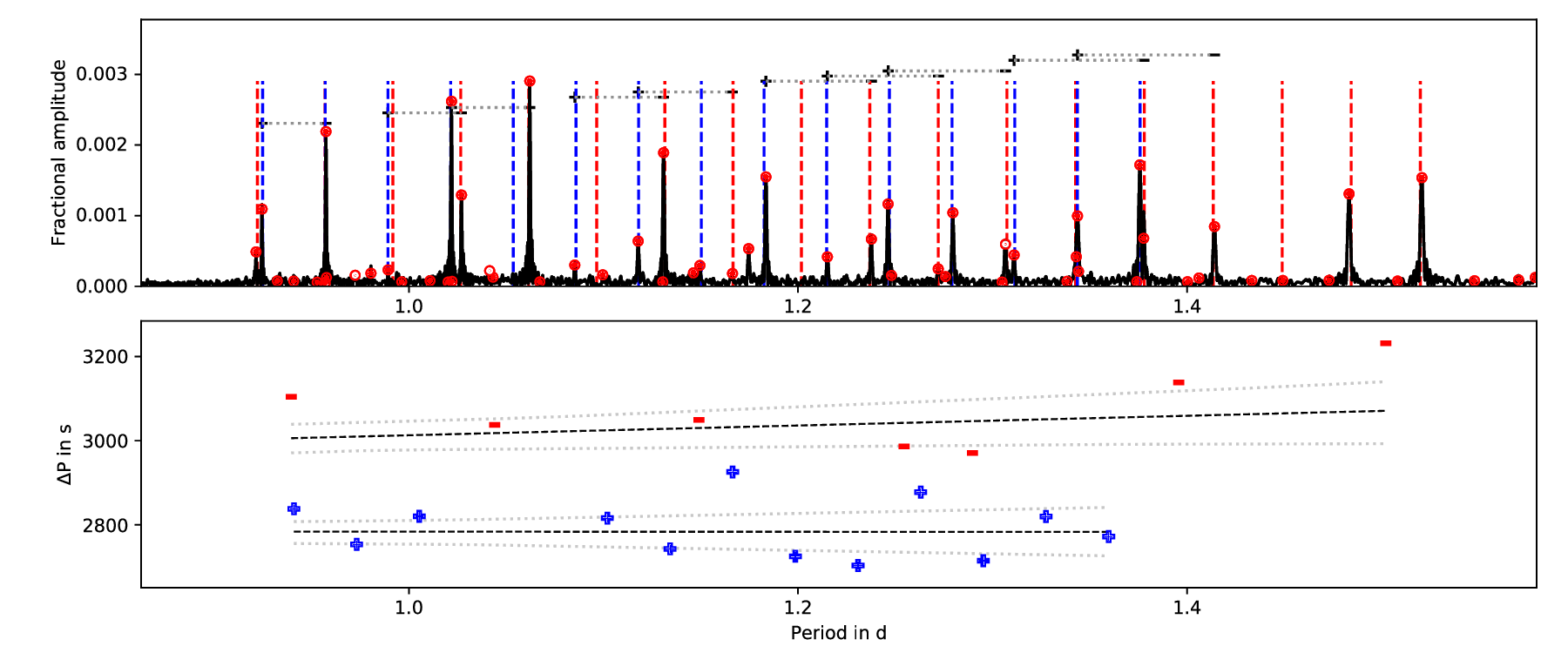}
\caption{The period spacing patterns of KIC\,7661054.}\label{fig:KIC 7661054}
\end{figure*}

\begin{figure*}
\centering
\includegraphics[width=0.85\textwidth]{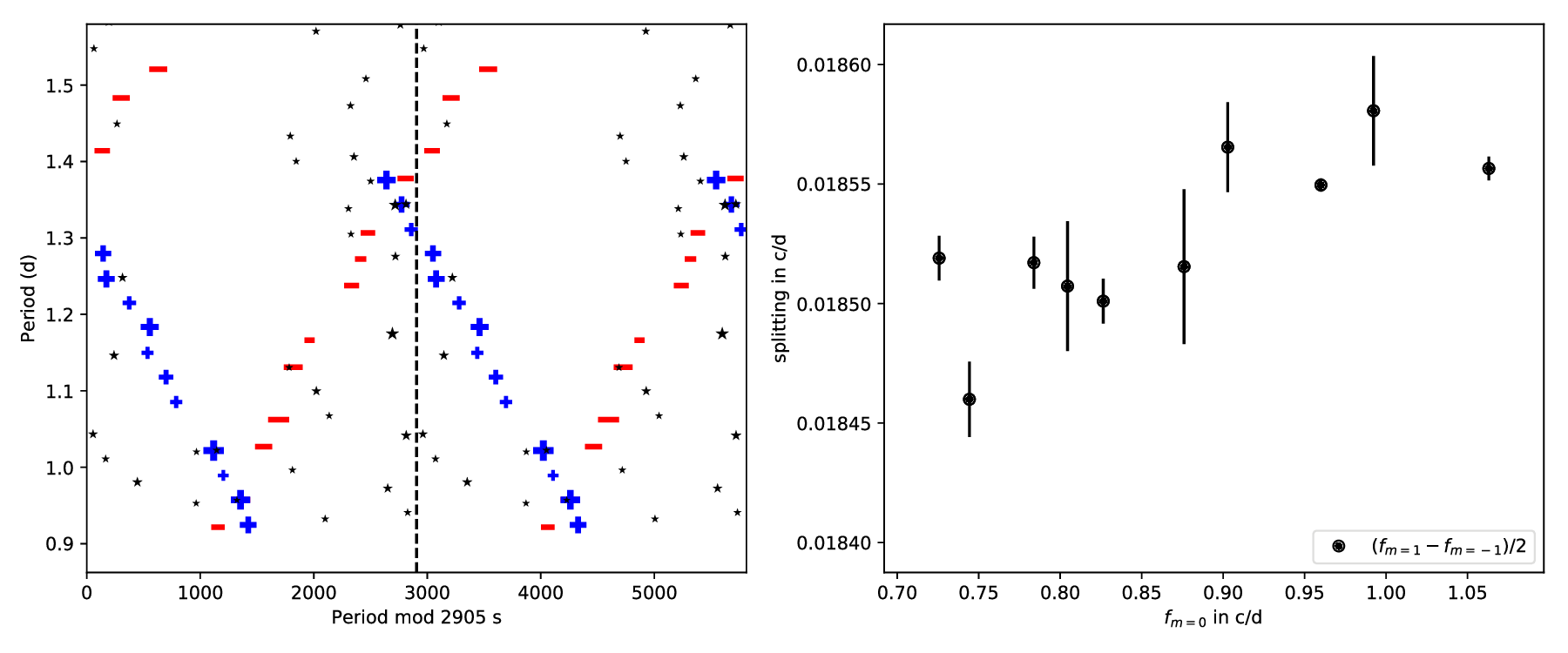}
\caption{The \'{e}chelle diagram and the splitting variations of KIC\,7661054.}\label{fig:KIC 7661054echelle_splitting}
\end{figure*}

\clearpage
\begin{figure*}
\centering
\includegraphics[width=0.9\textwidth]{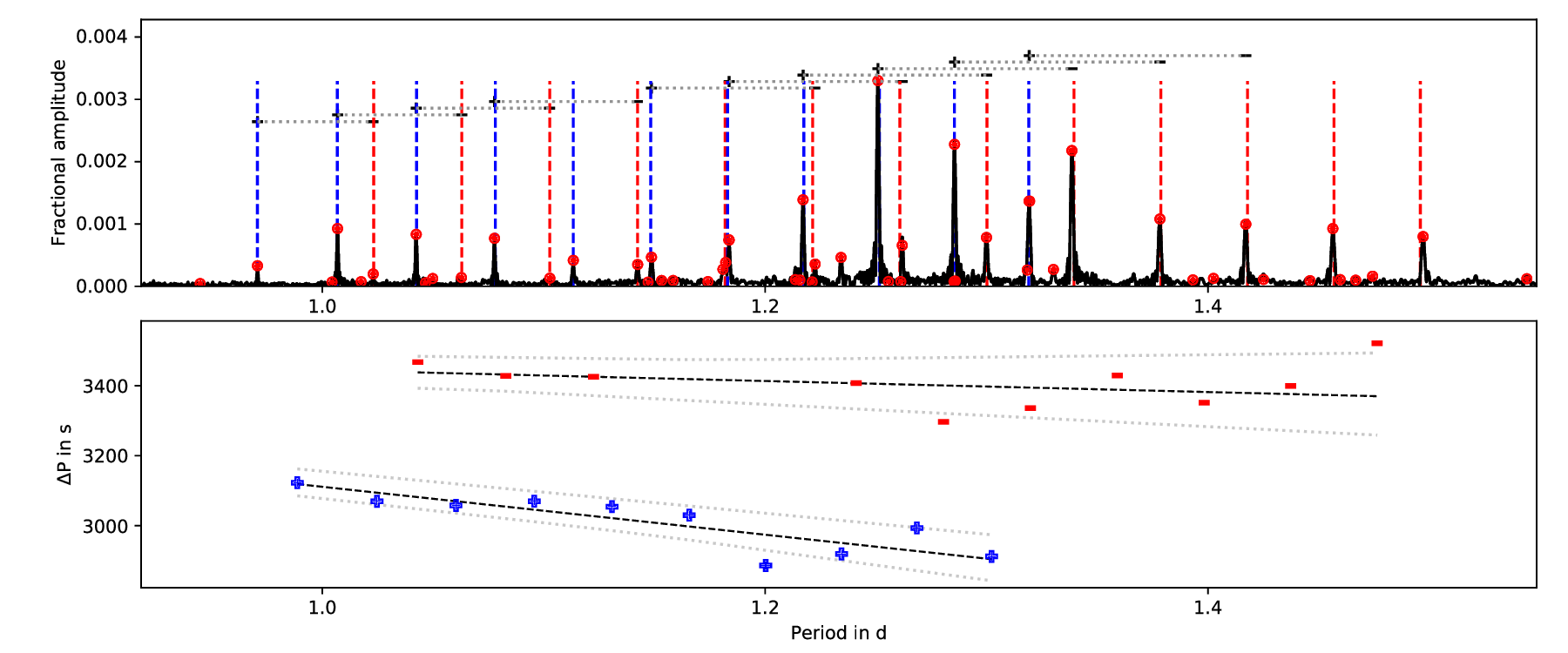}
\caption{The period spacing patterns of KIC\,7697861.}\label{fig:KIC 7697861}
\end{figure*}

\begin{figure*}
\centering
\includegraphics[width=0.85\textwidth]{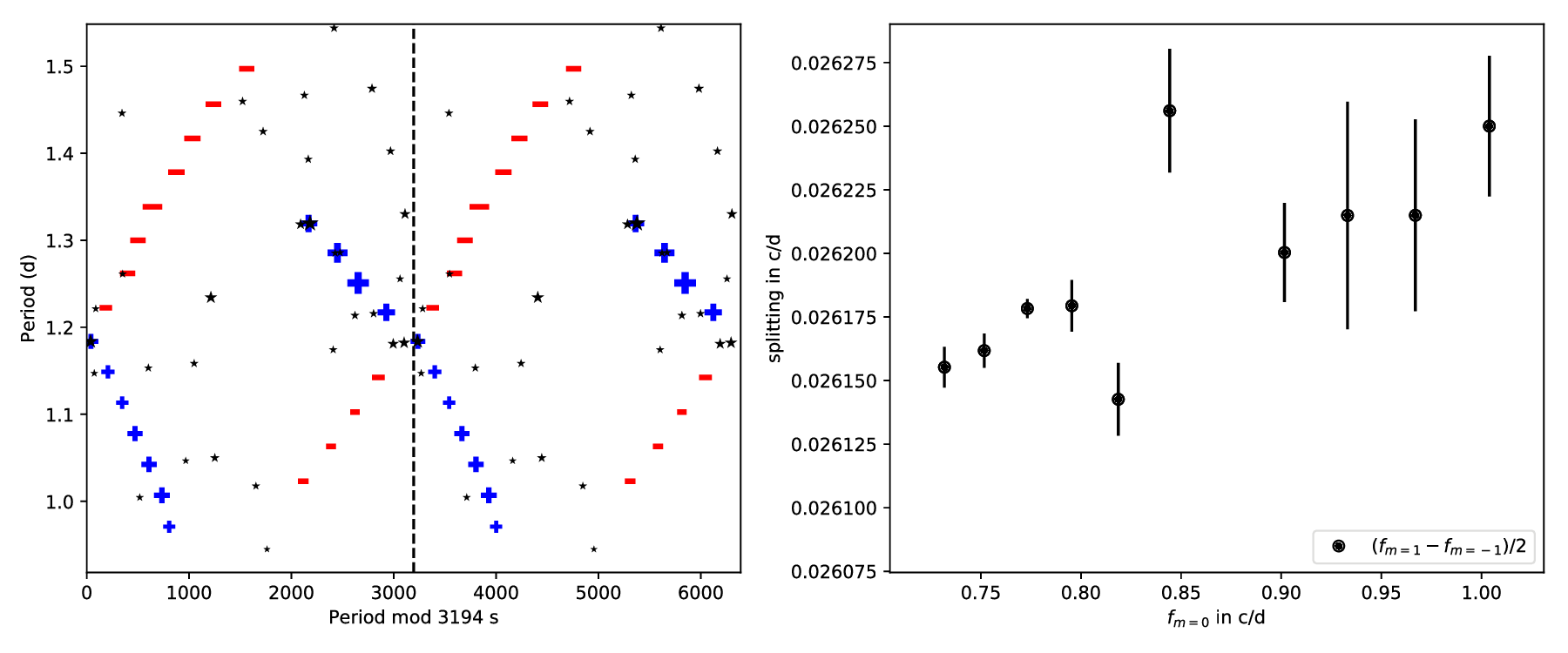}
\caption{The \'{e}chelle diagram and the splitting variations of KIC\,7697861.}\label{fig:KIC 7697861echelle_splitting}
\end{figure*}

\clearpage
\begin{figure*}
\centering
\includegraphics[width=0.9\textwidth]{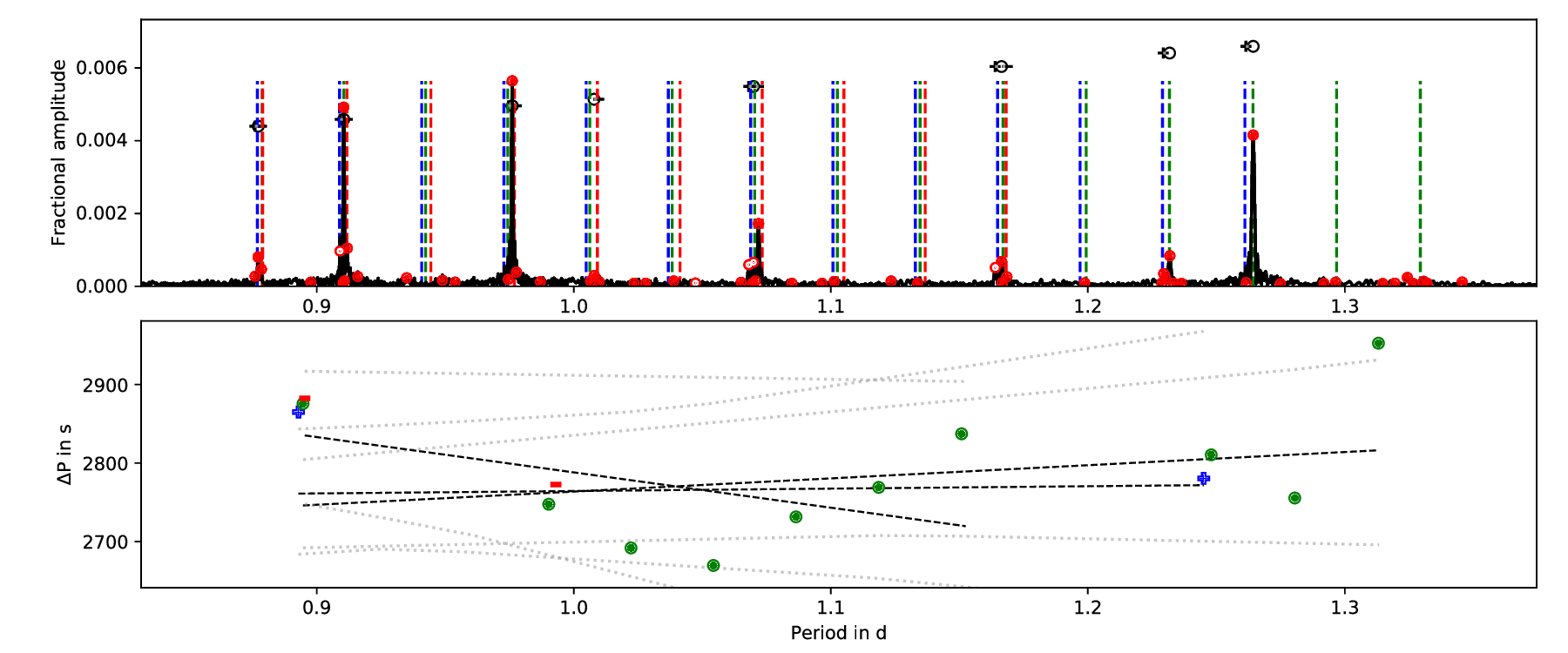}
\caption{The period spacing patterns of KIC\,8197761. Note the apparent dip shown in by the red circles.}\label{fig:KIC 8197761}
\end{figure*}

\begin{figure*}
\centering
\includegraphics[width=0.85\textwidth]{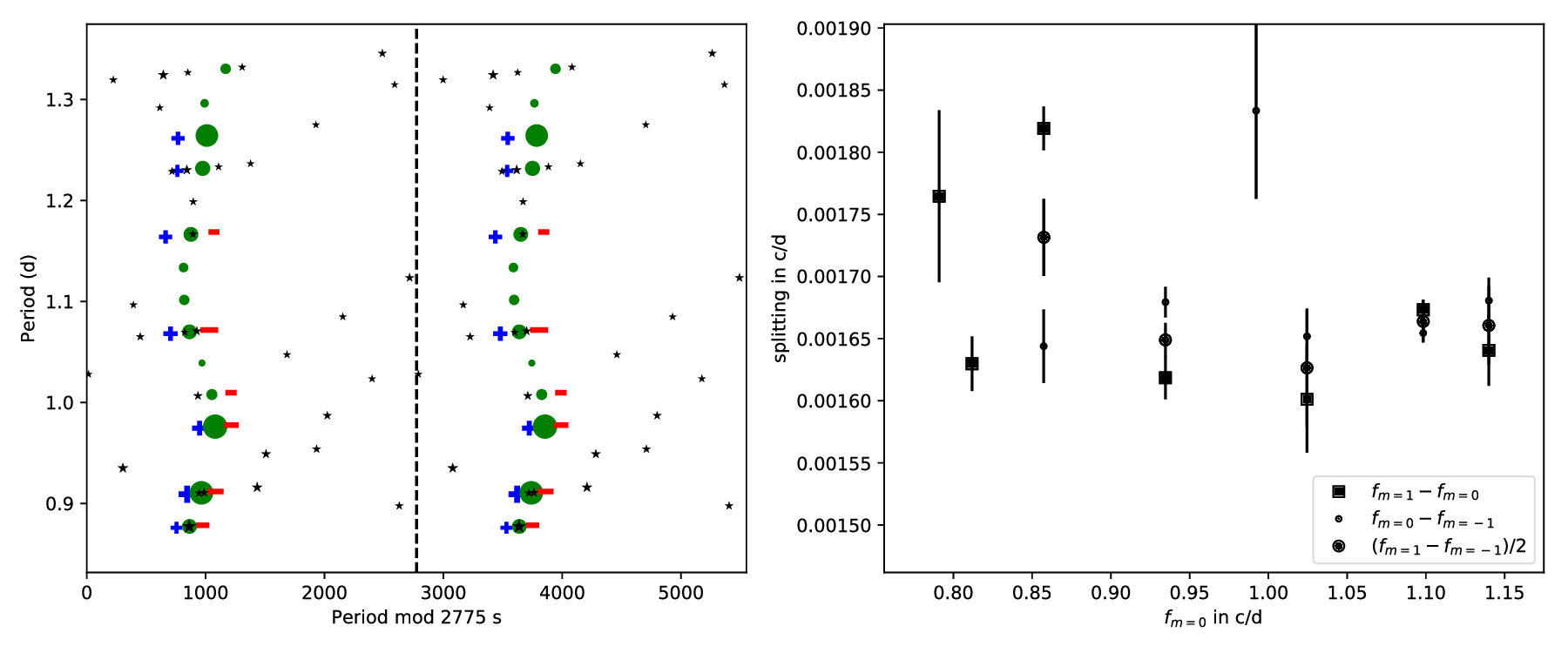}
\caption{The echelle diagram and the splitting variations of KIC\,8197761.}\label{fig:KIC 8197761vertical echelle}
\end{figure*}

\clearpage
\begin{figure*}
\centering
\includegraphics[width=0.9\textwidth]{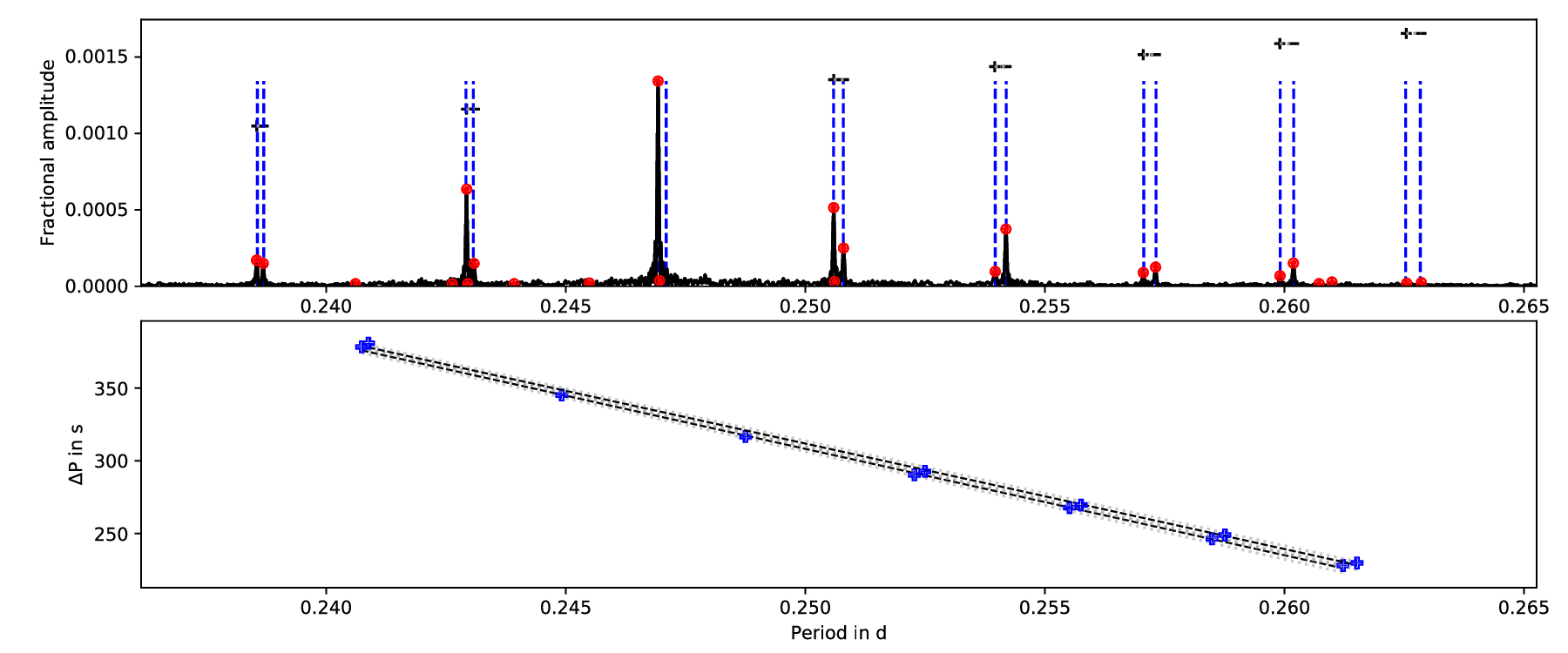}
\caption{The period spacing patterns of KIC\,8458690. Note that the splittings are not caused by the rotational effect.}\label{fig:KIC 8458690}
\end{figure*}

\begin{figure*}
\centering
\includegraphics[width=0.85\textwidth]{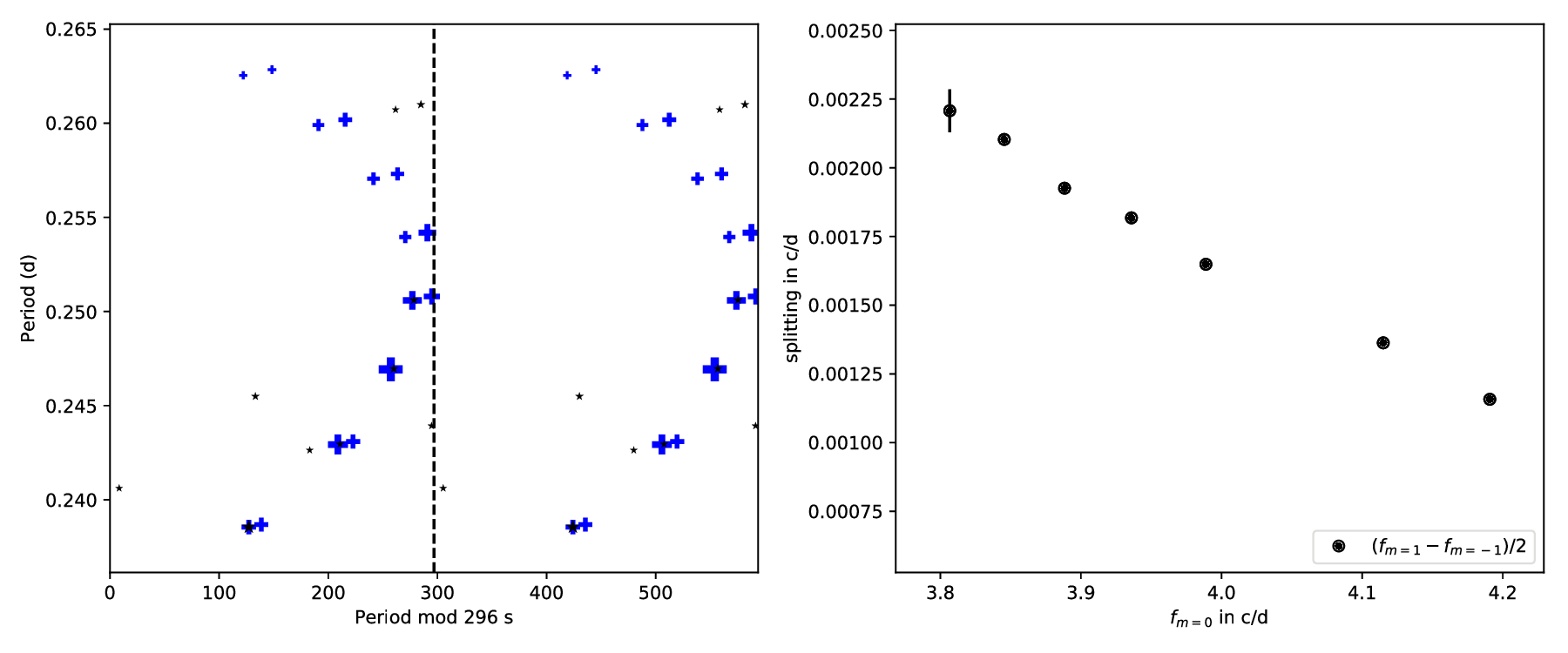}
\caption{The \'{e}chelle diagram and the splitting variations of KIC\,8458690. Note that the splittings are not caused by the rotational effect.}\label{fig:KIC 8458690echelle_splitting}
\end{figure*}

\clearpage
\begin{figure*}
\centering
\includegraphics[width=0.9\textwidth]{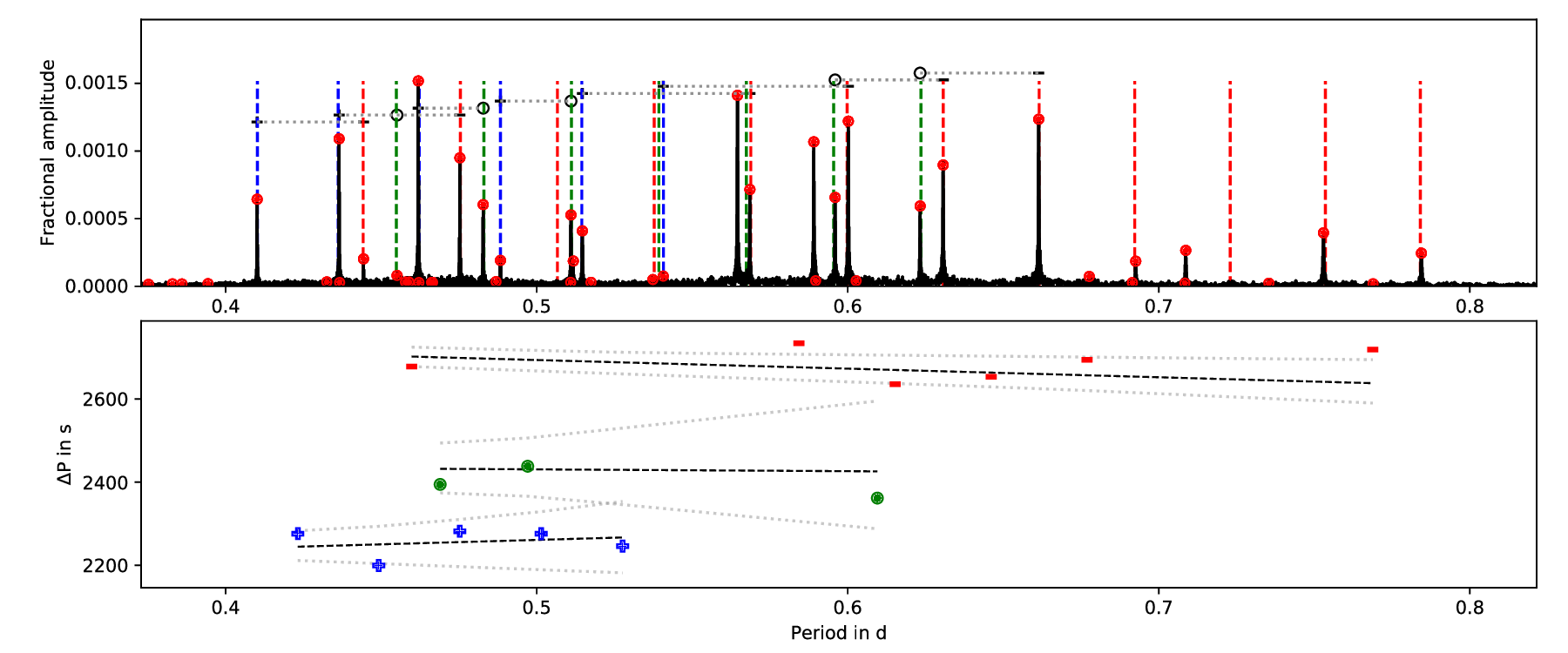}
\caption{The period spacing patterns of KIC\,9028134.}\label{fig:KIC 9028134}
\end{figure*}

\begin{figure*}
\centering
\includegraphics[width=0.85\textwidth]{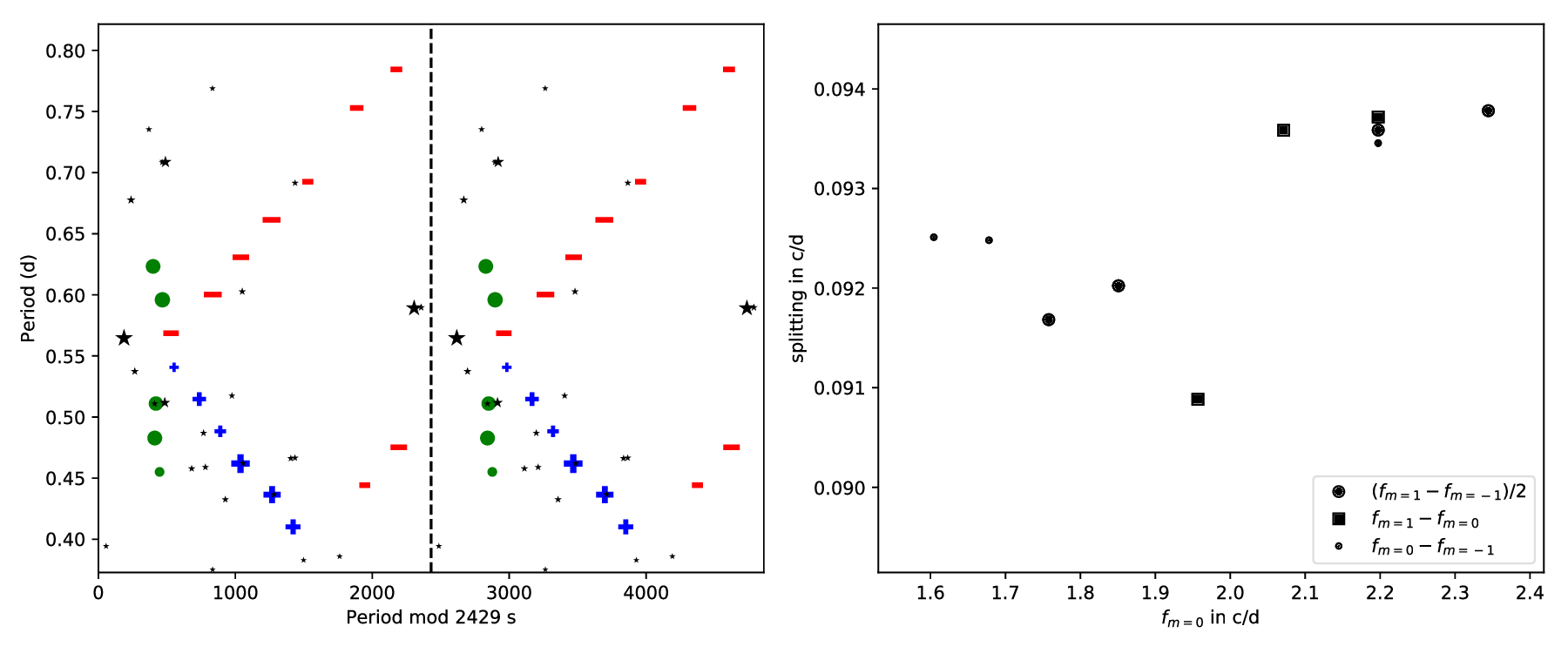}
\caption{The \'{e}chelle diagram and the splitting variations of KIC\,9028134.}\label{fig:KIC 9028134echelle_splitting}
\end{figure*}

\clearpage
\begin{figure*}
\centering
\includegraphics[width=0.9\textwidth]{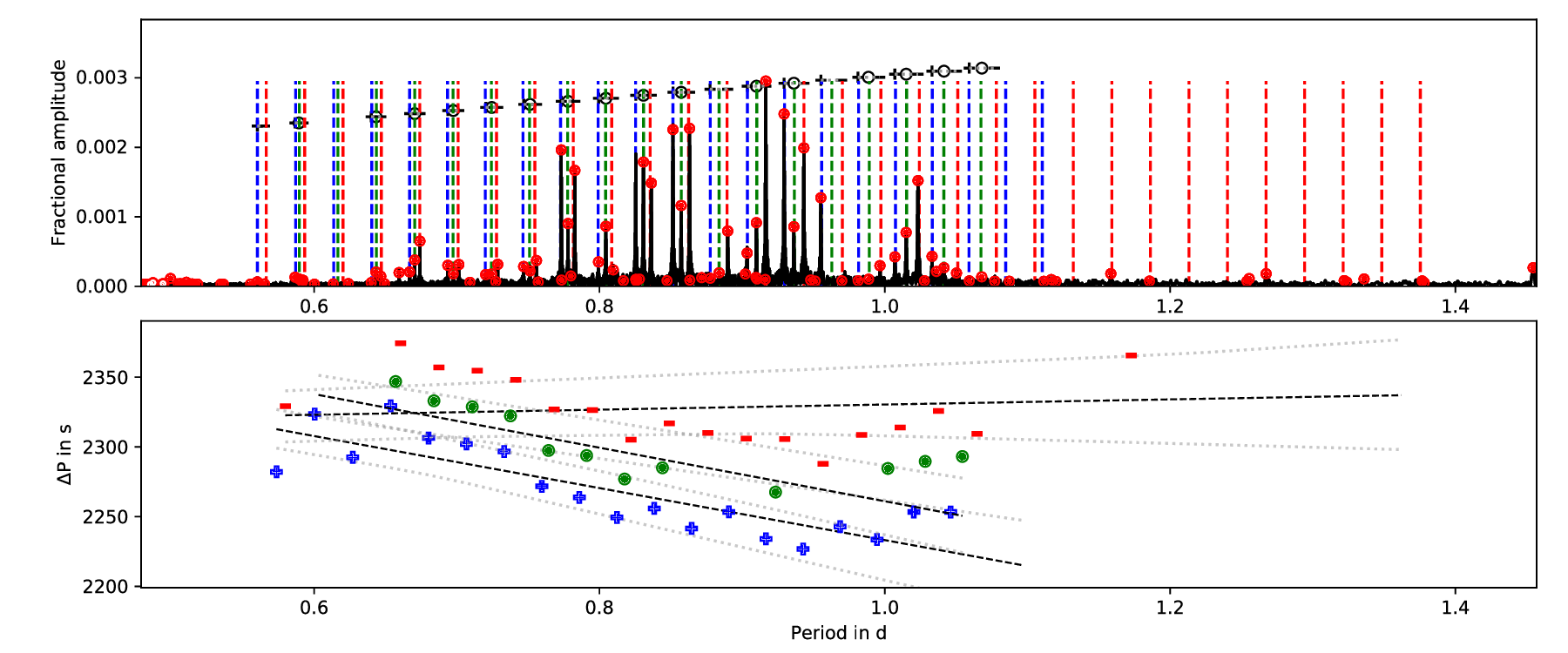}
\caption{The period spacing patterns of KIC\,9244992.}\label{fig:KIC 9244992}
\end{figure*}

\begin{figure*}
\centering
\includegraphics[width=0.85\textwidth]{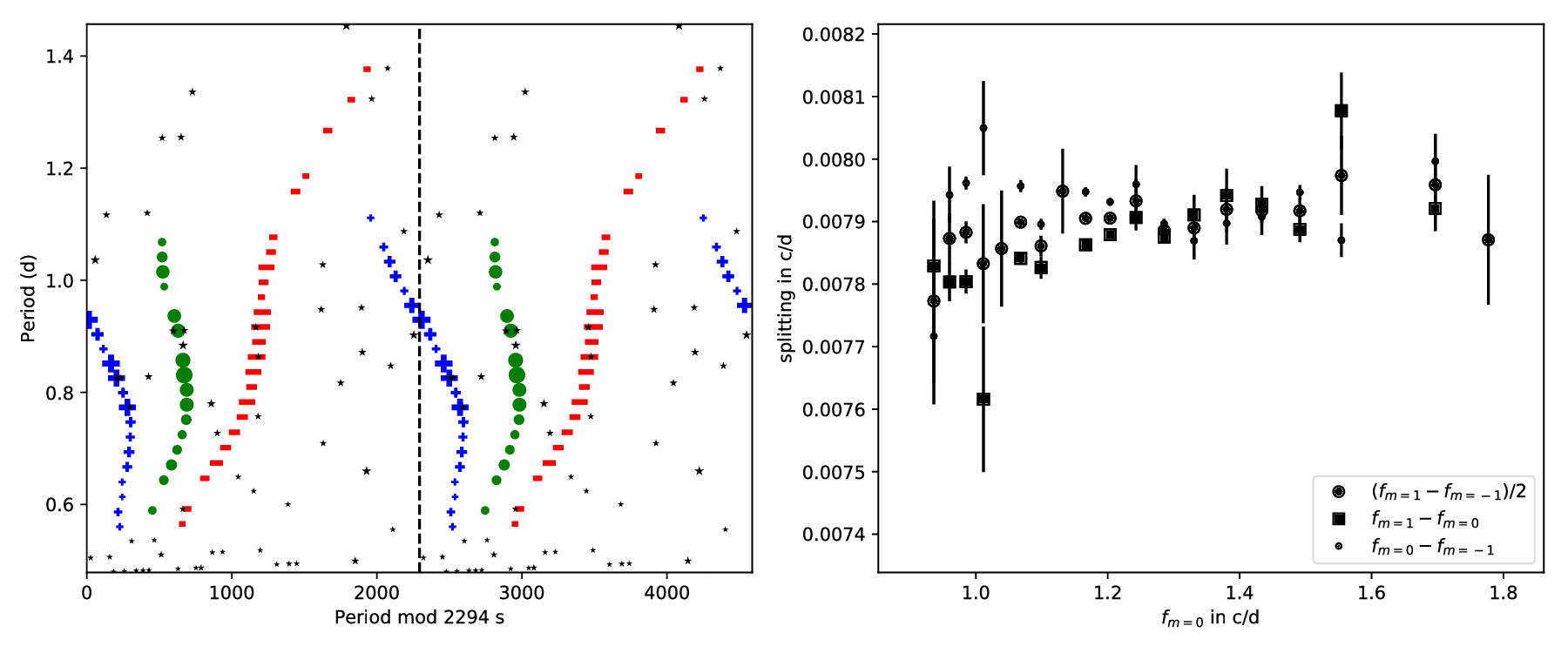}
\caption{The \'{e}chelle diagram and the splitting variations of KIC\,9244992.}\label{fig:KIC 9244992echelle_splitting}
\end{figure*}

\clearpage
\begin{figure*}
\centering
\includegraphics[width=0.9\textwidth]{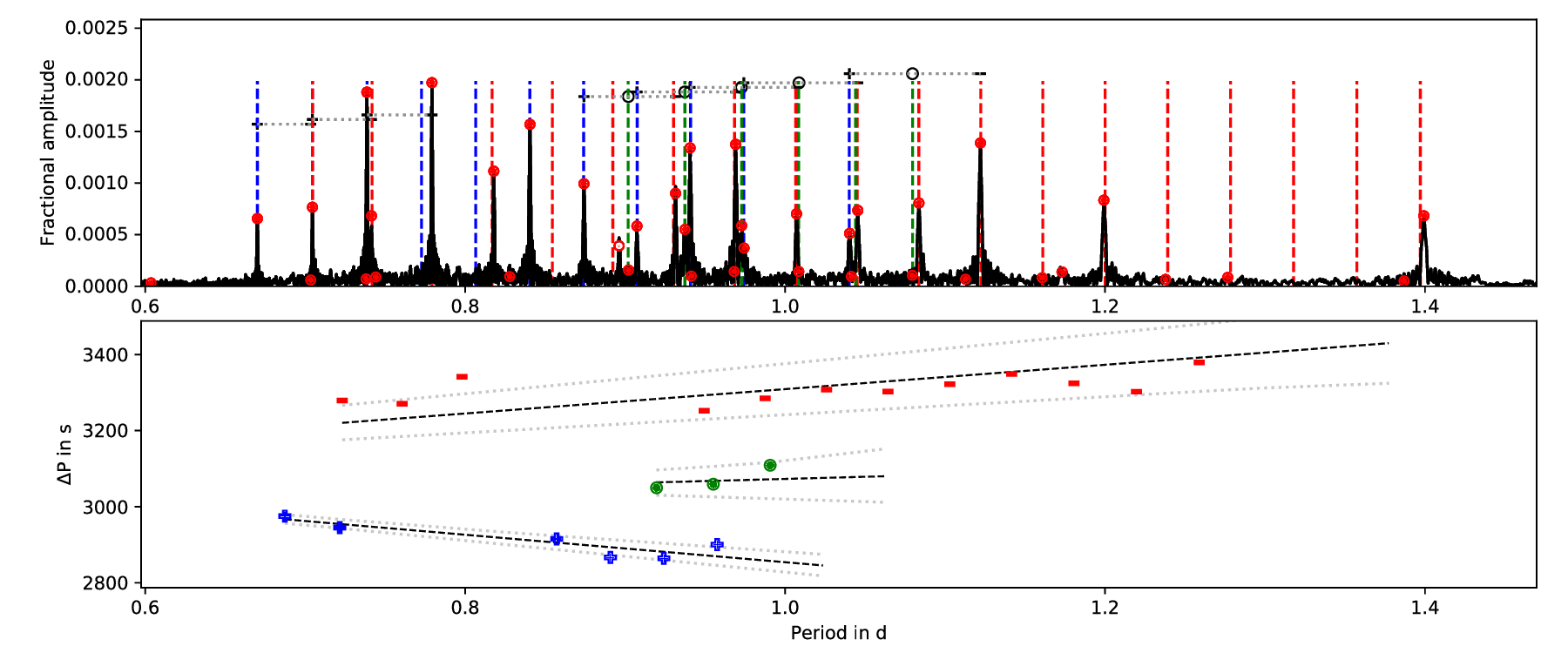}
\caption{The period spacing patterns of KIC\,9751996.}\label{fig:KIC 9751996}
\end{figure*}

\begin{figure*}
\centering
\includegraphics[width=0.85\textwidth]{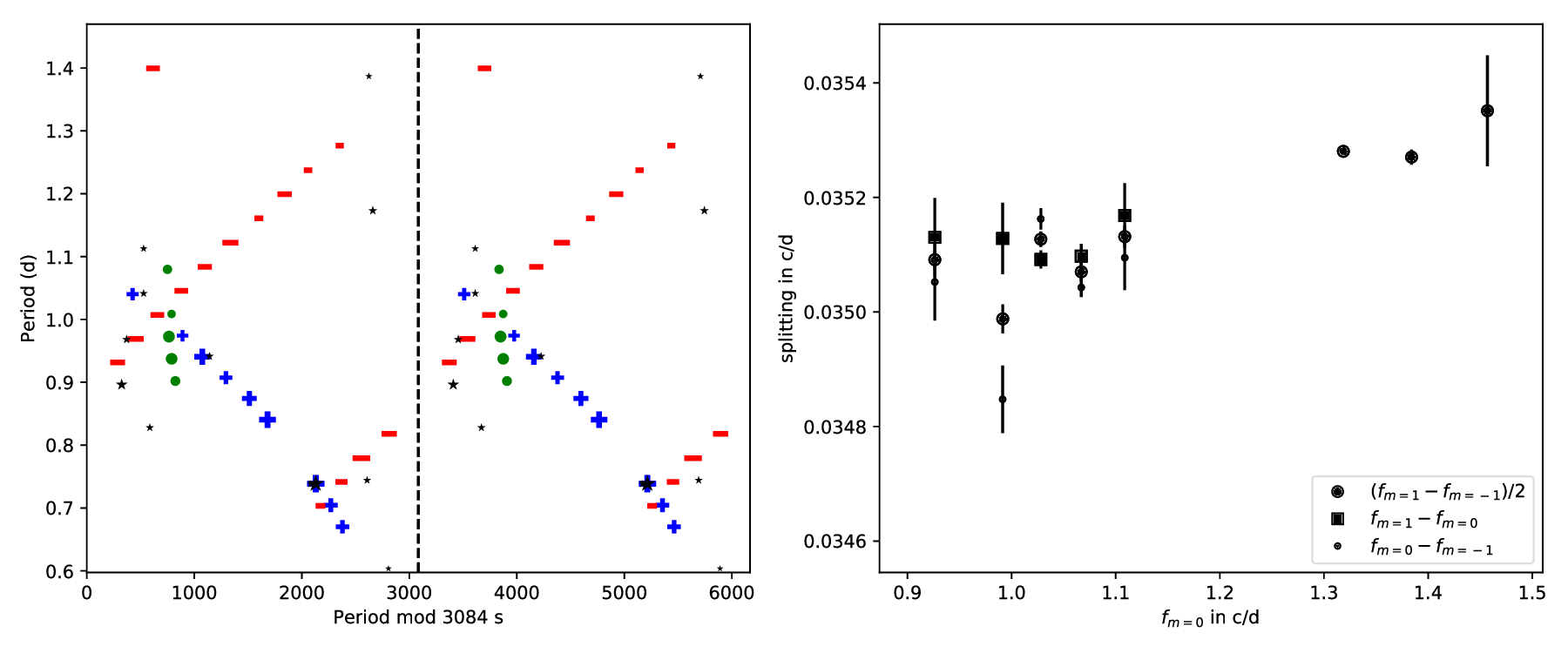}
\caption{The \'{e}chelle diagram and the splitting variations of KIC\,9751996.}\label{fig:KIC 9751996echelle_splitting}
\end{figure*}

\clearpage
\begin{figure*}
\centering
\includegraphics[width=0.85\textwidth]{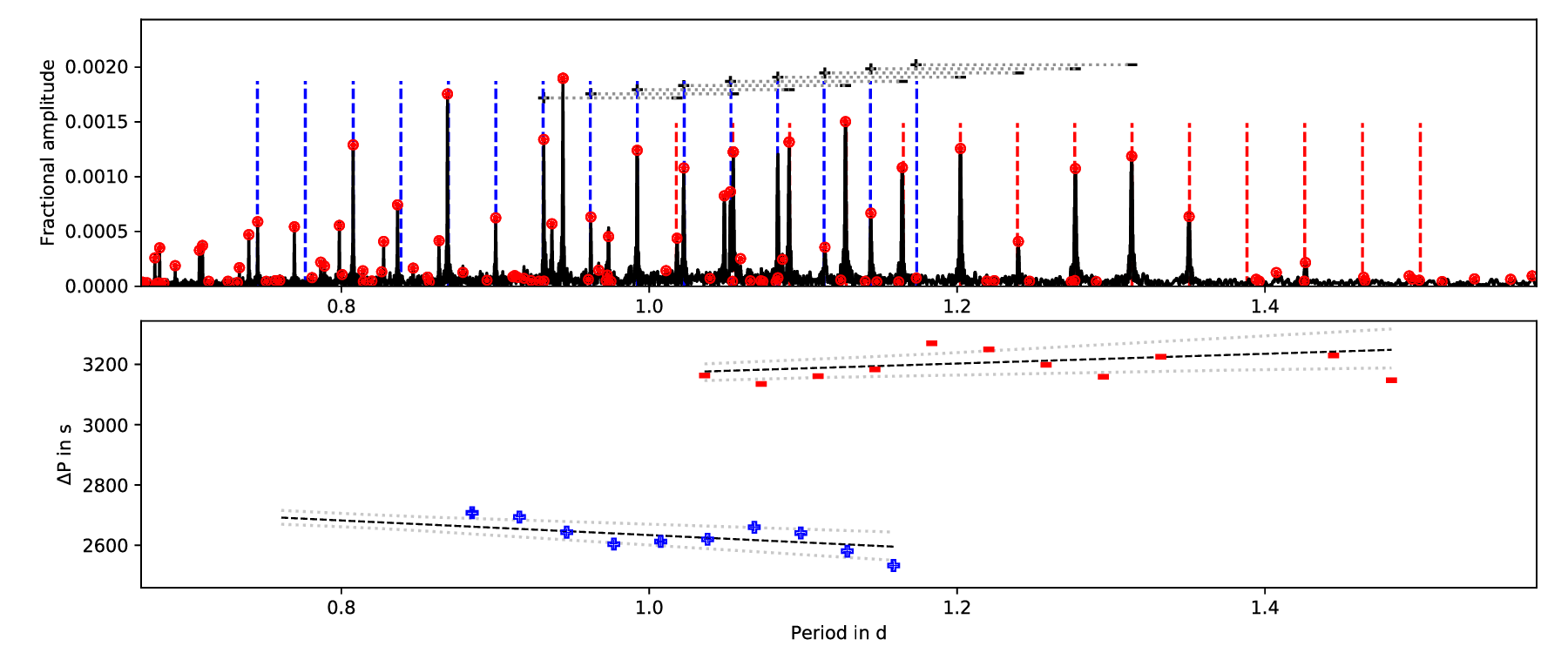}
\caption{The period spacing patterns of KIC\,10080943(B).}\label{fig:KIC 10080943Bechelle_splitting}
\end{figure*}

\begin{figure*}
\centering
\includegraphics[width=0.85\textwidth]{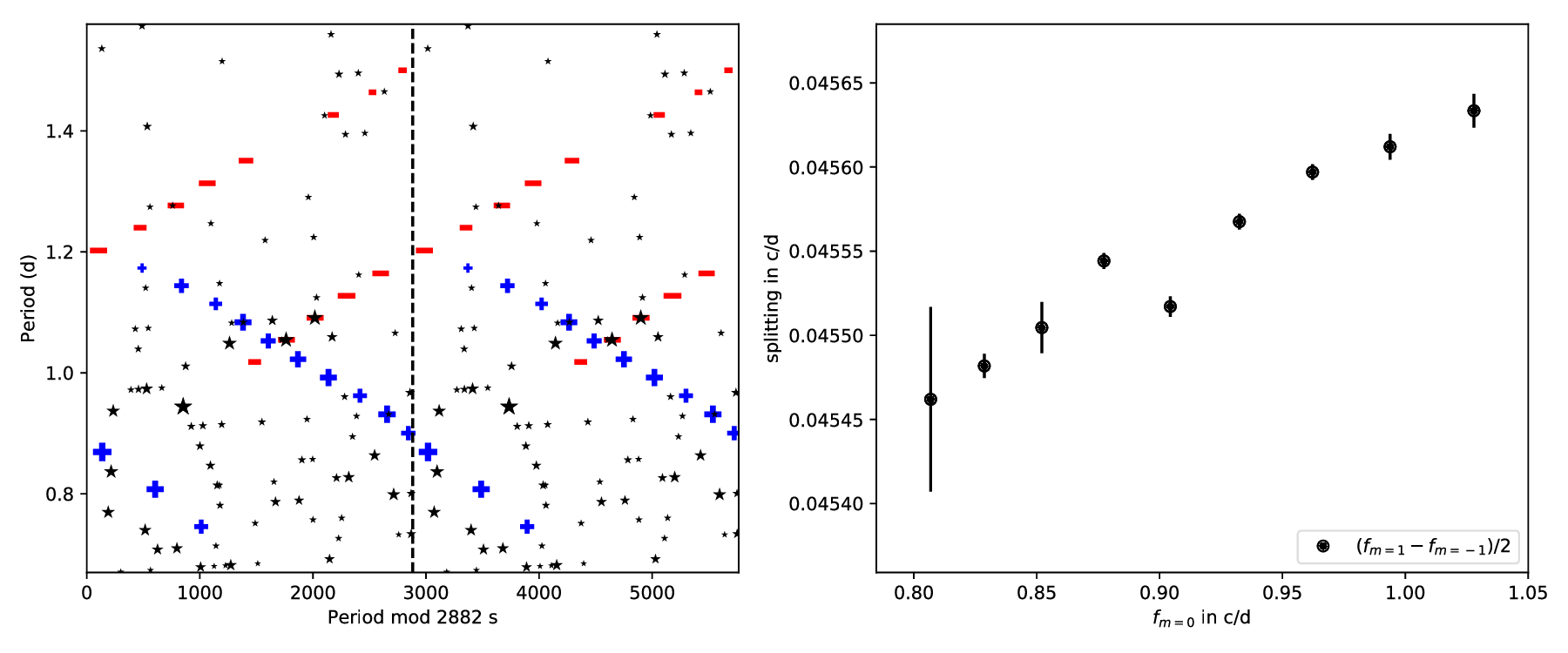}
\caption{The \'{e}chelle diagram and the splitting variations of KIC\,10080943(A).}\label{fig:KIC 10080943Aechelle_splitting}
\end{figure*}

\clearpage
\begin{figure*}
\centering
\includegraphics[width=0.85\textwidth]{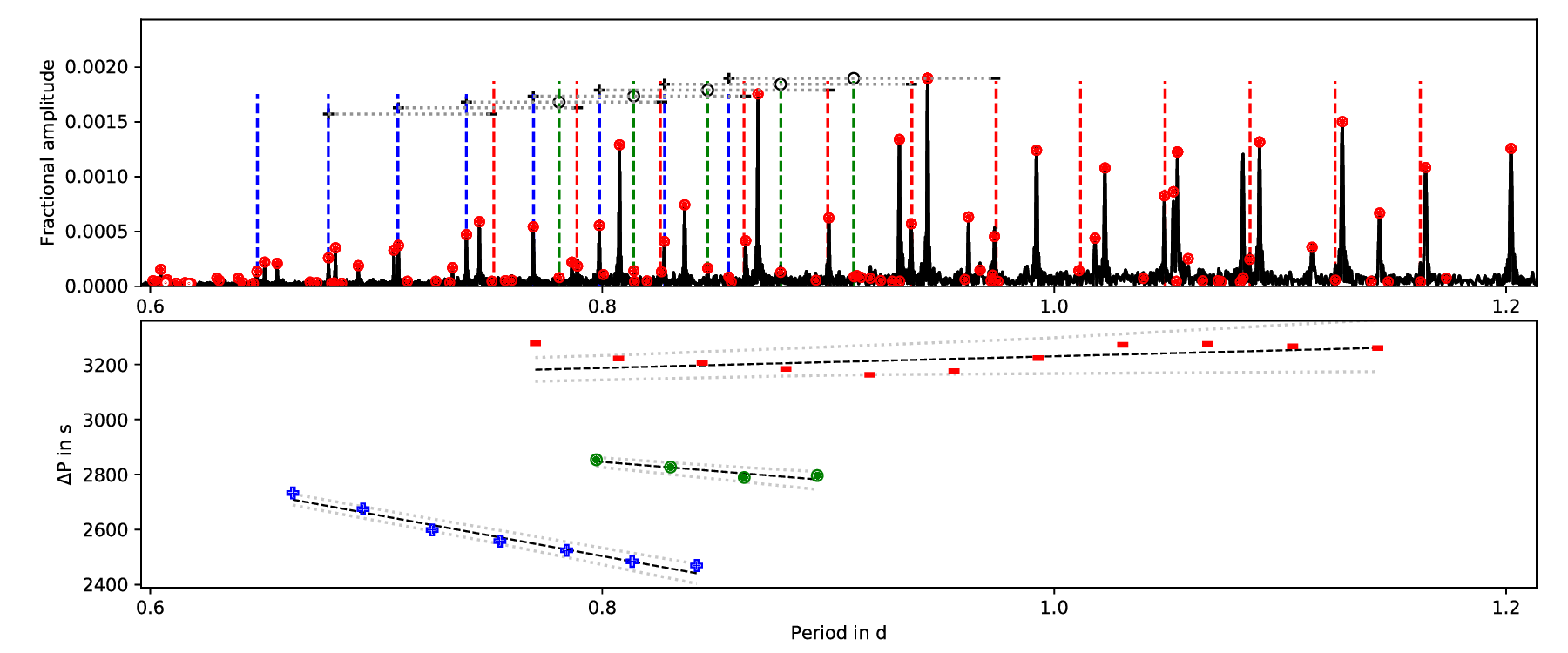}
\caption{The period spacing patterns of KIC\,10080943(B).}\label{fig:KIC 10080943Bechelle_splitting}
\end{figure*}

\begin{figure*}
\centering
\includegraphics[width=0.85\textwidth]{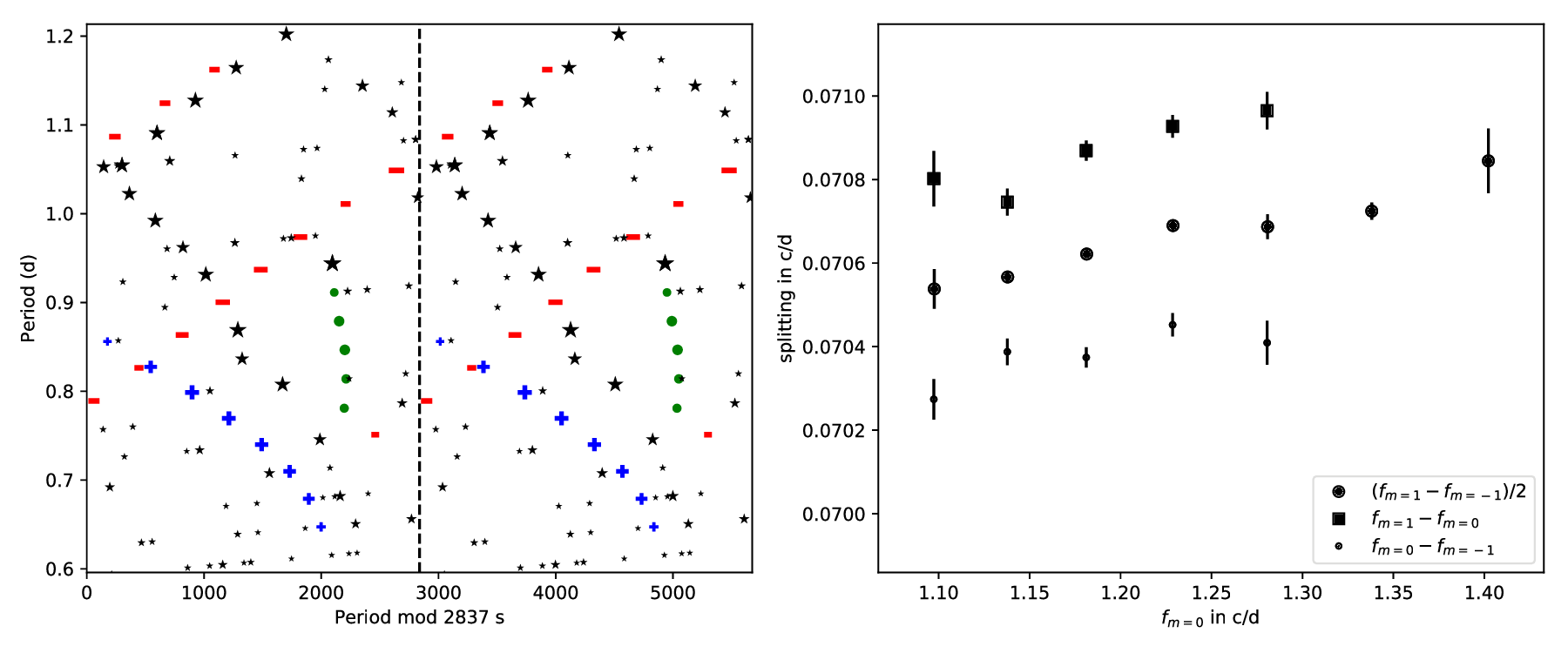}
\caption{The \'{e}chelle diagram and the splitting variations of KIC\,10080943(B).}\label{fig:KIC 10080943Aechelle_splitting}
\end{figure*}

\clearpage
\begin{figure*}
\centering
\includegraphics[width=0.9\textwidth]{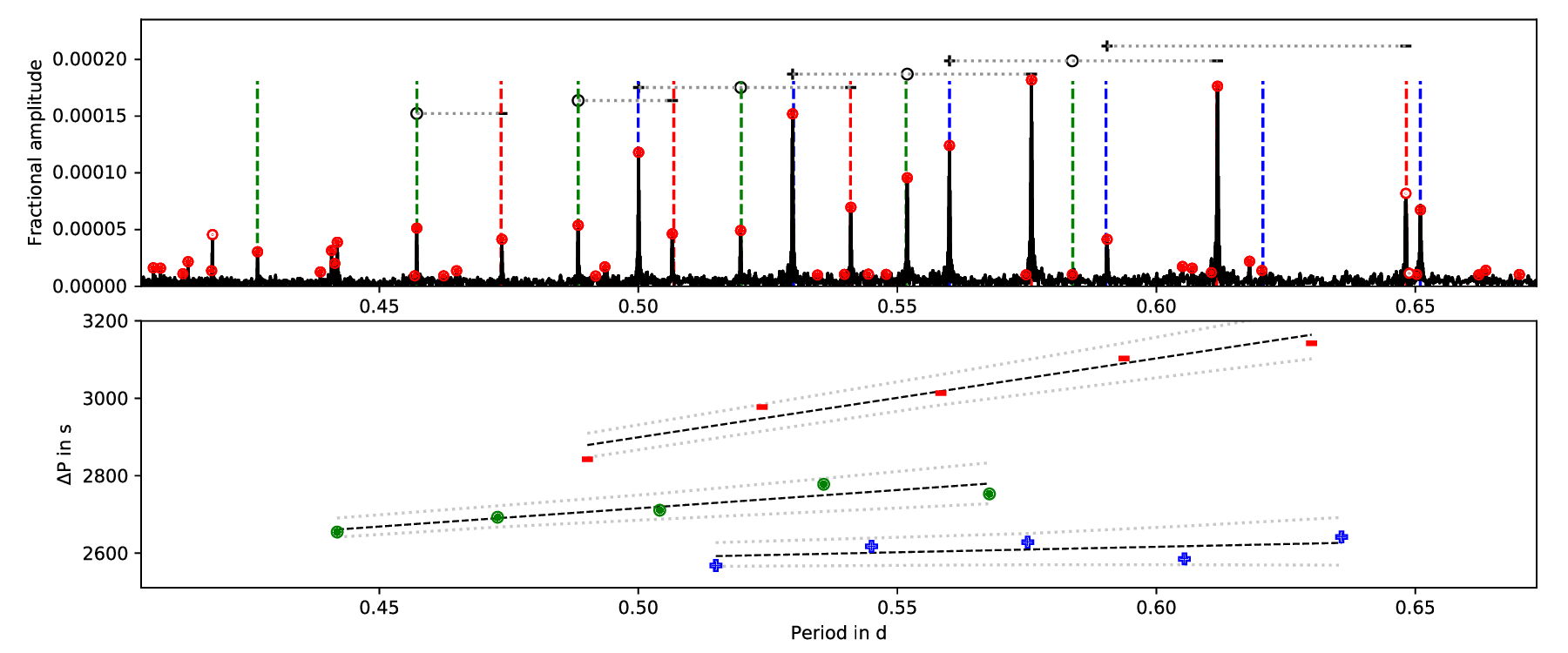}
\caption{The period spacing patterns of KIC\,10468883.}\label{fig:KIC 10468883}
\end{figure*}

\begin{figure*}
\centering
\includegraphics[width=0.85\textwidth]{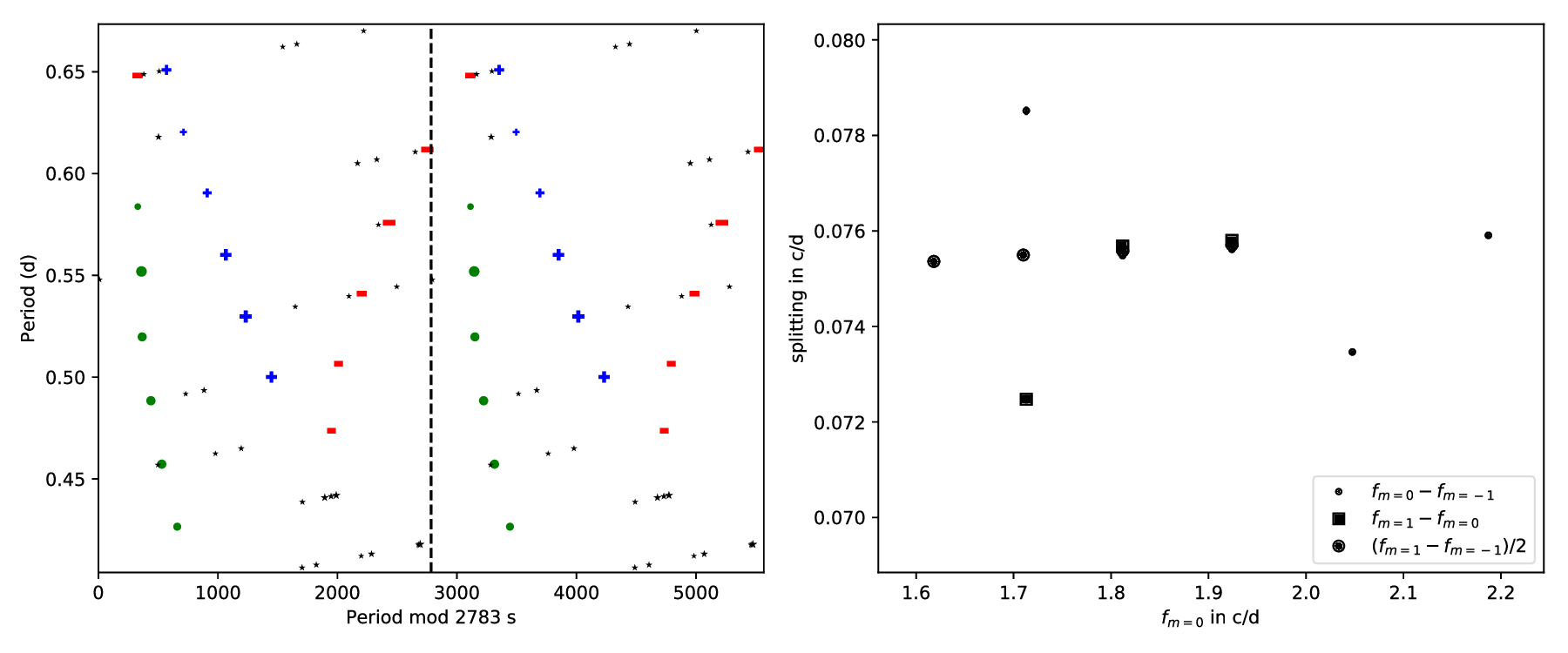}
\caption{The \'{e}chelle diagram and the splitting variations of KIC\,10468883.}\label{fig:KIC 10468883echelle_splitting}
\end{figure*}

\clearpage
\begin{figure*}
\centering
\includegraphics[width=0.9\textwidth]{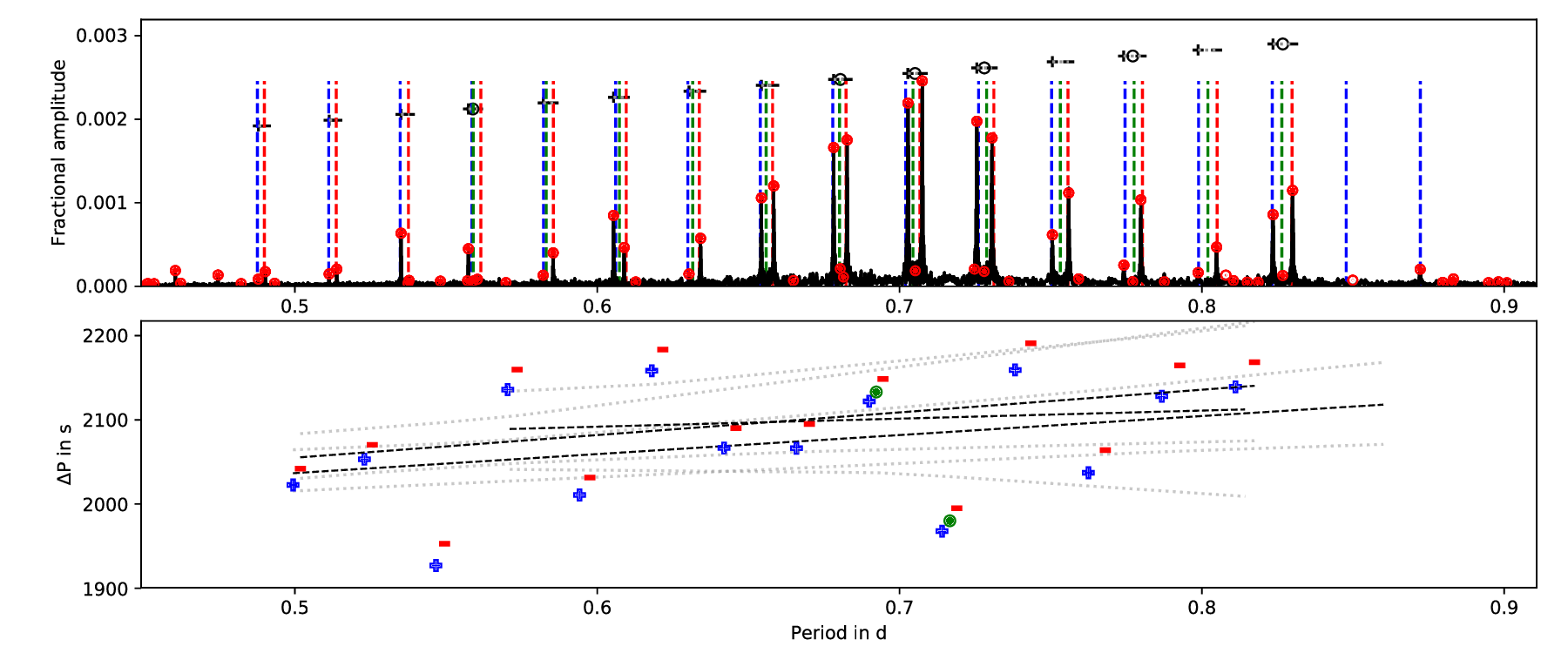}
\caption{The period spacing patterns of KIC\,11145123.}\label{fig:KIC 11145123}
\end{figure*}

\begin{figure*}
\centering
\includegraphics[width=0.85\textwidth]{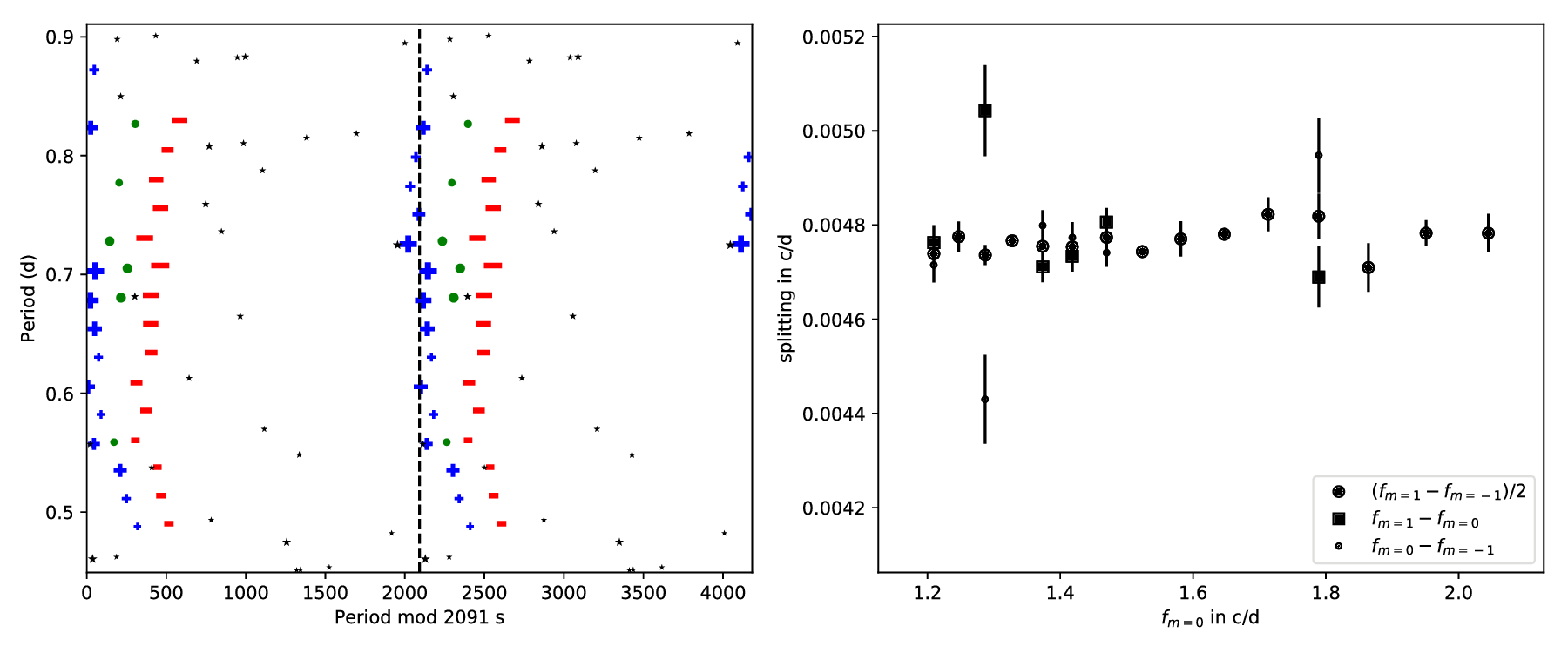}
\caption{The \'{e}chelle diagram and the splitting variations of KIC\,11145123.}\label{fig:KIC 11145123echelle_splitting}
\end{figure*}

\clearpage

\label{lastpage}

\end{document}